\documentclass[twocolumn,times,twocolappendix]{aastex631}
\usepackage{amsmath}
\usepackage{hyperref}

\graphicspath{{./}{figures_pdf/}}

\received{May 28, 2021}
\revised{August 9, 2021}
\accepted{August 16, 2021}
\submitjournal{Astrophysical Journal}

\shorttitle{Emission Line Diversity of Seyfert 2s and LINERs}
\shortauthors{Agostino et al.}

\begin{document}
\title{Physical Drivers of Emission Line Diversity of SDSS Seyfert 2s and LINERs After Removal of Contributions by Star Formation}

\author[0000-0002-9505-3310]{Christopher J. Agostino}
\affil{Department of Astronomy, Indiana University, Bloomington, IN, 47405, USA}

\author[0000-0003-2342-7501]{Samir Salim}
\affil{Department of Astronomy, Indiana University, Bloomington, IN, 47405, USA}

\author[0000-0003-4996-214X]{S.M. Faber}
\affil{Department of Astronomy and Astrophysics, University of California, Santa Cruz, CA, 95064, USA}

\author[0000-0002-0000-2394]{St\'ephanie Juneau}
\affil{NOIRLab, 950 N. Cherry Avenue, Tucson, AZ, 85719, USA}

\author[0000-0003-3385-6799]{David C. Koo}
\affil{Department of Astronomy and Astrophysics, University of California, Santa Cruz, CA, 95064, USA}

\author[0000-0003-2876-577X]{Yimeng Tang }
\affil{Key Laboratory for Research in Galaxies and Cosmology, Department of Astronomy,
University of Science and Technology of China, Hefei 230026, China}
\affil{School of Astronomy and Space Science, University of Science and Technology of China, Hefei 230026, China}

\author[0000-0001-7729-6629]{Yifei Luo}
\affil{Department of Astronomy and Astrophysics, University of California, Santa Cruz, CA, 95064, USA}
 
\author{Sofia Quiros}
\affil{Lakeridge High School, Lake Oswego, OR, 97034, USA}
\affil{Department of Astronomy, Columbia University, New York, NY, 10027, USA}

\author[0000-0002-4328-538X]{Pin-Song Zhao}
\affil{CAS Key Laboratory of Optical Astronomy, National Astronomical Observatories, Chinese Academy of Sciences, Beijing 100101, People’s Republic of China}
\affil{School of Astronomy and Space Science, University of Chinese Academy of Sciences, Beijing 100049, People’s Republic of China}

\begin{abstract}
Ionization sources other than HII regions give rise to the right-hand branch in the standard ([NII]) BPT diagram, populated by Seyfert 2s and LINERs. However, because the majority of Seyfert/LINER hosts are star forming (SF), HII regions contaminate the observed lines to some extent, making it unclear if the position along the branch is merely due to various degrees of mixing between pure Seyfert/LINER and SF, or whether it reflects the intrinsic diversity of Seyfert/LINER ionizing sources. In this study, we empirically remove SF contributions in $\sim$100,000 Seyfert/LINERs from SDSS using the doppelganger method. We find that mixing is not the principal cause of the extended
morphology of the observed branch. Rather, Seyferts/LINERs intrinsically have a wide range of line ratios. Variations in ionization parameter and metallicity can account for much of the diversity of Seyfert/LINER line ratios, but the hardness of ionization field also varies significantly. Furthermore, our $k$-means classification on seven decontaminated emission lines reveals that LINERs are made up of two populations, which we call soft and hard LINERs. The Seyfert 2s differ from both types of LINERs primarily by higher ionization parameter, whereas the two LINER types mainly differ from each other (and from star-forming regions) in the hardness of the radiation field. We confirm that the [NII] BPT diagram more efficiently identifies LINERs than [SII] and [OI] diagnostics, because in the latter many LINERs, especially soft ones, occupy the same location as pure star-formers, even after the SF has been removed from LINER emission.
 \end{abstract}


\section{Introduction} \label{sec:intro}

Active Galactic Nuclei (AGNs) may play an important role in the evolution of galaxies through feedback \citep{bower2006_feedback, cattaneo2009_feedback,harrison2017_feedback_review}, a process wherein the accretion of matter onto the central supermassive black hole (SMBH) imparts significant energy into the gas of the host galaxy and circumgalactic medium. AGN feedback has been proposed as a mechanism responsible for the triggering or quenching of star formation \citep{fabian2012, ishibashi2012, silk2013, zinn2013, dubois2016,chen2020, stemo2020, yao2021}, the exponential cutoff in the galaxy luminosity function \citep{best2006, bower2006_feedback, croton2006}, and regulating the accretion itself \citep{croton2006, cattaneo2009_feedback, fabian2012}. As a result, being able to securely identify AGN and study their intrinsic properties is of paramount importance for galaxy evolution and form the principal themes of the present study.

\begin{figure}[t!]
        \begin{center}
            \epsscale{1.25}
            \plotone{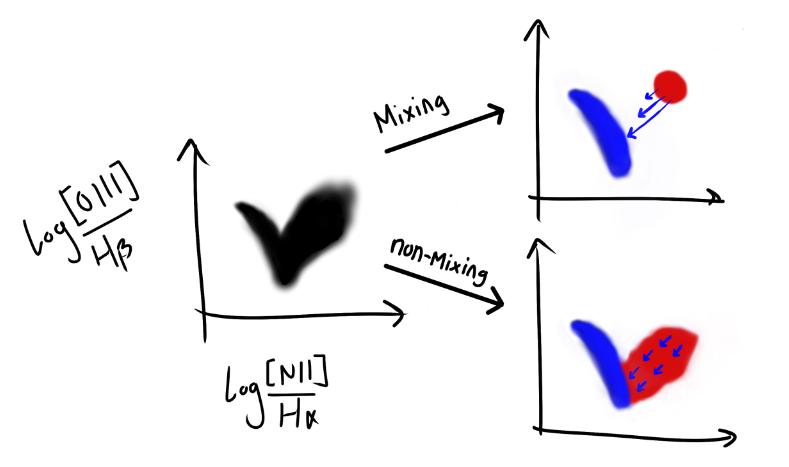}
            \caption{Cartoon showing the observed BPT diagram on the left and two scenarios for what an intrinsic BPT might look on the right. The intrinsic BPT is one where the contamination from SF is removed from the right branch. In the upper right panel, we show the mixing scenario in which galaxies on the right branch (in particular, Seyfert 2s) occupy a red patch at the top of the branch and with arrows of various lengths showing that increasing levels of contamination by star formation will push the Seyferts (and/or LINERs) by various degrees down the branch, producing the observed diagram. In the lower right panel, mixing does not dominate; Seyferts and/or LINERs intrinsically occupy most or entire right branch and are affected by mixing insomuch as they are systematically shifted towards the left (SF) branch.
            \label{fig:cartoon} Artistic rendering courtesy of MacCallum Robertson.}
        \end{center}
        \end{figure}

Type 1 AGNs, including Seyfert 1s, quasars and BL Lac objects, are relatively easy to identify owing to their broad Balmer lines and blue continua, but are relatively rare and therefore do not go a long way in filling the AGN census, at least locally. For the more common type 2 AGNs, which have neither the broad Balmer lines nor a blue continuum, a commonly used method of identification is to look at narrow-line ratios that may indicate the presence of some source of \emph{extra ionization}, one in addition to that arising from star formation (SF) in HII regions. \citet{bpt1981} introduced a variety of emission line diagnostics, but is best known for a diagram which is meant to separate the emission from HII regions from that where the source of ionization is distinct from HII regions (presumably from an AGN) or perhaps in addition to them. In particular, what is usually referred to as the Baldwin-Phillips-Terlevich (BPT) diagram uses the emission line ratios [OIII]$\lambda$5007/H$\beta$ and [NII]$\lambda$6583/H$\alpha$ in order to determine the dominant source of ionization in nebular gas. We show a cartoon version of the observed BPT diagram in Figure \ref{fig:cartoon}, with its two main features: the left branch which is dominated by ionization from HII regions and is therefore referred to as the SF branch, and the right branch, which must have some other source of  ionization, potentially in combination with HII regions. 

The right branch of the BPT diagram consists of Seyfert 2s, which are securely known to owe this extra ionization to an accreting SMBH, i.e, an AGN,  as well as objects objects known as Low Ionization Nuclear Emission Regions \citep[LINERs, ][]{heckman1980}, for which the mechanism which powers this extra ionization is still  debated. LINERs were originally proposed to be the result of low-luminosity AGN \citep{heckman1980}, but recently various researchers expressed doubts whether the low-luminosity accretion can actually provide enough energy to ionize the gas on the scales probed by fiber spectroscopy. This concern has prompted an alternative explanation wherein the emissions in LINERs are produced by a stellar mechanism, such as hot post-AGB stars (e.g., \citealt{stasinska2008,belfiore2016}). Regardless of the nature of the dominant sources of ionization in LINERs, they certainly contain what we refer to as the extra ionization (compared to  HII regions). This ionization ends up placing them on the right branch of the BPT diagram, which we will heretofore refer to as the Seyfert 2+LINER (S/L) branch. The goal of this work is to investigate the factors which control the observed and the intrinsic line ratios of galaxies on the S/L branch, regardless of whether the LINER emission is due to an AGN or not.

While the position along the SF sequence is well understood to be due to the intrinsic variety in the physical conditions of star forming regions (gas-phase metallicity, ionization parameter, electron density, and the hardness of the radiation field \citep{kewley2001b, nagao2006, kewley2006, stasinska2006, levesque2010, sanchez2015, kewley2019review}, an analogous clarity is absent for the S/L branch. Early studies of nearby Seyferts and LINERs demonstrated some variety among their line ratios (e.g., \citealt{vo87,  ho1993a, ho1993c}), and such diversity was in part attributed to intrinsic variations in the Seyferts/LINERs themselves. Modeling efforts (e.g., \citealt{ho1993a}) made attempts to pinpoint the physical drivers of such variations but degeneracies rendered definitive conclusions elusive. The advent of the Sloan Digital Sky Survey spectroscopic survey \citep[SDSS, ][]{york2000, strauss2002} has significantly increased the number of galaxies that can be studied using emission line diagrams, and has revealed the morphology of the S/L branch which is fully connected to the SF branch \citep{kauffmann2003}. The interpretation of this fact, however, has remained unclear. Namely, with the SDSS 3$''$ fiber, the typical physical size of an object covered at $z\sim0.1$ is $\sim$5 kpc, whereas the size of the NLR can range from 1-10 kpc depending on how luminous the AGN is \citep{bennert2006a_nlr, bennert2006b_nlr, greene2011_nlr, liu2013_nlr, liu2014_nlr, law2018_nlr, chen2019nlr}. Thus, the emissions from the NLR will be combined with the emissions of the star formation contained within the fiber, which on average contains $\sim$1/4-1/5 of the galaxy's total star formation \citep{puertas2017}. Given how centrally concentrated the star formation can be (e.g., \citealt{ellison2020, frasermck2020}), host galaxy contamination will be present to some degree even in slit spectra and the central pixels of IFU spectra, especially if taken without adaptive optics. Indeed, the typical resolution of a single MaNGA fiber is not much smaller than the original SDSS fiber size. The potential contamination of the emission lines of the objects on the S/L branch by stellar emission, and in particular by SF, poses a principal obstacle in the study of these objects, and its estimation and removal are the main aims of this study. We point out, and we will further highlight this fact, that both the Seyfert 2s and the LINERs can be found in hosts with significant SF, so that the majority of the objects on the S/L branch would benefit from the removal of such contamination.

The critical question can therefore be posed as: to what degree what we see as the S/L branch in modern spectroscopic surveys is the result of various amounts of contamination (mixing) of pure S/L emission lines with those from star formation as opposed to being the result of the intrinsic variation? We illustrate the two scenarios in Figure \ref{fig:cartoon}. In the mixing scenario (e.g., \citealt{kewley2019review}), one expects ``pure'' Seyferts 2s and LINERs (but in particular Seyferts) to be intrinsically confined to the top or the upper portions of the branch, but as the contribution of host star formation increases, so too would the displacement from the intrinsic region towards the SF branch, which is represented in Figure \ref{fig:cartoon} by the increasing lengths of the arrows. The result of the different degrees of SF contamination would be the observed (apparent) S/L branch (the ``mixing sequence'') that is connected to the SF branch at its base where most massive galaxies lie. Some basis for assuming the dominant role of the mixing scenario may be sought in pre-SDSS studies, such as \citet{ho1993a} and \citet{ho1993c}, in which the long-slit spectra of nearby AGN, less subject to mixing than SDSS fibers because the physical scales are smaller and include less non-nuclear light, appear to populate a portion of the BPT diagram separated from the star-forming locus. \citet{kewley2001a} considered emission line mixing as the principal cause of different line ratios in an analogy with the SF/AGN mixing of IR emission. Subsequently, the mixing picture\footnote{It should be noted that we are not focused here on spatially resolved individual galaxies and the mixing sequences which may be observed as a function of distance in IFU studies (e.g., \citet{davies2016, dagostino2019a, dagostino2019b, pilyugin2020}).} has been prevalent in the literature \citep{kewley2001a, kauffmann2003, kewley2006, stasinska2006, kh09, kewley2013cosmicbpt, kewley2013optlines}.

A conceptual alternative to the mixing scenario is that the diversity in line ratios seen on the S/L branch is to a great extent intrinsic. Seyfert 2s and/or LINERs could have a variety of intrinsic ionizing source parameters or NLR conditions that primarily determine the location on the BPT diagram, whereas the SF contamination that systematically shifts the galaxies in the S/L branch towards the SF branch would be a secondary effect (Figure \ref{fig:cartoon}).  A number of studies have shown that AGN photoionization models can adequately cover the span of observed line ratios on the S/L branch with no gap with respect to the star-forming branch \citep{ho1993a, ho1993b, ho1993c, groves2004, richardson2014, davies2016, feltre2016, perezmontero2019, flury2020,ji2020a, ji2020b}. Consequently, it would not be justified to exclude this scenario a priori.

While we know that contamination must affect to some degree the location of at the majority of SDSS Seyferts and LINERs on the BPT diagram, so far it has not been empirically established how large this effect is, and whether it is indeed responsible for producing the entire extent of the S/L branch, rather than just giving it some offset. In this work, we statistically disentangle the star-forming component in the emission lines and derive pure Seyfert/LINER emissions by empirically removing the contributions of the host galaxy due to star formation and diffuse ionized gas, if present, in order to test the mixing picture and/or study the intrinsic properties of Seyfert/LINERs.

In Section \ref{sec:sample}, we describe our galaxy samples and selections. We describe our method for removing the stellar ionizing emission in Section \ref{sec:matching} and our method for determining the dust corrections in Section \ref{sec:dust}. We present the BPT diagram for pure Seyfert 2s and LINERs in Section \ref{sec:bpt_gen} and evaluate the mixing scenario versus the non-mixing scenario. In Section \ref{sec:agntypes} we present evidence that the population on the S/L branch may be composed of three populations with distinct line ratios, rather than just Seyferts and LINERs. We then explore how the position on the BPT diagram might depend on its ionizing source properties in Section \ref{sec:ion}. We present the positions of ``pure'' (with host contributions removed from emission lines) Seyferts and LINERs  on other commonly used emission line diagnostic diagrams in Section \ref{sec:otherdiags}. In Section \ref{sec:discussion}, we discuss the results of our study in the context of the literature and the implications for future studies of SDSS AGNs.

\section{Sample and Data}  \label{sec:sample}
In this work, we aim to study the optical emission line properties of pure Seyfert 2s and LINERs in SDSS. By pure, we mean the ones in which the contribution from a host galaxy's SF to their emission lines is removed, revealing only the emission lines of the extra ionization source (an AGN, or some other phenomenon that can place a galaxy on the S/L branch of the BPT diagram). We will continue to refer only to the [NII]-based emission line diagram as the BPT diagram. We empirically estimate the contamination due to SF by using the emission lines of non-S/L galaxies of similar global properties. Matching of galaxies on the S/L branch, or various types of AGN, to their non-active doppelgangers based on host properties has been used successfully in many SDSS studies (e.g., \citealt{pasquali2005,pace2014}). Note that due to its large sample size, the legacy SDSS spectroscopic survey is much better suited for our method than the MaNGA sample.

\subsection{Data}
The measurements pertaining to the spectrum or the fiber---line fluxes, the mass contained within the spectroscopic fiber, and stellar velocity dispersion---come from the MPA/JHU catalog \footnote{https://wwwmpa.mpa-garching.mpg.de/SDSS/DR7/}, based on the DR7 of SDSS \citep{abazajian2009}, and derived following \citet{tremonti2004}. The global galaxy properties used in this work come from the second release of the GALEX-SDSS-WISE Legacy Catalog \citep[GSWLC\footnote{https://salims.pages.iu.edu/gswlc/}, ][]{salim2016, salim2018}, a catalog of $\sim$700,000 optically selected galaxies with SFRs, stellar masses, and dust attenuations determined based on the SED fitting that combines UV/optical photometry with IR luminosities. IR luminosities were derived from mid-IR photometry that includes an empirical correction for AGN emission if present. The GSWLC provides galaxy parameters based on three levels of depth for GALEX UV observations: Shallow (GSWLC-A2), Medium (GSWLC-M2), and Deep (GSWLC-D2), and the three catalogs cover 88\%, 49\%, and 7\% of SDSS, respectively. Here we utilize the medium depth catalog, GSWLC-M2, because the SED fitting afforded by the deeper UV coverage provides more accurate estimates of SFR than available for the shallow coverage survey, GSWLC-A2, while still providing a much larger sample than the deep coverage survey GSWLC-D2. We limit our sample to those objects with good SED fit and remove type 1 AGNs (including Seyfert 1s), i.e., galaxies classified by the SDSS pipeline as `QSOs' based on their spectra, which effectively eliminates galaxies with emission line FWHMs greater than 500 km s$^{-1}$ \citep{sdssdr12}. The resulting galaxy sample contains $\sim$340,000 galaxies, and spans redshifts between 0.01 and 0.3.

\subsection{Selection of target and matching pool samples} \label{sec:agn_selection}

In order to derive the intrinsic emission line properties of Seyferts and LINERs (our target sample), we also need to identify galaxies for our matching pool (non-Seyfert/LINERs). In short, we select Seyferts and LINERs using the ([NII]) BPT diagram, and non-Seyfert/LINERs as galaxies other than them, including ones that have weak or no detectable emission lines.

\begin{figure}[t!]
        \begin{center}
            \epsscale{1.1}
             \plotone{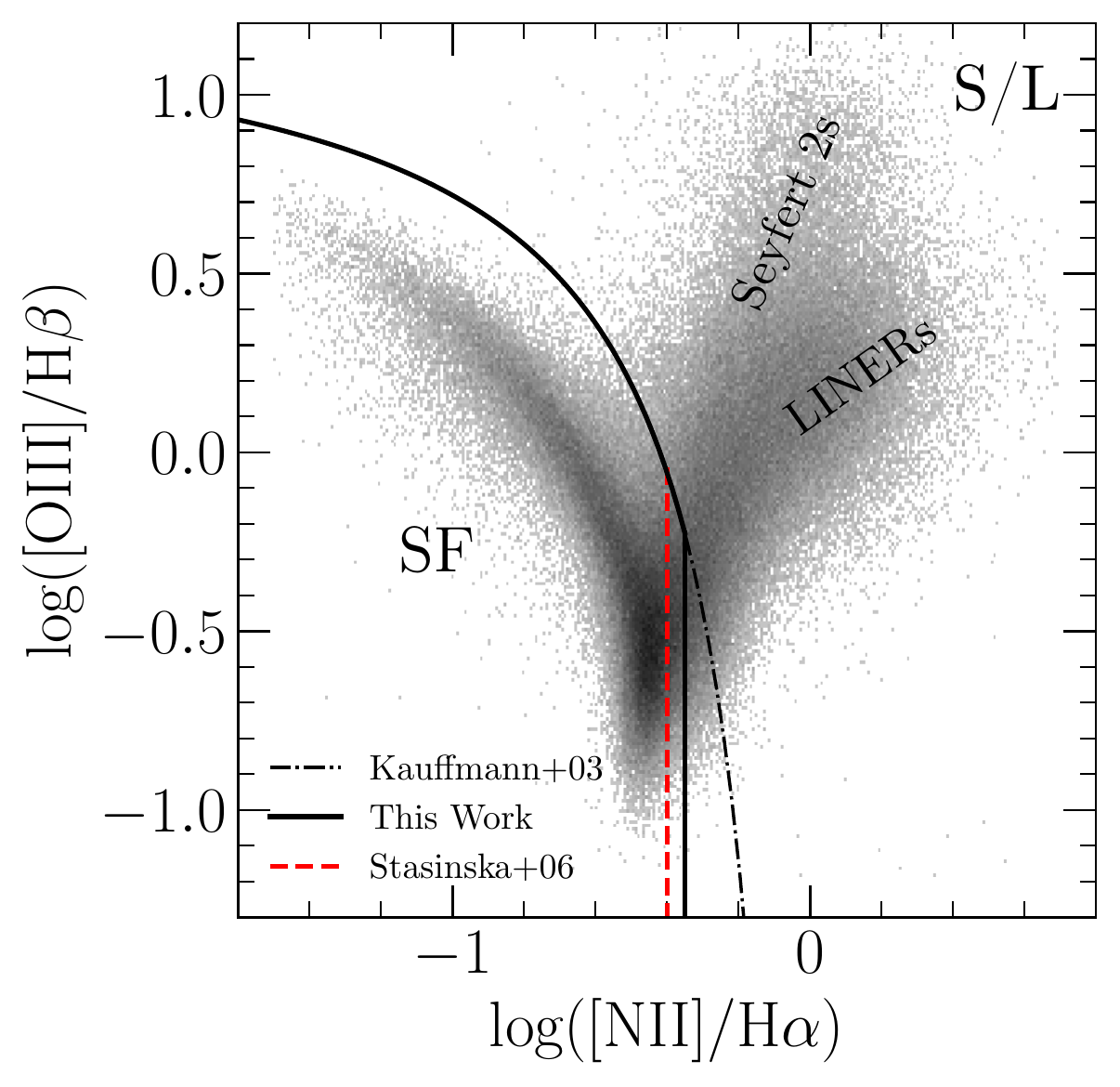}
            \caption{BPT diagram and the sample selection. Shown are $\sim200,000$ classifiable (S/N$>2$ in four emission lines) galaxies from GALEX-SDSS-WISE Legacy Catalog (GSWLC-M2). The solid black line is used to select our target sample (Seyferts and LINERs) and is the combination of the \citet{kauffmann2003} demarcation line in the upper part and a vertical cut at log([NII]/H$\alpha$)=-0.35 in the lower part. The dash-dotted line shows the continuation of the \citet{kauffmann2003} line that we do not use for sample selection. The log([NII]/H$\alpha)=-0.4$ cut proposed by \citet{stasinska2006} (red dashed line), again combined with the \citet{kauffmann2003} line is used to select a portion of the matching pool sample.  We point out that labeling the galaxies below the modified Kauffmann line as non-Seyfert/LINERs is more precise than labeling them SF, since SF galaxies are present on both branches, whereas the left branch are almost exclusively non-Seyfert/LINERs.  To create the greyscale distribution in this and other figures, we utilize a two-dimensional histogram, with the shading indicating the number density raised to the 0.3 power, in order to highlight outliers. \label{fig:bptplus_select}}
        \end{center}
        \end{figure}

The \citet{kauffmann2003} line, based on an empirical analysis of SDSS galaxies, separates galaxies between the two branches on the BPT diagram. We show GSWLC-M2 galaxies on the BPT diagram in Figure \ref{fig:bptplus_select} and with the \citet{kauffmann2003} line shown as a dash-dotted black line for log([NII]/H$\alpha)>-0.35$ and as a solid black line below that threshold. To be placed on the BPT diagram, we require S/N$>2$ in all four BPT lines. S/N is calculated based on the formal flux errors listed in MPA/JHU without the addition of calibration errors, since these would be largely correlated for lines close in wavelength \citep{juneau2014}. 

        \begin{figure}[t!]
        \begin{center}
            \epsscale{1.15}
             \plotone{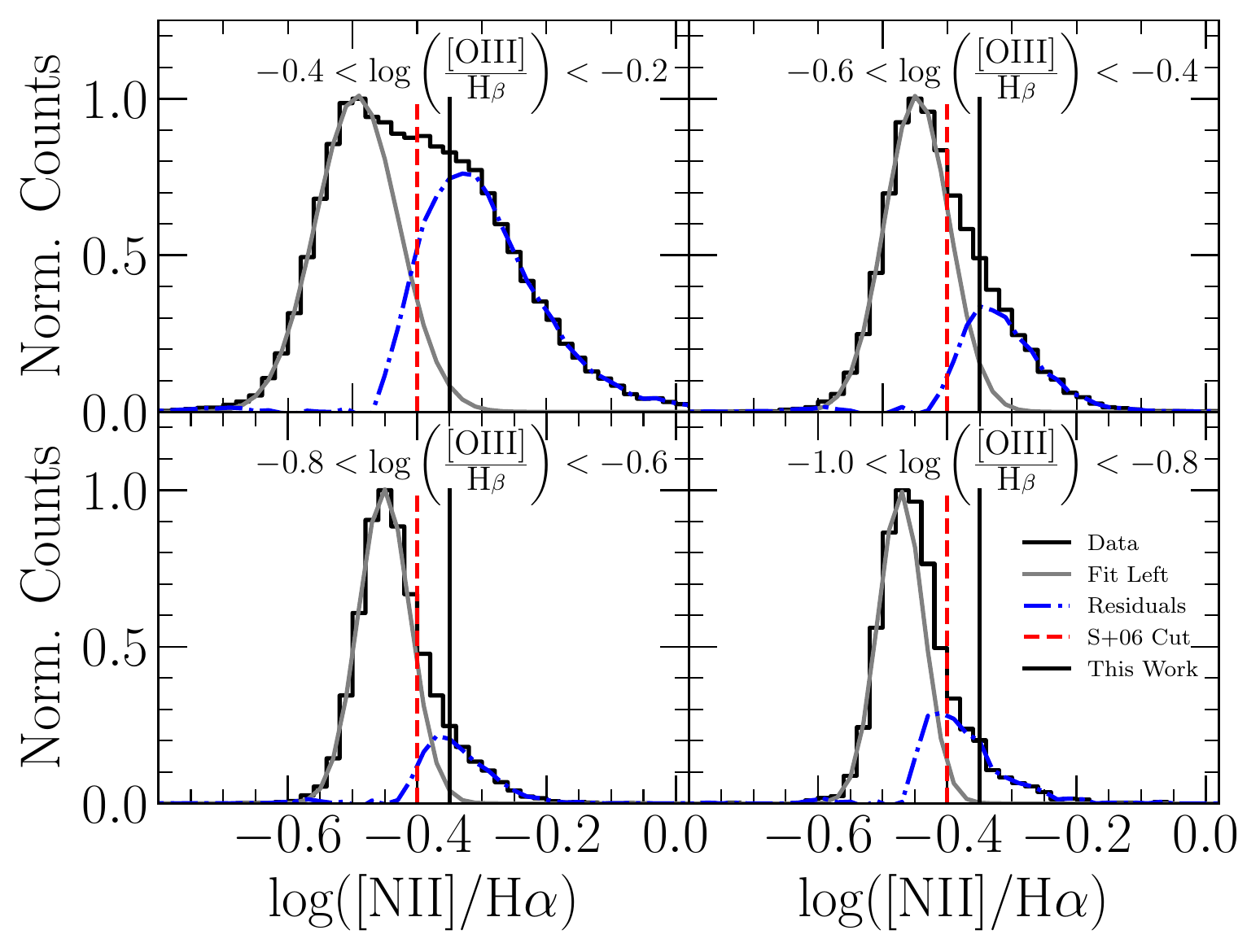}
            \caption{
            Justification for the adopted selection cuts in the lower portion of the BPT diagram. log([NII]/H$\alpha$) distributions in four bins of log([OIII]/H$\beta$) are shown as black histograms. In each a Gaussian is fitted to the primary component corresponding to the star-forming sequence and is shown as a gray line. The residuals to that fit are shown in blue, and correspond to objects with [NII]/H$\alpha$ excess (Seyferts or LINERs). The vertical \citet{stasinska2006} cut is shown as a dashed red line at $\log($[NII]/H$\alpha) = -0.4$, which we use for selecting the portion of the matching pool (i.e., the non-Seyferts/LINERs). The solid black vertical line at $\log($[NII]/H$\alpha) = -0.35$ shows the adopted demarcation above which objects are selected to be Seyferts/LINERs and where the contamination from non-Seyfert/LINERs (grey curve) is quite small.
            \label{fig:gauss_decomp}}
        \end{center}
        \end{figure}

In this study we select Seyferts and LINERs using a modification of the \citet{kauffmann2003} demarcation line that increases the completeness of the S/L branch without introducing significant contamination from the SF branch. The \citet{kauffmann2003} demarcation line follows the curvature of the SF branch. On the other hand, \citet{stasinska2006} proposed a simpler demarcation based on the x-axis of BPT diagram (log([NII]/H$\alpha$)) alone, with S/L having log([NII])/H$\alpha>-0.4$ (other works have adopted the same or a similar cut when [OIII] or H$\beta$ is weak, e.g., \citealt{miller2003, brinchmann04, andrews2013}). We present the \citet{stasinska2006} vertical cut on the BPT diagram in Figure \ref{fig:bptplus_select} as a red dashed line. We potentially wish to select our target sample using the combination of these lines that would bring in more galaxies close to the base than just the \citet{kauffmann2003} selection, but we first have to evaluate the potential contamination of moving the selection closer to the SF branch. To estimate what fraction of objects in the star-forming branch might lie to the right of the \citet{stasinska2006} cut, we perform a decomposition of the observed distribution. Here, we assume that in the absence of extra ionization that gives rise to the S/L branch the log([NII]/H$\alpha$) distribution would be symmetric. In four bins of log([OIII]/H$\beta$) ranging from $-$1.0 to $-$0.2 in steps of 0.2, we fit a Gaussian to the main (star-forming) peak of the distribution and show the results in Figure \ref{fig:gauss_decomp}. The gray lines show a Gaussian fit to the star-forming peak of the distribution and the blue dash-dotted lines the residuals from that fit, which we attribute as belonging to the S/L component. The red dashed line shows the \citet{stasinska2006} cut at $-$0.4. As one can see, the star-forming Gaussians do cross over somewhat substantially into that region. Altogether, at log([OIII]/H$\beta$) $<-0.2$, the star-forming component comprises $\sim$20\% of the objects above log([NII]/H$\alpha$)$=-0.4$, which is not negligible. As most of the contamination is just above the cut, with a slightly higher cutoff at log([NII]/H$\alpha)=-0.35$, the contamination to the S/L component drops to only $\sim$3\%. We adopt it as a cut in the lower portion of the BPT diagram (vertical black line in Figure \ref{fig:bptplus_select}). This new cut also seems to agree with where [NII]/H$\alpha$ turns over in star-forming photoionization models for fairly high ionization parameter of log$(U) = -2$ \citep{ji2020b}.  To summarize, we adopt a new selection for Seyferts/LINERs as galaxies above the \citet{kauffmann2003} line or to the right of log([NII]/H$\alpha)=-0.35$.

Of the $\sim$340,000 galaxies in our initial sample, $\sim$200,000 can be placed on the BPT diagram with S/N ratios of at least 2 in each of the four lines. An additional $\sim$63,000 have weak [OIII] or H$\beta$ but can be still be reasonably well classified based on [NII]/H$\alpha$ ratio alone. The remaining $\sim$76,000 are not classifiable by either method because of weak or non-existing emission lines. Because our target sample (Seyferts and LINERs) may contain various levels of SF (down to practically zero), the galaxies to which we match do not necessarily have to be star-forming or have detectable emission lines. Consequently our matching sample, hereafter referred to as \textit{non-Seyferts/LINERs}, is comprised of three parts: 

\indent (a) galaxies from the SF branch on the BPT diagram, 

\indent (b) galaxies with weak [OIII] or H$\beta$ but with [NII]/H$\alpha$ ratio that rules out a LINER/Seyfert, and

\indent (c) galaxies with essentially no lines. 

Non-Seyfert/LINERs in the BPT diagram are selected to lie below the \citet{kauffmann2003} line and to have log([NII]/H$\alpha)<-0.4$. We leave out objects with log([NII]/H$\alpha$) in between $-0.4$ and $-0.35$ from both the target and the matching samples in order to reduce a few S/L that may be found there. For objects with low S/N in [OIII] or H$\beta$ but with detectable [NII] and H$\alpha$ we select the matching pool sources using the log([NII]/H$\alpha)<-0.4$ criterion, which will remove LINERs/Seyferts. Objects with low SNR [OIII]/H$\beta$  and log([NII]/H$\alpha)<-0.4$ can potentially lie above the \citet{kauffmann2003} line, but because the density of objects with good SNR in this region is low, this is unlikely. Finally, we include in the matching pool all galaxies that could not be classified using above methods due to weak or non-existent lines. We emphasize here that by excluding objects with weak [OIII] or H$\beta$ (both which tend to be lower S/N) and with high S/N [NII]/H$\alpha$ indicative of a Seyfert/LINER from the matching pool, we are ensuring that we will not simply match to a Seyfert/LINER which has just one weak line. By contrast, the galaxies which make up the unclassifiables tend to have one low S/N line in the two different regions of the spectrum probed by the BPT diagram and so the non-detection of lines (and thereby an extra ionization source in addition to SF) can be considered robust. In such unclassifiable objects, the remaining line contributions (if any) may be a result of the low levels of SF or other stellar processes. In any case, these contributions are typically small, and though in some cases they may originate from stellar processes that are not strictly related to SF, we refer to them as `SF contributions'.

To summarize, our target sample consists of $\sim$94,000 galaxies on the S/L branch of the BPT diagram. The matching pool consists of $\sim$90,000 galaxies from the SF region of BPT diagram, 11,000 additional galaxies with weak [OIII] or H$\beta$ but with well measured log([NII]/H$\alpha$) below $-0.4$, and 76,000 galaxies which are not classifiable by either method. 

\begin{figure}[t!]
        \begin{center}
            \epsscale{1.2}
            \plotone{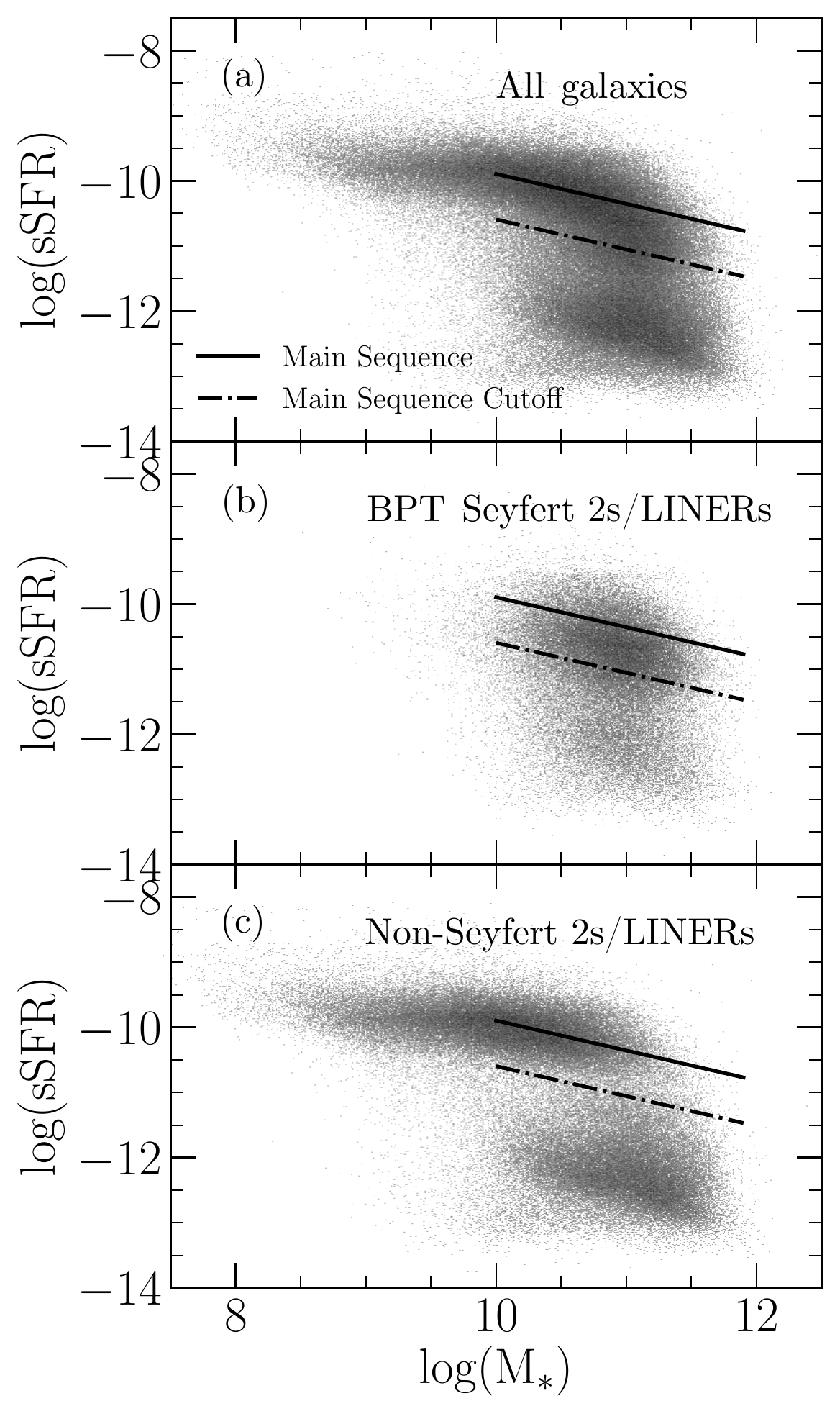}
            \caption{sSFR-M$_{*}$ diagram for (a) all galaxies in the parent sample (GSWLC-M2), (b) our target sample of Seyferts and LINERS selected by the BPT diagram a (Figure \ref{fig:bptplus_select}), and (c) the non-Seyfert/LINERs used in our matching procedure. The solid black line in the diagrams is the main-sequence fit based on the median of log(M$_{*})>10$ galaxies with log(sSFR)$>$-11 and the dash-dotted line shows a parallel line shifted by 0.7 dex to signal a region below which are the green valley and quiescent galaxies.  \label{fig:mainseq}}
        \end{center}
        \end{figure}

In Figure \ref{fig:mainseq}, we show the specific star-formation rate (sSFR; star-formation rate normalized by stellar mass) versus the stellar mass for all of the galaxies in GSWLC-M2, as well as for our target sample (Seyferts and LINERs) and non the matching pool (non-Seyferts/LINERs). The solid black line in Figure \ref{fig:mainseq} shows the star-forming main sequence based on the median fit to star-forming galaxies with log$M_{*}>$10 and log sSFR$>-11$. Most Seyfert/LINERs lie along the massive end of the star-forming main sequence and make up a significant fraction of the objects in the green valley \citep[$-12<$log sSFR$<-10.8$; ][] {salim2007, salim2014}. Importantly for our matching method, no host type (in terms of sSFR and $M_*$) is exclusive to Seyferts and LINERs.

\section{Methods}\label{sec:methods}
\subsection{Emission flux decomposition}\label{sec:matching}
In order to remove the star formation contribution to the emission lines of the galaxies on the S/L branch, we match the galaxies in the target sample to those in the non-Seyfert/LINER sample and then compute the flux of the pure Seyfert or LINER as:
\begin{equation}
    f_{\operatorname{pure\ S/L}} = f_{\operatorname{S/L}} - f_{\operatorname{non-S/L}}
\end{equation}
We use fluxes, rather than luminosities, because the match is required to be at the same redshift as the Seyfert/LINER. Ideally, the match should have the same star formation rate in the fiber, HII gas-phase metallicity and ISM properties as the Seyfert/LINER being matched.  In practice, we require that the properties most relevant to the star formation and stellar population be as similar as possible for the Seyfert/LINER and the match. Specifically, we match by the global SFR (determined through SED fitting, not emission lines), stellar mass, stellar mass contained within the spectroscopic fiber, and the global dust attenuation. The SFR is a necessary criterion as it is directly related to the amount of ionizing radiation produced by star-forming regions. We consider the stellar mass as well as the mass contained within the spectroscopic fiber because it is important for ensuring that the contributions of the star-forming region relative to the host galaxy's light are similar, and because spectral properties vary as a function of fiber mass for a fixed host mass. A lower-mass galaxy with a SFR of 1 solar mass per year will have more easily detectable emission lines than a much more massive galaxy with the same SFR as the contributions from SF may be drowned out by stellar emissions. The similar masses also will provide that the metallicities of the galaxies are not too disparate, which is relevant for the production of nebular emission lines. By matching both the total stellar mass and the stellar mass within the fiber, we implicitly match the galaxy mass profile, an important determinant of SMBH growth \citep{chen2020}. We include the stellar continuum dust attenuation in the matching criteria as it should constrain the amount of dust attenuation in the star-forming regions, and as a result, of emission lines. Note that we cannot correct for the dust emission using the observed Balmer decrement before we subtract the SF contribution because the extra ionization source (e.g., the AGN) and the stars within the same SDSS aperture may be obscured by different amounts of dust and have different dust-free Balmer decrements. Similarly, we cannot rely on the observed Balmer decrement of a Seyfert/LINER for matching to a non-Seyfert/LINER in terms of their dust content as the observed lines used in the Balmer decrement (H$\alpha$, H$\beta$) consist of some unknown combination of Seyfert/LINER and non-Seyfert/LINER contributions. By including the SFR, mass, and dust attenuation, we effectively also ensure that the target galaxy and its match have similar molecular gas masses \citep{yesuf2019}, and thereby will have similar amounts of cold gas available for fueling star formation. In addition to the physical parameters, we also match by redshift, so that the spectroscopic fiber covers similar physical scales in the target and matching galaxy. The above matching criteria implicitly yield a match with similar quality of a spectrum. We construct a distance metric composed of the five quantities, defined as
    \begin{eqnarray} \label{eq:dist}
    d = \Bigg[\left(\log\operatorname{SFR}_{\operatorname{S/L}}-\log\operatorname{SFR}_{\operatorname{non-S/L}} \right)^2
     &+\nonumber\\
     \left(\log M_{*,\operatorname{S/L}}-\log M_{*,\operatorname{non-S/L}} \right)^2 &+ \nonumber\\ 
    \left(\log M_{\operatorname{fib},\operatorname{S/L}}-\log M_{\operatorname{fib},\operatorname{non-S/L}}\right)^2  &+ \nonumber \\
        (A_{V, \operatorname{S/L}} - A_{V, \operatorname{non-S/L}})^2
     &+ \nonumber \\
    \left(\dfrac{z_{\operatorname{S/L}}-z_{\operatorname{non-S/L}}}{\sigma_{z}} \right)^2
    \Bigg]^{1/2} 
    \end{eqnarray}
Redshift is normalized by the standard deviation of redshifts for the ensemble ($\sigma_z$) in order that all five quantities would have a comparable range of values (within a factor of few) and therefore similar weights in the metric. For each Seyfert/LINER (S/L) we identify its non-Seyfert/LINER doppelganger by finding a match with a minimal $d$. We confirm that the S/L and the match quantities correlate well, with the scatter similar to the uncertainties of the measurements (except the redshift). Note that we scale the fluxes of the match to account for any (small) difference in the redshifts.

Seyferts and LINERs have line contributions from sources other than HII regions (extra ionization), which is why they end up on the S/L branch in the first place. While the above matching should ideally identify a match with emission line fluxes that are smaller than that of the Seyfert/LINER,``predicting'' the line fluxes due to SF based on the global SFR and other properties will be subject to some stochasticity. As a result, most (78\%) of the initial matches end up having at least one of the fluxes too similar or sometimes even larger than in the target galaxy. We have established that we can effectively refine the matching by introducing an additional criterion that the differences in fluxes between the target and match galaxy must be positive and $>2$ times the flux error. We have extensively compared the results obtained with and without the application of this criterion and find that there are no systematic differences (see Appendix \ref{app:no_sn} for details). Beyond this, we investigated averaging the emission line fluxes of multiple non-S/Ls, testing a variety of $n_{\mathrm{avg.}}$ from 3-100, finding that $n_{\mathrm{avg.}} =5$ produces the greatest fraction of objects with fainter matches ($24\%$), which is not a significant improvement over using only the closest match alone ($22\%$). The median match distance with $n_{\mathrm{avg.}}=5$ is similar to that when we require positive fluxes in the matching, but without the advantage of guaranteed positive fluxes.

In addition to comparing various schemes for the doppelganger method, we have investigated a different method for subtracting SF contributions which is not based on matching to a single object, but where the flux due to the SF component is estimated based on a linear regression with the parameters we use in the nominal matching as well as additional parameters ($A$(FUV), H$\delta$, and D$_{4000}$). Unlike in the matching procedure, there is no penalty incurred by adding more parameters to the regression. The percentage of S/Ls which have predicted match fluxes that are lower in all four lines is similar to that of the single-match method, $26\%$. The resulting pure S/L branch on the BPT diagram using this regression method is remarkably similar to that found using the doppelganger method, lending additional credence to our nominal method.

Note that we expect the majority of the stellar population-related contamination of S/L emission lines to be due to star formation. In reality, some fraction of the contamination may arise from the  diffuse ionized gas (e.g., \citealt{sanders2017}), especially if the SF itself is relatively low. While the exact powering mechanism behind  the diffuse ionized gas is not known, \citet{jones2017dig} find that the properties of such emission are dependent on the host properties, e.g., SFR, stellar mass, of the galaxy. As a result, we can expect that any such component that may be present in a target galaxy will be present to a similar degree in its match and would therefore be subtracted together with the SF contribution.  

One concern regarding our matching procedure is that it is based on global SFR, whereas the SF within the fiber may differ between the S/L and its match. In Appendix \ref{app:altmatch}, we describe an alternative matching procedure which uses the fiber-based quantities D$_{4000}$ and H$\delta$ Lick index instead of the global SFR. The results are essentially the same, suggesting that the adopted matching method produces robust results. We opt to present results from the matching based on the global SFR so that we can consider our results in the context of general host properties. 

Overall, we find that both the Seyferts and the LINERs have doppelgangers among the galaxies in the matching pool. This is the consequence of the fact that a galaxy with a given host properties (in particular, a given stellar mass, SFR and central mass density) is never found exclusively only on the S/L branch. This is certainly expected for Seyferts due to the episodic nature of AGN activity, but seems also to be the case for LINERs. Details on matching quality are given in Appendix \ref{app:dists}.

Please note that the existence of good matches for Seyferts among our matching pool consisting of non-Seyferts and non-LINERs is necessary for our goal of the removal of SF contribution to Seyfert lines, as we are specifically focused on testing the mixing scenario, but this does not imply any specific evolutionary scenario. Seyferts, may still, in principle, evolve from LINERs.

\subsection{Emission Line Dust Corrections}\label{sec:dust}
While the BPT line ratios are reddening independent, the resulting emission line fluxes---those which have had the star formation component removed---need to be corrected for internal dust attenuation in order to derive properties such as the ionization parameter or the metallicity. Following established methods, we determine flux corrections from the Balmer decrement (observed H$\alpha$ to H$\beta$ ratio), assuming the dust-free ratio of 3.1 for narrow-line regions \citep{agn_squared}, and assuming the \citet{cardelli1989} attenuation curve. The  Balmer decrement may be poorly constrained when the S/N of H$\beta$ is low. In those cases we estimate the dust correction of the Seyfert/LINER emission lines indirectly, via the global (stellar continuum) dust attenuation.  To arrive at the calibration, we fit a linear regression between the global dust attenuation available from GSWLC-M2, $A_{V,\operatorname{stellar}}$, and that derived from the Balmer decrement when the S/N of H$\beta>10$ ($A_{V,\operatorname{Balmer-Pure\ S/L}}$). 

        \begin{figure}[t!]
        \begin{center}
            \epsscale{1.1}
            \plotone{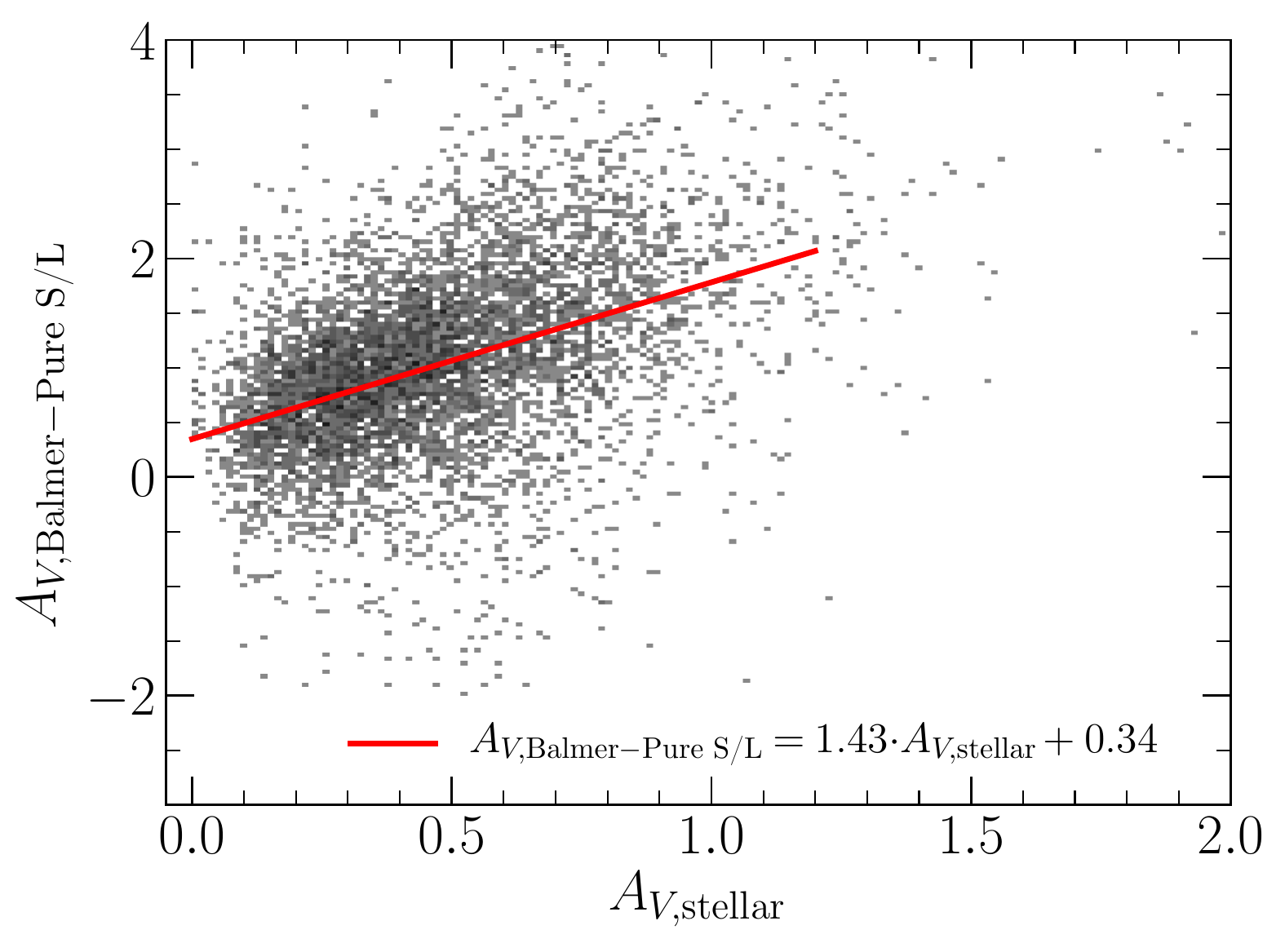}
            \caption{
            The relation between the nebular dust attenuation of the Seyfert/LINER emission from Balmer decrement, and the stellar continuum attenuation for entire galaxies (from the SED fitting). The relation is based on Seyfert/LINERs with relatively well determined Balmer decrements (SF-corrected H$\beta$ S/N$>10$). The calibration (red line) is used to estimate nebular dust attenuation of Seyfert/LINER lines when H$\beta$ is too weak to yield reliable Balmer decrement. We verify that the calibration would be similar to the one shown if Seyfert 2s and LINERs were treated separately.
            \label{fig:dust_calibration}}
        \end{center}
        \end{figure}
        
We show $A_{V,\operatorname{Balmer-Pure\ S/L}}$ versus $A_{V,\operatorname{stellar}}$ in Figure \ref{fig:dust_calibration} for Seyfert/LINERs that have S/N $>10$ in their H$\beta$ line. The derived linear fit is given as 
\begin{equation} \label{eq:dustcal_sub}
    A_{V,\operatorname{Balmer-Pure\ S/L}} = 1.43\cdot  A_{V,\operatorname{stellar}} + 0.34
\end{equation}
We investigated the possibility to refine this calibration by including SFR and stellar mass in the linear regression, but did not find that it substantially improved the ability to predict A$_{V,\operatorname{Balmer-Pure\ S/L}}$. Furthermore, we tested using the attenuation derived from the SED fitting of just the continuum in the fiber (taken from the MPA/JHU catalog) and it too does not improve the ability to predict nebular attenuation compared to the global dust attenuation. We also tested if the calibration was different for Seyfert 2s and LINERs separately, but found it to be quite similar. \citet{cidfernandes2011} used a similar calibration based on $A_{V,\operatorname{stellar}}$ to estimate the nebular dust. The $A_{V,\operatorname{stellar}}$ in that work was based on optical spectrum continuum fits assuming a \citet{cardelli1989} attenuation curve, and so their calibration should be similar to ours because massive galaxies in GSWLC-M2 tend to have an extinction curve similar to the Milky Way \citep{salim2018}. Indeed, \citet{cidfernandes2011} find the coefficient the slope of their line to be 1.56 and the intercept to be 0.53, with the (minor) differences likely being due to a difference in how we choose which galaxies to use in the calibration and because they use a fit to the median trend whereas we simply fit a line to all of the data.

With the calibration in place, we use equation \ref{eq:dustcal_sub} to compute the nebular attenuation for all objects with H$\beta$ S/N $<10$. For objects with H$\beta$ S/N $>10$, we correct the pure S/L line fluxes with the pure Balmer decrement (SF fluxes removed for both H$\alpha$ and H$\beta$), except that we set a small number of galaxies with formally negative $A_{V,\operatorname{Balmer-Pure\ S/L}}$ to 0 and those greater than 3 to be 3. For clarity, we show the following equation to present the different cases:

\small
\begin{equation}
 A_{V,\operatorname{B-P\ S/L}} =
\begin{cases}
  1.43\cdot A_{V,\operatorname{S}} + 0.34 \qquad \text{if H$\beta$ S/N$<10$} \\
0 \qquad \text{if H$\beta$ S/N$>10$ and $A_{V,\operatorname{B-P\ S/L}}<0$} \\
3 \qquad \text{if H$\beta$ S/N$>10$ and $A_{V,\operatorname{B-P\ S/L}}>3$} \\
A_{V,\operatorname{B-P\ S/L}} \qquad \text{otherwise} \\
\end{cases}
\end{equation}
where $A_{V,\operatorname{S}}$ = $A_{V,\operatorname{stellar}}$, and $ A_{V,\operatorname{B-P\ S/L}} = A_{V,\operatorname{Balmer-Pure\ S/L}}$.
\normalsize

\section{Results} \label{sec:bpt_gen}
In this section, we present BPT diagrams before and after the removal of the SF contribution to evaluate the mixing scenario. We then revisit the classification of Seyfert 2s and LINERs suggested by their emission lines. We then investigate how the position of Seyfert 2s and LINERs on the BPT diagram depends on ionizing source parameters and conditions in the ISM that gives rise to the emission lines.

        \begin{figure}[t!]
        \begin{center}
            \epsscale{1.0}
            \plotone{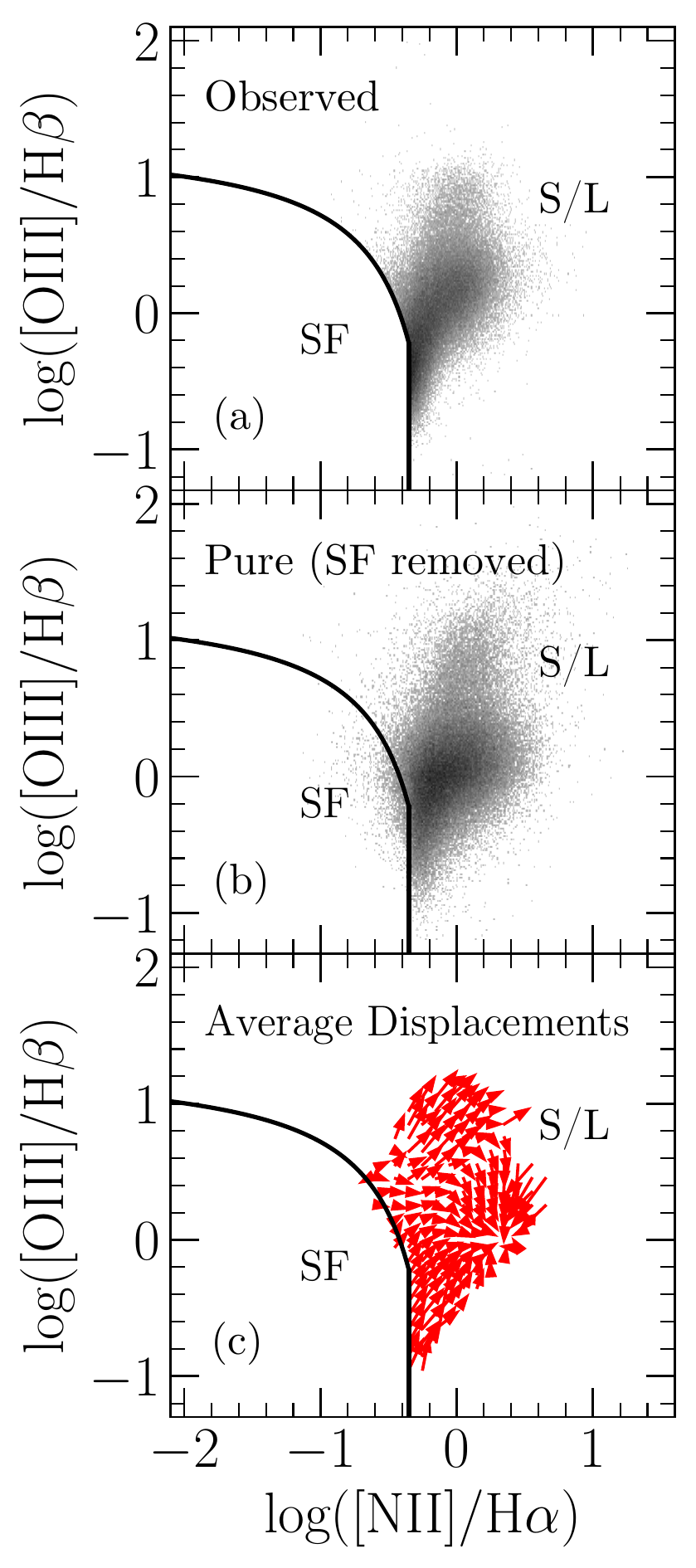}
            \caption{BPT diagrams of $\sim$94,000 Seyfert 2s and LINERs before the removal of SF flux in top panel (a), after the removal of SF in middle panel (b), and the average displacements due to SF removal in 0.1 dex by 0.1 dex cells in bottom panel (c). The solid line is our modified \citet{kauffmann2003} demarcation. Even after the contributions from star formation are empirically removed, both the Seyfert and the LINER ``sub-branches'' that are typically associated with the upper and mid-right portions of the branch retain their span, suggesting that the dominant reason for the diversity of line ratios in the observed BPT is intrinsic. The slight spilling of sources into the SF portion of the diagram is due to noise. 
             \label{fig:bpt_gen}}
        \end{center}
        \end{figure}

\subsection{Testing the mixing sequence scenario}

To properly understand how the removal of the SF component affects the morphology of S/L branch, we first show the observed branch before the subtraction of the SF component is performed (Figure \ref{fig:bpt_gen}(a)). If the mixing is the principal cause for the observed extent of the S/L branch (or at least its Seyfert portion), we would expect the branch to become significantly confined after the subtraction (Figure \ref{fig:cartoon}). We find that this is not the case (Figure \ref{fig:bpt_gen} (b)). Collectively, pure Seyferts and LINERs intrinsically span the full range, which is inconsistent with the mixing scenario (Figure \ref{fig:cartoon}).

In Figure \ref{fig:bpt_gen}(c), we show the average displacement in BPT diagram due to the removal of SF contamination. The tail of each arrow represents the original position and the head shows the position after the subtraction of SF fluxes. Under the mixing scenario, the relative contributions from SF to the emission lines should decrease from the base of the S/L branch to its top, especially in the Seyfert sub-branch \citep{kewley2001b, thomas2018}, so the arrows should get shorter the farther the tail is from the demarcation line. In contrast, the objects in the uppermost portion and in the base of the branch tend to move up and to the right with a similar magnitude. The distribution of the absolute displacements peaks at 0.11 dex with a tail towards higher values, giving an average of 0.2 dex.  At each position there is a wide range of displacements, as expected if S/L themselves have a range of intrinsic properties.

From the analysis here, we conclude that the mixing scenario cannot provide an adequate explanation for the variety of line ratios along the S/L branch, which then must be intrinsic.

\subsection{Classification of objects on the Seyfert/LINER branch} \label{sec:agntypes}

Focusing on the Seyfert 2 and LINER sub-branches on the BPT diagram, we notice that both are actually more distinct after the removal of SF contribution (Figure \ref{fig:bpt_gen}(b)). Nevertheless, the two populations still overlap to some extent and it is widely known that other emission line ratios ([SII]/H$\alpha$, [OI]/H$\alpha$ \citealt{vo87, kewley2006}) need to be considered to achieve a cleaner separation, at least in the upper branch. Furthermore, it is not clear whether the objects near the base of the S/L branch are more similar to the Seyferts or the LINERs, or whether they form a distinct group.  Our preliminary investigation of various emission line diagrams other than the [NII] BPT diagram suggested that the latter may indeed be the case -- that there is a population of objects that are distinct from either Seyferts and traditional LINERs by having weaker [OI] and therefore overlapping with pure star formers in diagrams other than the ([NII]) BPT even when the star forming component has been removed. The presence of a distinct population is further supported by a somewhat bimodal distribution among the non-Seyferts in both [SII]/H$\alpha$ and [OI]/H$\alpha$.

However, rather than defining the types based on some hard cuts in a particular two-dimensional line ratio plot, we choose to perform a multi-dimensional classification based on all emission lines at our disposal. Specifically, we use the k-means classification algorithm and apply it to [OII] (both lines in the doublet), [OI], [NII], [SII] (both lines in the doublet), H$\alpha$, and H$\beta$, all relative to [OIII]. The lines had the SF contribution removed by subtracting the SF line fluxes of the matched galaxy. [OIII] is chosen for normalization as it always has a high S/N in our sample and is related to the strength of the extra ionization source \citep{heckman2005, lamassa2010, azadimosdef2017, glikman2018}. We perform k-means clustering using Euclidean distance of the logarithms of flux ratios. Requiring that all seven lines have S/N$>2$ (after the removal of SF) reduces the sample from $\sim$94,000 S/Ls to $\sim$ 45,000 S/Ls, primarily because many objects have weak [OI]. However, we confirm that while having [OI] increases the accuracy of classification, it is quite robust ($>90$~\% agreement) even without [OI]. In what follows, we present analysis based on the sample with detections in all seven lines, to some extent because of the somewhat better classification when [OI] is retained, but primarily so that we can discuss and compare diagnostic diagrams that involve and do not involve [OI] using consistent samples. Importantly, we confirm that all of the trends that involve quantities that do not require [OI] (ionization parameter, metallicity, electron density) are unchanged whether we use the sample and clustering with or without [OI]. For the classification that includes [OI], 16\% are classified as Sy2, 51\% as H-LINERs and 33\% as S-LINERs. Without [OI], we would have 13\%, 56\% and 31\% respectively. In Table 1 we provide classifications of all 339,805 SDSS galaxies considered in this work, both at the general level, allowing one to distinguish between a SF galaxy, a S/L or a galaxy with weak or no lines, and for S/L, allowing one to distinguish between Sy2, H-LINERs and S-LINERs. More information is provided in the note to the table.

 \begin{deluxetable}{c|ccc}
    
    \tablecaption{Classification of SDSS galaxies
      \label{tab:classes} }
     \tablehead{\colhead{\texttt{SDSS\_ObjID}}
     &\colhead{\texttt{gen\_el\_class}} &\colhead{\texttt{sl\_class1}}
     &\colhead{\texttt{sl\_class2}}}
         \startdata
 1237645942905634957 &       0 &                0 &              0 \\
  1237645942905634975 &       5 &                3 &              3 \\
  1237645942905635004 &       0 &                0 &              0 \\
  1237645942905700478 &       5 &                2 &              2 \\
  1237645942905700561 &       1 &                0 &              0 \\
  1237645942905831669 &       0 &                0 &              0 \\
  1237645942905897107 &       3 &                0 &              0 \\
 1237645942905897117 &       3 &                0 &              0 \\
  1237645942905897152 &       0 &                0 &              0 \\
  1237645942905897218 &       3 &                0 &              0 \\
  1237645942905962547 &       1 &                0 &              0 \\  
  \enddata
   \tablenotetext{a}{
      In \texttt{gen\_el\_class}, there are six categories: 0 = unclassifiable due to weak [NII] or H$\alpha$, 1 = star-forming with SNR $>2$ in all four BPT lines, 2= star-forming according to [NII]/H$\alpha$ alone, 3 = S/L based on [NII]/H$\alpha$ alone, 4 = objects with SNR $>2$ in all four BPT lines and lying between the \citet{stasinska2006} cut (log([NII]/H$\alpha=-0.4$)) and our adopted cut (log([NII]/H$\alpha=-0.35$)), and 5 = the objects which are S/Ls (SNR$>2$ in all four BPT lines and above our modified \citet{kauffmann2003} line) that make up our target sample.       
      In \texttt{sl\_class1}, the classifications based on k-means with all seven emission lines (including [OI]), there are five categories: 0 = not a S/L, 1 = Sy2,  2 = H-LINER, 3 = S-LINER, and 9 for the S/Ls which had weak SNR in [OI], [SII], or [OII] and so could not be included. 
      The final column, \texttt{sl\_class2}, gives a classification based on six lines if [OI] is not detected, otherwise is based on seven lines and is identical to \texttt{sl\_class1}. Full table is available in the electronic version of the journal and at https://salims.pages.iu.edu/agn}
 \end{deluxetable}

  \begin{figure}[t!]
        \begin{center}
            \epsscale{1.1}
            \plotone{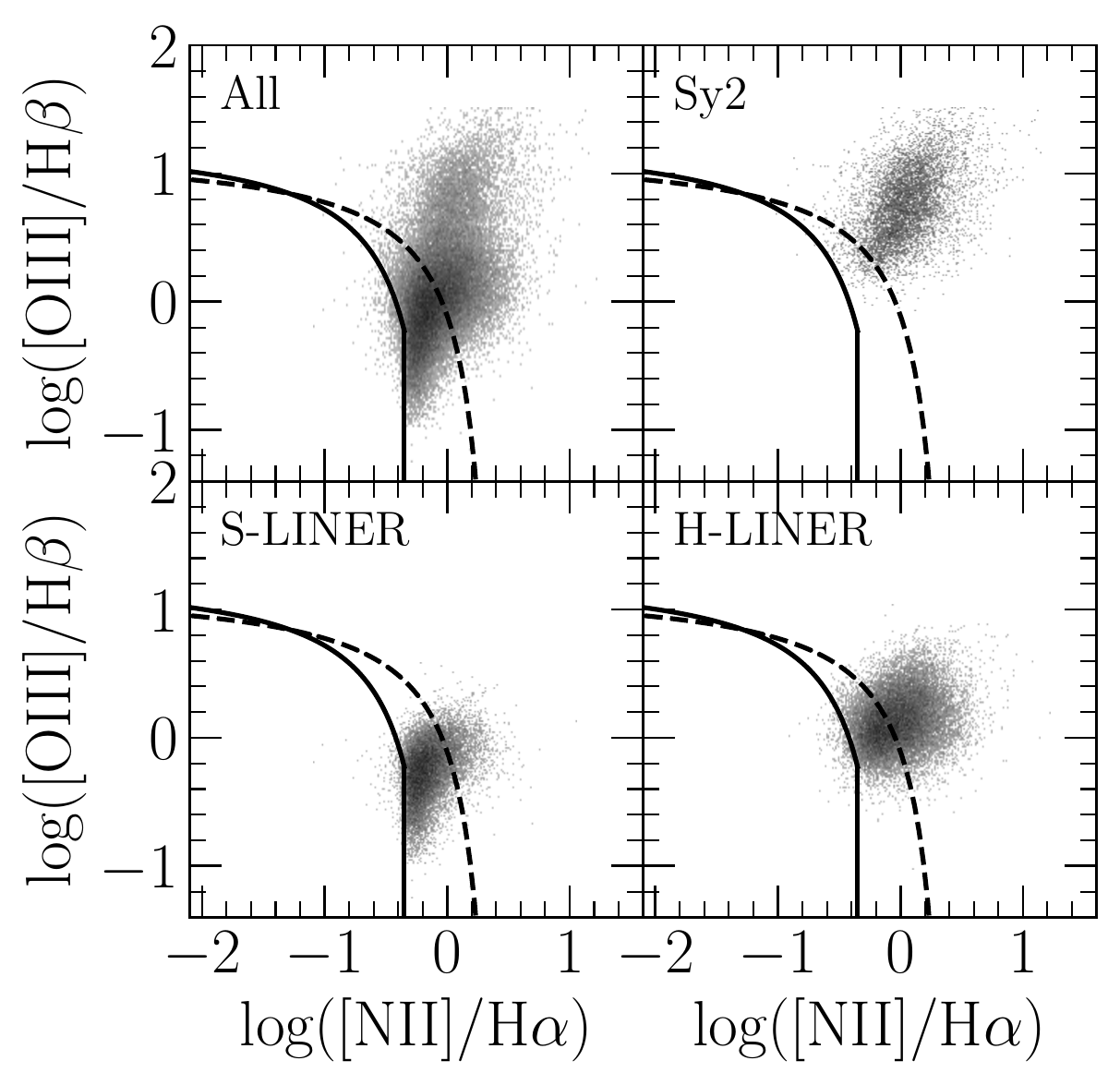}
            \caption{All classified galaxies on the Seyfert/LINER (S/L) branch of the BPT diagram, plus individual types: Seyfert 2 (Sy2), Soft LINER (S-LINER), and Hard LINER (H-LINER). The types are assigned using k-means clustering on the emission-line ratios relative to [OIII]. Emission lines have SF or stellar emission contribution empirically removed using the matching of S/L galaxies to non-S/L doppelgangers. Seyfert 2s primarily occupy higher [OIII]/H$\beta$ values. S-LINERs lie further down the branch but overlap with H-LINERs. H-LINERs occupy the region of BPT typically associated with traditional LINERs. S-LINERs and H-LINERs primarily are separated by the hardness of their ionizing radiation, whereas both types of LINERs separate from Sy2s mainly in their ionization parameter. In this and all subsequent plots we show $\sim$45,000 S/Ls for which the classification into sub-types using seven emission lines is possible. We show the \citet{kewley2001b} extreme starburst demarcation as a dashed line, though we argue that this or any other demarcation of the so called composite region has little physical importance.
             \label{fig:bpt_clusters}}
        \end{center}
        \end{figure}

  \begin{figure}[t!]
        \begin{center}
            \epsscale{1.1}
            \plotone{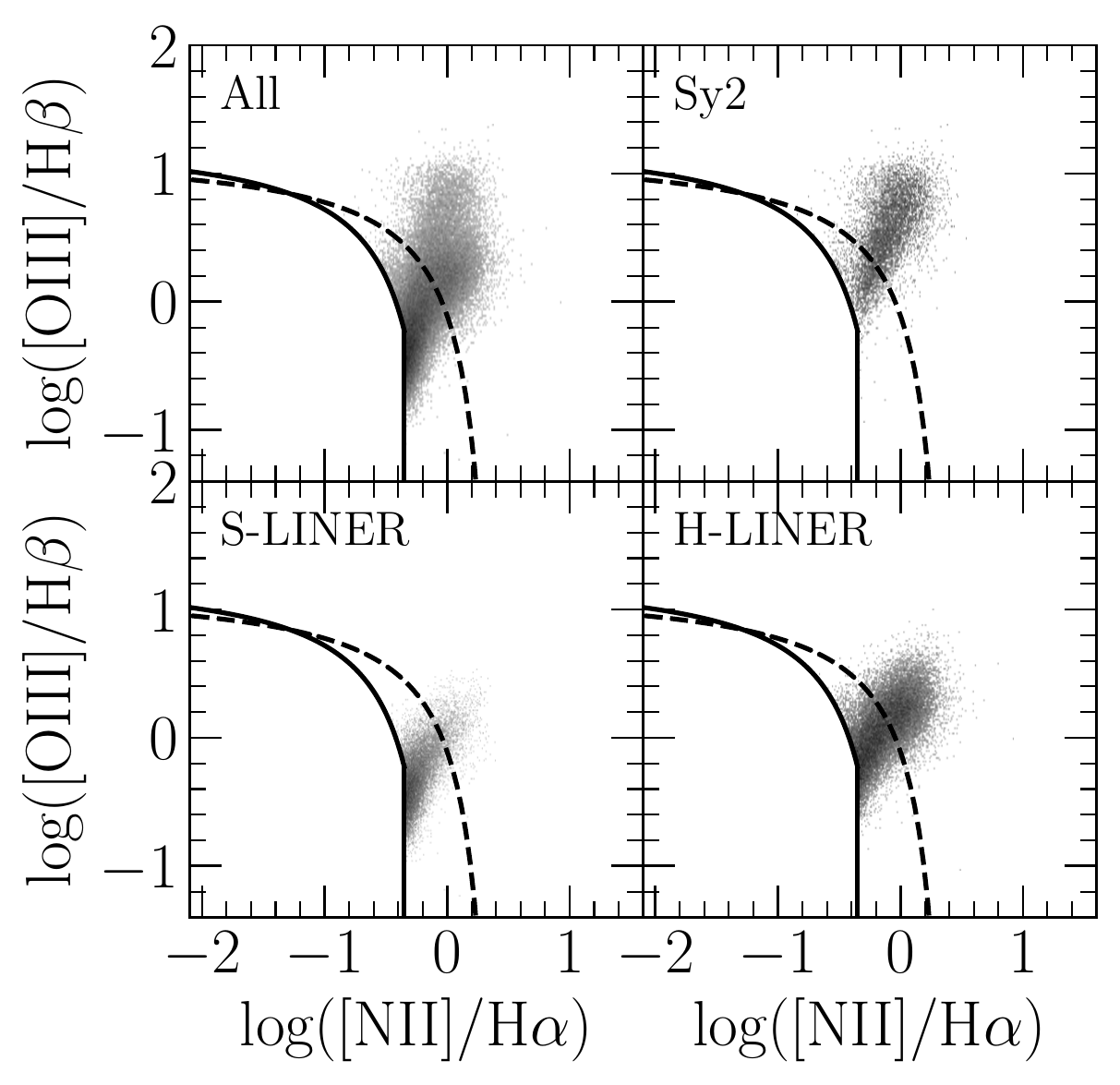}
            \caption{All classified galaxies on the S/L branch of the BPT diagram, plus individual types: Sy2s, S-LINERs, and H-LINERs, now shown with line ratios before the removal of the SF flux. There is more significant overlap between the three groups than after SF removal (Figure \ref{fig:bpt_clusters}).
             \label{fig:bpt_clusters_presub}}
        \end{center}
        \end{figure}

We designate the three types of objects classified using the k-means algorithm as Sy2, hard and soft LINERs (H-LINER and S-LINER). We refer to Seyferts identified using our method as Sy2 to distinguish them from objects identified as Seyferts using standard classification plots and the observed line ratios (Section \ref{sec:otherdiags}). The rationale for the naming of the latter two is based on the fact that the two types of LINERs differ primarily in [NII]/H$\alpha$, [SII]/H$\alpha$ and [OI]/H$\alpha$, quantities which are linked to the hardness of the radiation field \citep{vo87, kewley2006, kewley2019review}, as we will see further in Section \ref{sec:otherdiags}. We show the location of these three types in the BPT diagram in Figure \ref{fig:bpt_clusters}. As expected, Sy2s occupy the uppermost portion of the branch and the H-LINERs lie in the region typically considered to be occupied by traditional LINERs. The S-LINERs on the other hand overlap somewhat with the H-LINERs but extend lower down the branch, spanning a wider range in [OIII]/H$\beta$ than the H-LINERs. Comparing the location of the three types to where they were in the original BPT diagram (Figure \ref{fig:bpt_clusters_presub}), we see that the Sy2 sub-branch has shifted somewhat, but still has a similar extent as it did before the removal of SF contributions, attesting to a wide range of intrinsic properties of Seyferts. The two LINER types are more distinct after the removal of SF.

Some studies use the \citet{kewley2001b} theoretical upper envelope of line ratios due to HII regions (extreme starbursts), to split the S/L branch in two, with the objects in the lower portion called SF/AGN composites, or just the ``composites''. An objection that any division should be associated with the maximum starburst line, which no galaxies in the local universe attain, has already been raised by \citet{cidfernandes2010}. More importantly, this distinction is rooted in the mixing scenario, in the sense that the galaxies in the composite region have high SF contributions, whereas those in the upper part of the S/L branch have little or none. Based on this idea, many studies of S/Ls remove the galaxies from the composite region as being just the contaminated versions of the purer S/Ls found outside of the composite region. Our results cast a new light on this practice. We present the \citet{kewley2001b} line on Figure \ref{fig:bpt_clusters} to demonstrate that even after the removal of the SF contributions, many galaxies (in particular the LINERs) intrinsically lie below this line.  Altogether, we conclude that this (or any other) dividing line and the associated identification of a composite region are of little physical importance for the study or the selection of Seyferts or LINERs, and should not be used.  We show the \citet{kewley2001b} line on Figure \ref{fig:bpt_clusters_presub} to indicate the portion of the objects that would have been left out if that line was used to select Seyferts/LINERs. The removal of these objects is insufficiently motivated because they are not affected by SF contamination much differently than the objects above the line and the primary reasons for an object to lie above or below the line are intrinsic.

\subsection{Properties of ionizing source }\label{sec:ion}
The previous analysis has indicated that SF mixing is not the main cause of the emission-line diversity along the S/L branch.  However, before that view is firmly adopted, it is wise to ascertain that the new SF-subtracted line ratios are physically plausible. To that end, we now investigate the drivers of the spread in intrinsic line ratios along the S/L branch by fitting standard emission-line relations to the new ratios. The derived parameters are found to be reasonable, and interesting new trends in physical properties emerge along the two relations as a byproduct.  In the next section we look at quantities related to the ionizing source. In the following section, \ref{sec:nlr_cond}, we investigate how the conditions of the ISM or NLR itself affect the position on the BPT diagram. As a summary of this and the following section, in Table \ref{tab:quant} we provide qualitative comparisons and quantitative measurements of each of the quantities for the Sy2s, S-LINERs, and H-LINERs. It needs to be pointed out that any emission line characterization is inevitably resolution dependent. Here, the resolution corresponds to 3 arcsec diameter of SDSS fibers. Higher spatial resolution may reveal nuclear line ratios that differ from the ones in the fiber due to the spatial variability of ionizing source properties or mechanisms (e.g., \citealt{ma2021}).

    \begin{deluxetable*}{c|cccc|cccc|cccc|c}
     \tablecaption{Representative physical parameters for Sy2, S-LINERs, and H-LINERs. First column for each type presents a qualitative description of the median value of each group relative to the others, and is followed by the median, the 16th, and 84th percentiles for each quantity. \label{tab:quant}}
\tablehead{\colhead{Quantity} &\multicolumn{4}{c}{Sy2}& \multicolumn{4}{c}{S-LINER} & \multicolumn{4}{c}{H-LINER} & \colhead{Figure} \\
 \omit & \colhead{Qual.}& \colhead{Median} &\colhead{16th} &\colhead{84th} & \colhead{Qual.}&\colhead{Median} &\colhead{16th} &\colhead{84th} &\colhead{Qual.}&\colhead{Median} &\colhead{16th} &\colhead{84th} & \omit}    
    \startdata
  log($U$) & High& -2.89 &-3.29 & -2.36  & Lowest & -3.83& -3.91 &-3.69 &  Low &-3.67 & -3.81 & -3.43 & \ref{fig:bpt_U_ccode}\\
  log([OI]/[SII]) & High &-0.66 &-0.86 & -0.45  & Low &-0.79 & -1.02 &-0.56 & Highest & -0.61 & -0.84 & -0.36& \ref{fig:bpt_oi_sii_ccode} \\
  log(O/H)+12 & High & 8.77 &8.59 & 8.93  & High & 8.76 & 8.62 & 8.91 & Lower & 8.66 & 8.50 & 8.83& \ref{fig:bpt_logoh_ccode} \\
  log($n_{e}$) & High & 2.83 & 1.64 & 3.42 & Lowest & 2.28  & 0.55 & 2.96 & Low & 2.46 & 0.30 & 3.31 & \ref{fig:siidoub}\\
    \enddata
    \end{deluxetable*}

        \begin{figure}[t!]
        \begin{center}
            \epsscale{1.2}
            \plotone{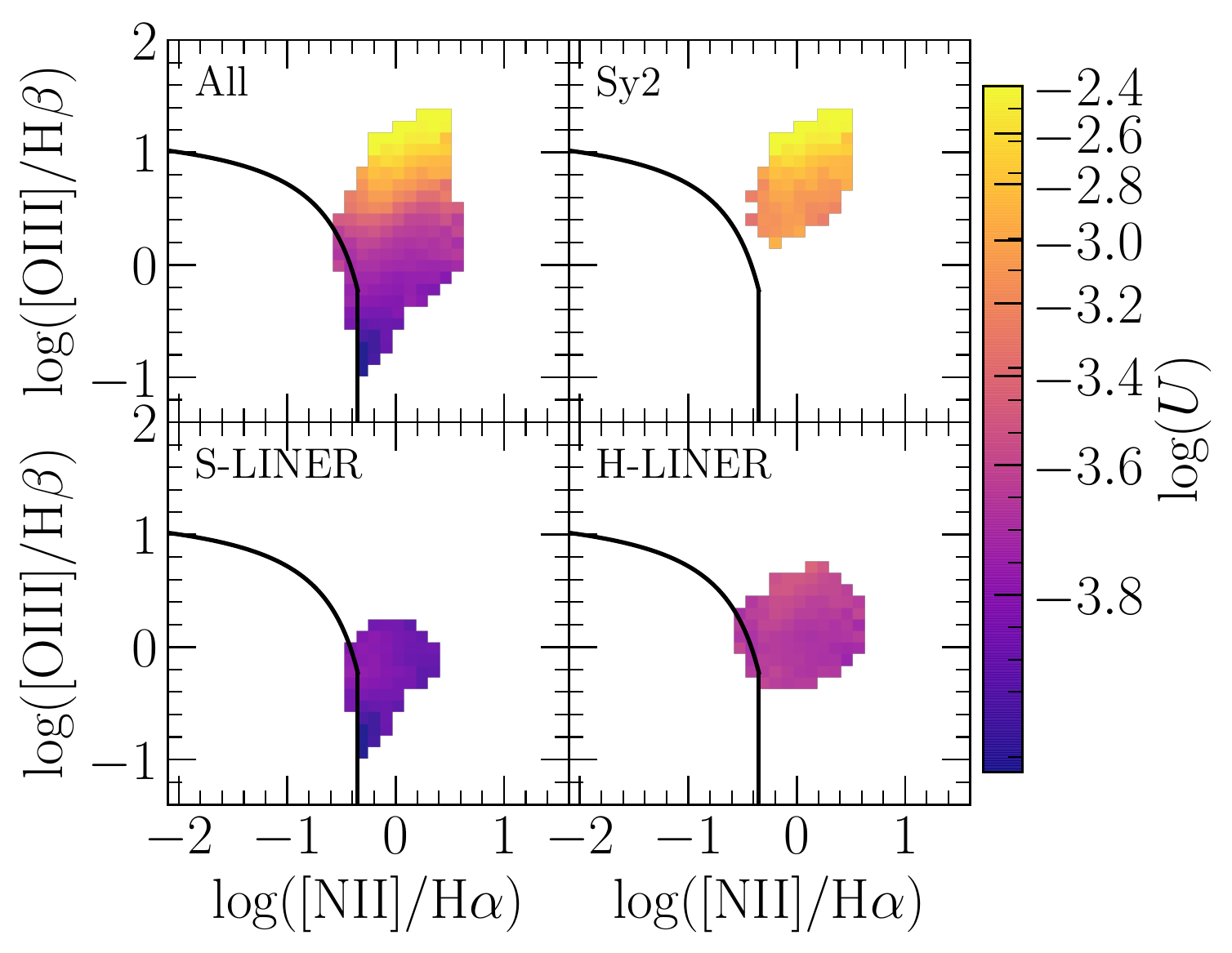}     
            \caption{Pure Seyferts and LINERs (without the SF contribution to emission lines) on the BPT diagram color-coded by ionization parameter (log($U$)). The ionization parameter is strongly correlated with vertical position for Seyferts and LINERs taken together and for the Sy2s. The S-LINERs and H-LINERs do not have as significant of variation in ionization parameter with the position on the AGN branch as the Sy2s do, but differ on average from one another, H-LINERs having higher ionization parameters. The results would be qualitatively the same with the more complete sample where [OI] is allowed not to be detected.
            \label{fig:bpt_U_ccode}}
        \end{center}
        \end{figure}
\subsubsection{Ionization parameter} \label{sec:logu}

The number of ionizing photons produced by a source will critically affect the amount of heating which takes place in the NLR or nuclear ISM, and will thus affect the observed emission line ratios \citep{agn_squared}. Here, we probe the ionization parameter via the line ratio [OIII]/[OII], converted into $\log (U)$ via the calibration provided by \citet{carvalho2020}. We show the BPT diagram color-coded by average $\log (U)$ in Figure \ref{fig:bpt_U_ccode}. For the Seyferts and LINERs taken together, there appears to be a trend of ionization parameter along [OIII]/H$\beta$ direction, with higher ionization parameter on the top. When the objects are split into the different types of S/Ls, we still find a trend for the Sy2s, with this type having the highest ionization parameter values in general. We find, however, that the dependence is not so strong for S-LINERs or H-LINERs. Additionally, we note that while they occupy similar positions on the branch, S-LINERs appear to have ionization parameter that is on average 0.2 dex lower than that of the H-LINERs and H-LINERs have ionization parameter $\sim$0.5 dex lower than that of the Sy2s at the same position(though the overlap between these two groups is smaller than that between S-LINERs and H-LINERs). This highlights the point that the BPT diagram is not sufficient for the characterization of emission line properties.

It needs to be pointed out that similar trends to the ones shown here for the ionization parameter are also obtained for [OIII] luminosity or the specific [OIII] luminosity ($L_{\mathrm{[OIII]}}$/$M_{*}$). However, the trend of the ionization parameter is the strongest, and we consider the trends with other parameters to merely reflect the correlations that exist between them and the ionization parameter.

\subsubsection{Hardness of radiation field} \label{sec:hardness}
As the proportion of higher-energy photons from an ionizing source increases, the size of the partially ionized zone will also increase \citep{bpt1981, vo87, agn_squared, cann2019}, and as a result, the emission of low-ionization lines with ionization energies similar to that of hydrogen will increase. Two such lines are [OI] and [SII]. The ionization energy of [OI] is closely matched to that of hydrogen, and so its emission is primarily found within the partially ionization zones whereas [SII] can also be found within the H$^{+}$ zone, and so it is not as sensitive to the size of the partially ionized zone \citep{vo87}. Due to this difference, the ratio [OI]/[SII] provides a rough probe of the hardness of the ionizing radiation. \citet{malkan2017} found that [OI]/[SII] correlates with ionization parameter for Seyfert galaxies in SDSS, suggesting its viability for probing the accretion.

        \begin{figure}[t!]
        \begin{center}
            \epsscale{1.2}

            \plotone{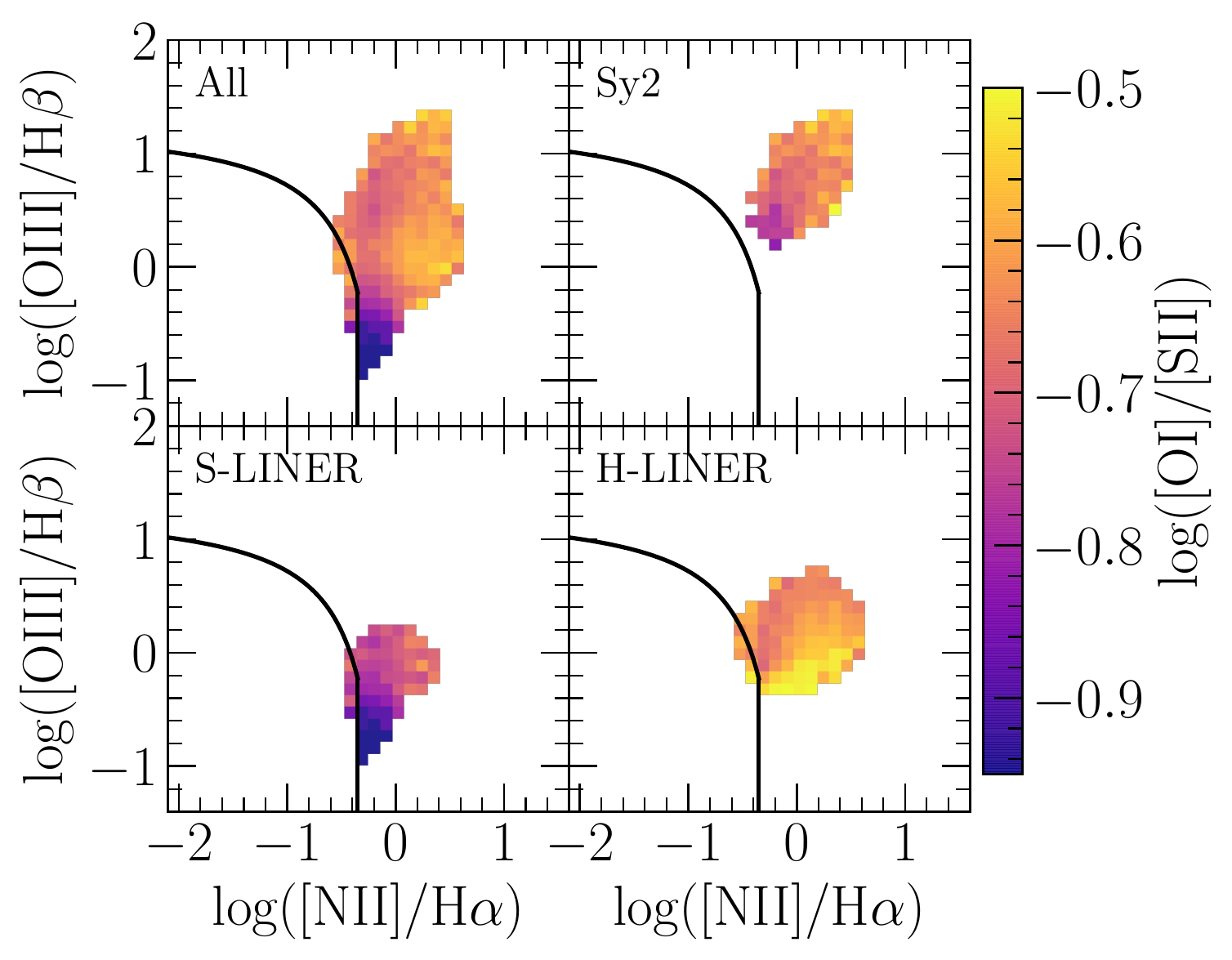}  
            \caption{
            Pure Seyferts and LINERs (without the SF contribution to emission lines) on the BPT diagram color-coded by the hardness of radiation field (log([OI]/[SII])), which increases as [OI]/[SII] increases. Sy2s and S-LINERs become harder along the direction of the branch but are both softer than the H-LINERs.
            \label{fig:bpt_oi_sii_ccode}}
        \end{center}
        \end{figure}
We show the pure S/L branch color-coded by [OI]/[SII] in Figure \ref{fig:bpt_oi_sii_ccode}. There is not a definitive trend for the Seyferts and LINERs taken together, but it appears that hardness increases along the direction of the branch for the Sy2s and the S-LINERs, agreeing with \citet{ho1993a} model tracks. H-LINERs are generally harder than Sy2s and especially S-LINERs, which provides the basis for our naming scheme.

\citet{kewley2006} also suggest that Sy and LINERs differ in ionization parameter and hardness, which we verify in this work, finding Sy2s have much higher ionization parameter than H-LINERs and that the H-LINERs have a harder radiation field than the Sy2s.

\subsection{Physical conditions of the narrow-line region/nuclear ISM}\label{sec:nlr_cond}
In addition to the properties of the ionizing source, the properties of the narrow-line region or the nuclear ISM will also affect how the radiation from the ionizing source(s) is processed and re-radiated. In this section, we explore how the position of S/L on the BPT diagram depends on the oxygen abundance (log(O/H)+12) and electron density ($n_{\mathrm{e}}$) of the NLR/ISM.

\subsubsection{Oxygen Abundance} \label{sec:logoh}
       \begin{figure}[t!] 
        \begin{center}
            \epsscale{1.2}
            \plotone{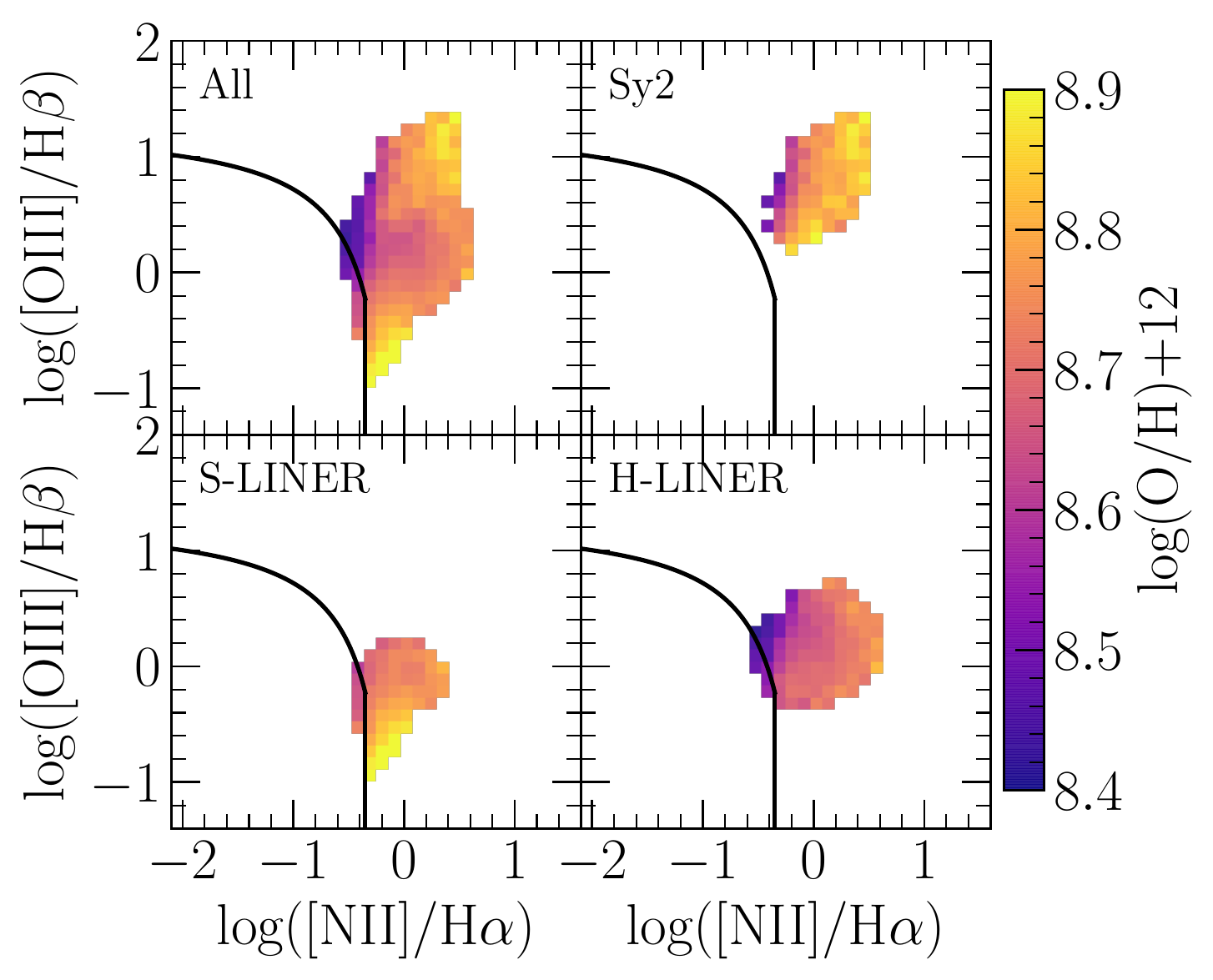}            
            \caption{Pure Seyferts and LINERs (without the SF contribution to emission lines) on the BPT diagram color-coded by oxygen abundance. Oxygen abundance increases from the left to the right side for all three classes of objects.
            \label{fig:bpt_logoh_ccode}}
        \end{center}
        \end{figure}
 
Increases in the abundance of O reduces the amount of [OIII] emission in star-forming galaxies \citep{agn_squared}. It is reasonable to expect that a similar effect may occur for NLR or nuclear ISM, as the increases in O should aid in the cooling of the gas and thus reduce the amount of emission in [OIII]. Here we estimate the metallicity using the line ratio [NII]/[OII], which is then converted into oxygen abundance, log(O/H)+12, using the strong-line calibration for AGNs provided by \citet{castro2017_logoh}. We use [NII]/[OII] rather than some other strong-line indicator because of its smaller dependence on the ionization parameter \citep{kd02, zhang2017}.

The pure S/L branch color-coded by oxygen abundance is shown in Figure \ref{fig:bpt_logoh_ccode}. For the Seyferts and LINERs taken together, there is not a consistent trend but such trends emerge when we look at individual types. For the Sy2s, there is a clear increase in log(O/H)+12 perpendicularly to the direction of the branch. This trend is consistent with [NII]/H$\alpha$ being a rough indicator of metallicity for galaxies on the SF branch \citep{vanzee1997}. For the S-LINERs and H-LINERs, it also appears that the oxygen abundance increases from the left side of the branch to the right, though the trend is not as strong. The H-LINERs appear to on average have lower abundances than the other two groups.

\subsubsection{Electron Density}
Higher electron densities should increase the amount of collisional excitation and radiative de-excitation for forbidden lines, until a critical density is reached and collisional de-excitation is favored over radiative de-excitation \citep{agn_squared}. We investigate the electron densities of the gas subject to extra ionization (the NLR/nuclear ISM) at different position on the BPT diagram, as probed via the [SII]$\lambda\lambda$6717,6731 doublet. In order to convert the [SII] ratio into an electron density, a temperature must be determined. To do so using optical nebular emission lines requires the auroral line [OIII]$\lambda$4363 in combination with the [OIII]$\lambda \lambda$ 4959,5007 doublet (e.g., \citet{kewley2019review} and \citet{flury2020}). However, [OIII]$\lambda$4363 is typically weak in most galaxies so we instead assume a temperature for each type of object. A recent study suggests Seyfert 2s on average have $T_{e} \approx 14000$ K \citep{flury2020} which we adopt for our Sy2s. Because H-LINERs and S-LINERs have lower ionization parameters than Sy2s, we expect they should have successively lower nebular temperatures, and for these we assume 12000 K and 11000 K, respectively. We then use the formulation from \citet{proxauf2014} to convert the [SII] doublet into electron density, $\log n_{e}$. 

\begin{figure}[t!]
        \begin{center}
            \epsscale{1.2}

            \plotone{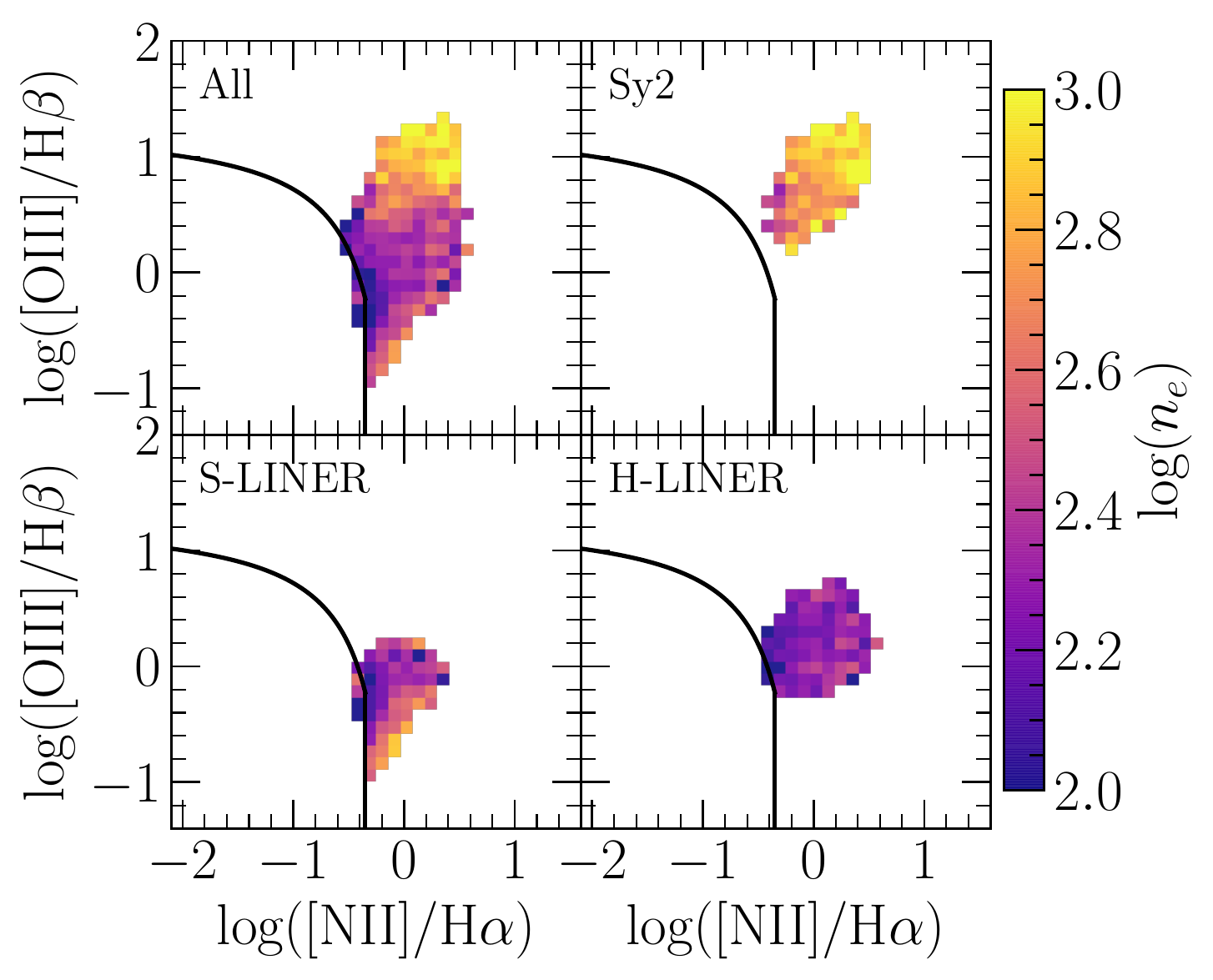}            
            \caption{Pure Seyferts and LINERs (without the SF contribution to emission lines) on the BPT diagram color-coded by electron density probed via the [SII] doublet. Sy2s have higher electron densities than LINERs. Electron density appears to increase slightly up the branch for the Sy2s whereas no clear trend emerges for the S-LINERs and H-LINERs. For this plot alone we require that both components of [SII] doublet be detected with S/N$>2$.
            \label{fig:siidoub}}
        \end{center}
        \end{figure}

We show the pure S/L branch color-coded by electron density in Figure \ref{fig:siidoub}. For the Sy2s, there is a general increase in the electron density with increasing [OIII]/H$\beta$ and [NII]/H$\alpha$. With S-LINERs and H-LINERs, there is not a clear trend. The median electron densities for Sy2s, S-LINERs, and H-LINERs are $\approx$ 620 cm$^{-3}$, 170 cm$^{-3}$, and 220 cm$^{-3}$, respectively.

\vspace{8mm}
\subsubsection{Is the removal of SF contributions critical?}

While we relegate SF mixing to a secondary role, it is important to note we are not arguing that SF contamination should be ignored. To demonstrate this point, we present BPT diagrams before the removal of SF contributions color-coded by the ionization parameter and metallicity in Appendix \ref{app:presub}. The ionization parameter trend is tilted in the version before removal of SF contribution whereas it is primarily dependent on [OIII]/H$\beta$ when SF is removed. The trends for metallicity before removal of SF contributions for the three AGN types are not as strong as afterwards, and this is most apparent when considering the Sy2s. Furthermore, as pointed out, the separation into Seyferts and LINERs may be cleaner after the removal of SF contribution. Altogether, this shows that SF removal is preferable, but is not absolutely critical for understanding the demographics of the S/L branch.        
       
\subsection{Other emission line diagrams} \label{sec:otherdiags}
In this section we consider the implications that our selection of Seyfert/LINERs based on the usual ([NII]) BPT diagram, the removal of SF contribution, and the classification into three types have on the use of other common emission-line diagnostics. We focus on [SII]/H$\alpha$ and [OI]/H$\alpha$ diagnostics \citep{vo87}, the [OIII]/[OII] versus [OI]/H$\alpha$ \citep{kewley2006}, and the WHAN diagram \citep{cidfernandes2011}.
      \begin{figure}[t!]
        \begin{center}
            \epsscale{1.1}
            \plotone{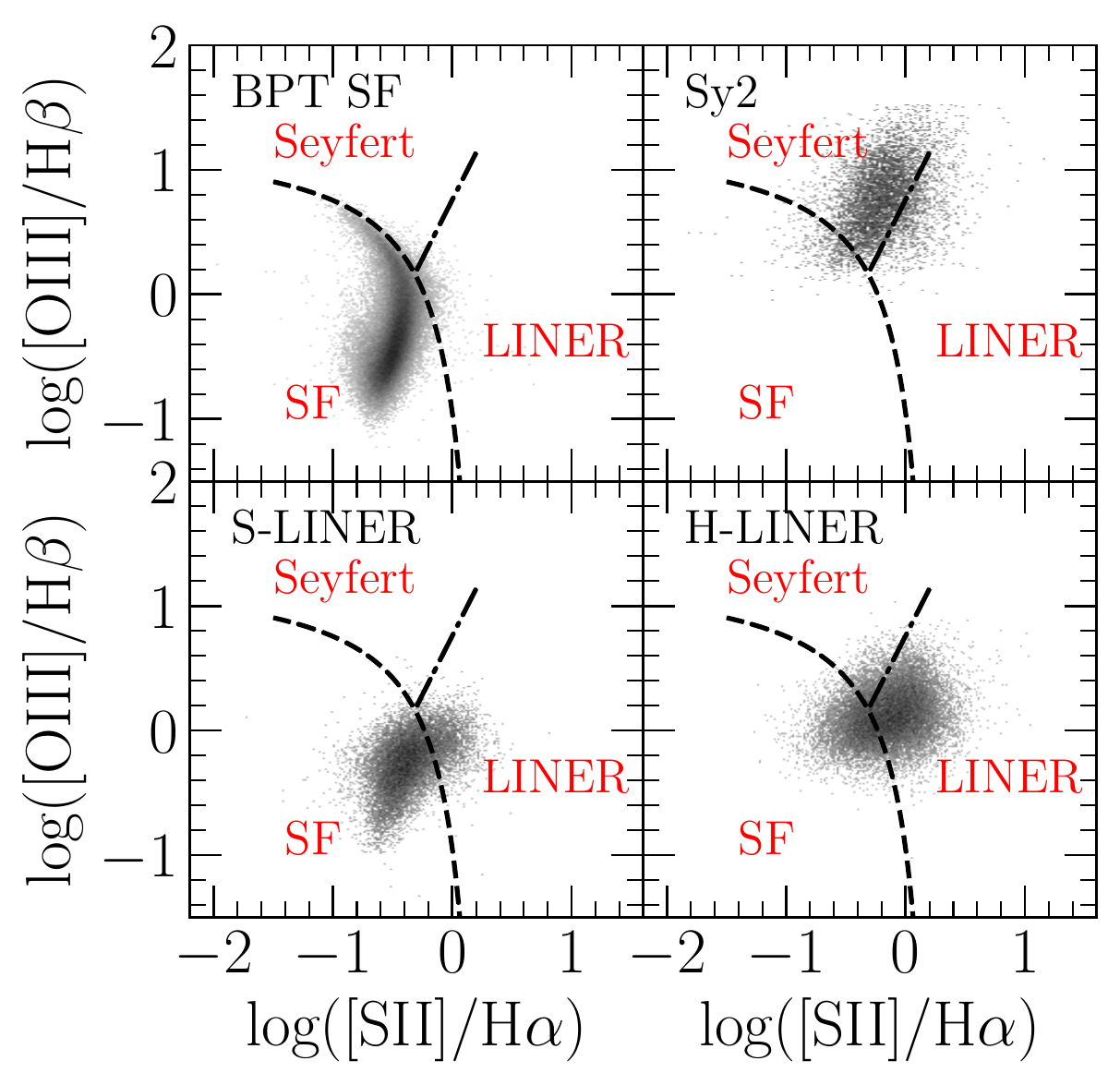}            
            \caption{Pure Seyferts and LINERs (without the SF contribution to emission lines) on the [SII]/H$\alpha$ VO87 diagram. Division into regions labeled ``SF'', ``Seyfert'', and ``LINER'' (red) is according to \citet{kewley2006}, performed on observed (uncorrected) emission lines. Many objects selected as belonging to the Seyfert/LINER branch using the [NII] BPT diagram are found in the SF region, overlapping with the galaxies from the [NII] BPT star-forming branch (upper left panel). This is true both before and after the SF removal, confirming that [NII] is essential for high completeness of sources with extra ionization, especially LINERs.  \label{fig:siiha}}
        \end{center}
        \end{figure}
        
                \begin{figure}[t!]
        \begin{center}
            \epsscale{1.1}
            \plotone{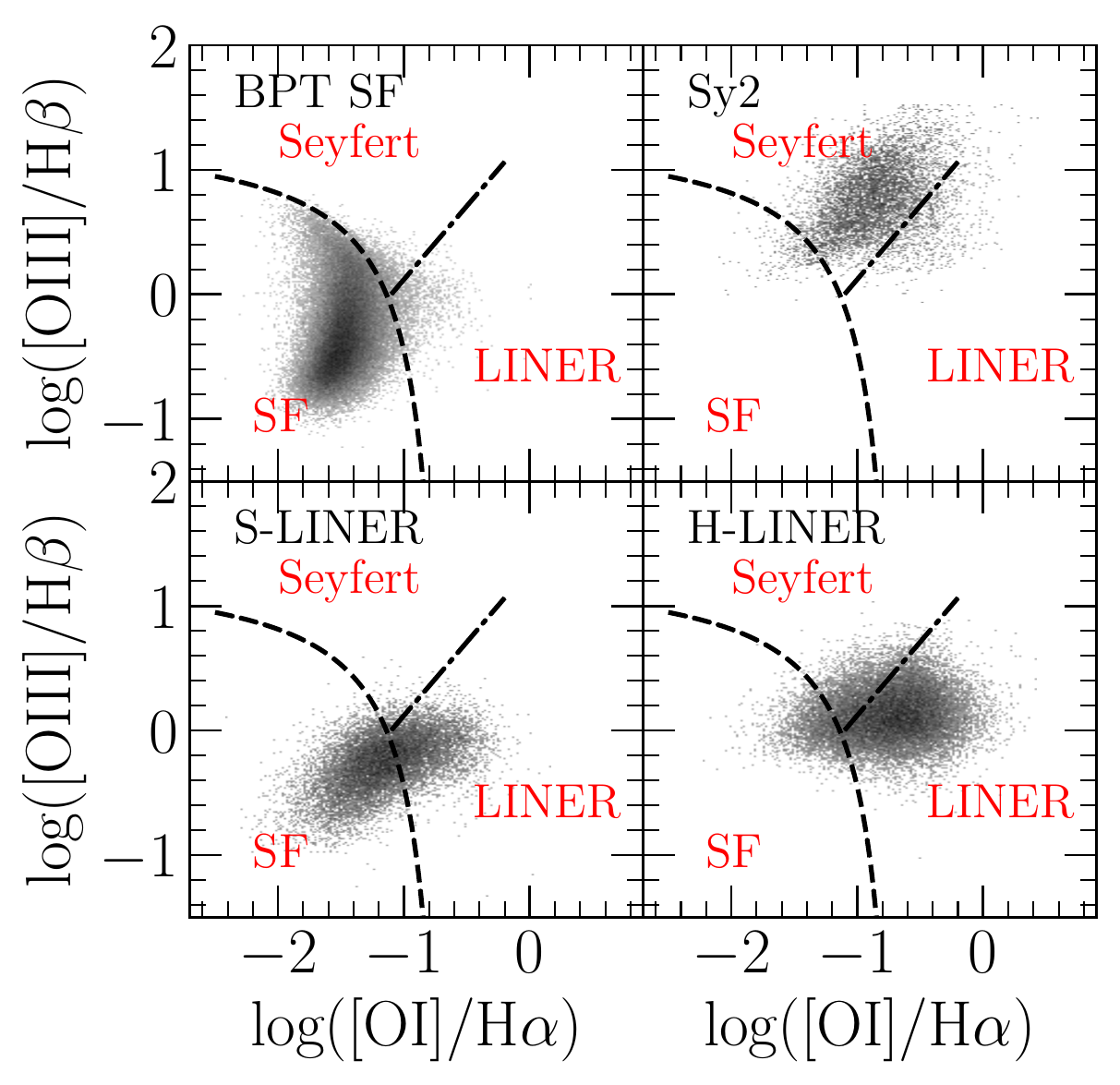}  
            \caption{Pure Seyferts and LINERs (without the SF contribution to emission lines) on the [OI]/H$\alpha$ VO87 diagram. Division into regions labeled ``SF'', ``Seyfert'', and ``LINER'' (red) is according to \citet{kewley2006}, performed on observed (uncorrected) emission lines. The correspondence between our sub-types and \citet{kewley2006} regions, and the overlap of LINERs and [NII] BPT starformers (upper left panel) is similar as for [SII]/H$\alpha$ diagram (Figure \ref{fig:siiha}).
            \label{fig:oiha}}
        \end{center}
        \end{figure}

As previously pointed out by \citet{stasinska2006}, diagnostic diagrams that replace [NII] with [SII] or [OI] do not select objects with extra ionization (Seyferts and LINERs) as effectively as the usual [NII] BPT diagram. Consequently, when placed on the [SII]/H$\alpha$ or [OI]/H$\alpha$ diagnostic plots, many of the objects that are clearly on the S/L branch in the [NII] diagram end up overlapping with the objects from the star-forming branch \citep{tanaka2012a, vogt2014, agostino2019}. Previously it has been proposed (e.g., \citealt{kewley2006}) that the reason for the overlap is the contamination by SF. In  Figures \ref{fig:siiha} and \ref{fig:oiha} we show [SII] and [OI] diagrams of our Seyfert/LINERs from which the SF has been removed. We see that a large fraction of LINERs, and especially S-LINERs, still cross into a region populated by star formers. We conclude that many pure LINERs intrinsically have low [SII]/H$\alpha$ or [OI]/H$\alpha$ ratios like non-S/L (SF) galaxies. Finally, we note that the Sy2/LINER split introduced by \citet{kewley2006} follows the Sy2/H-LINER split in Figure \ref{fig:siiha} and \ref{fig:oiha} reasonably well, though a somewhat less tilted line may be better. Soft and hard LINERs overlap to some extent, except when [OIII]/H$\beta$ is low, a unique signature of S-LINERs.

\begin{figure}[t!]
        \begin{center}
            \epsscale{1.1}
            \plotone{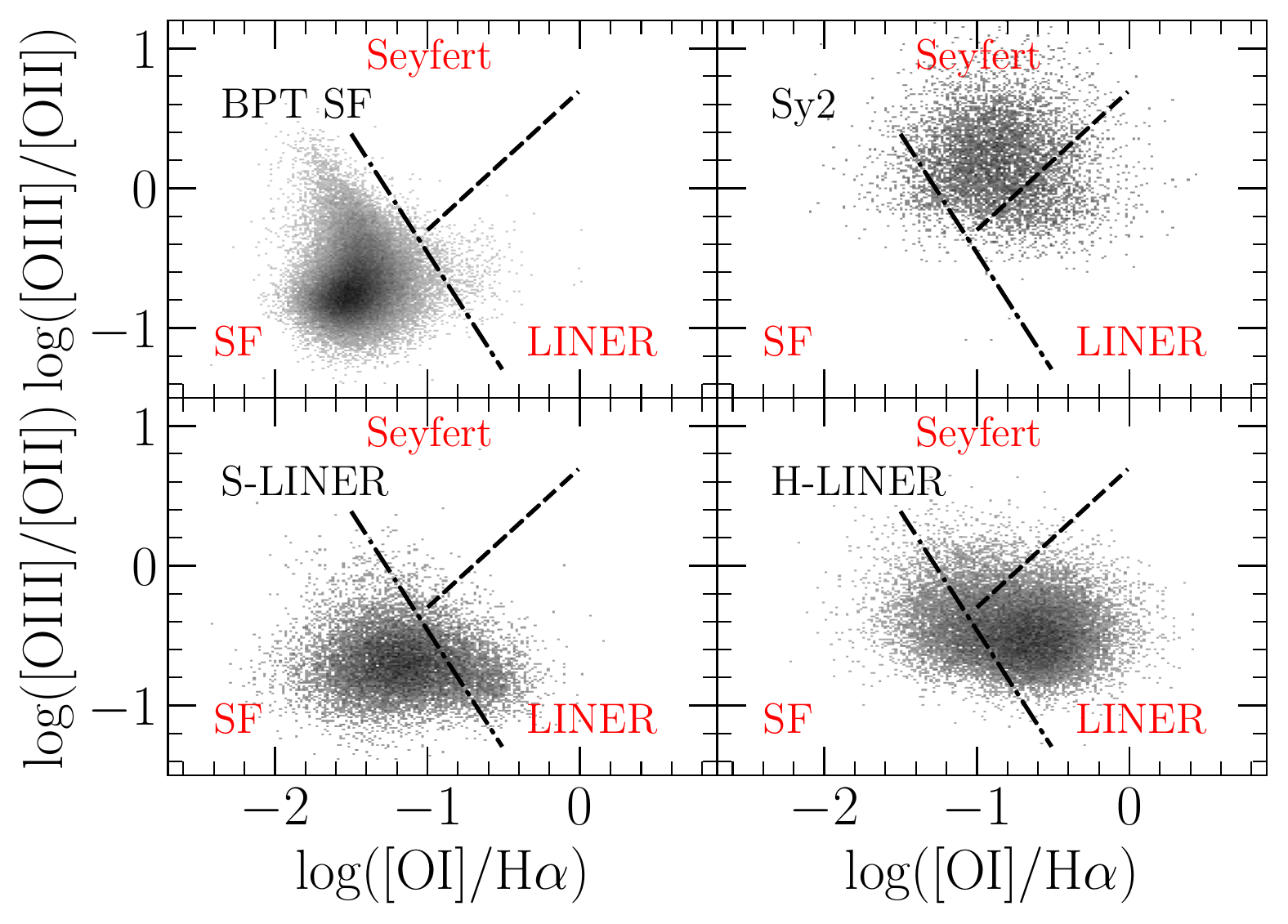}  
            \caption{Pure Seyferts and LINERs (without the SF contribution to emission lines) on the [OIII]/[OII] versus [OI]/H$\alpha$ emission line diagnostic from \citet{kewley2006}. The y-axis represents the ionization parameter and the x-axis probes the hardness of the radiation field. We show the demarcations from \citet{kewley2006} into regions labeled ``SF'', ``Seyfert'', and ``LINER'' (red). The results are similar to those we see with the [SII]/H$\alpha$ and [OI]/H$\alpha$ BPT diagnostics.
            \label{fig:ooo}}
        \end{center}
        \end{figure}

Next, we turn to the [OIII]/[OII] versus [OI]/H$\alpha$ (Figure \ref{fig:ooo}). As expected based on the plots we just discussed, we again find that most S-LINERs fall in the region where they overlap with galaxies selected from [NII] BPT as star-formers. We again argue that such line ratios for S-LINERs are intrinsic. Sy2s and H-LINERs are mostly confined into the Seyfert/LINER regions. We note that the \citet{heckman1980} criteria for separating Seyferts and LINERs, at log([OIII]/[OII])$=0$, works better for our decontaminated  Seyfert/LINERs than the slanted cut of \citet{kewley2006}. 

In Tables \ref{tab:class_pre} and Tables \ref{tab:class_pure}, we present the percentage of each type (Sy2/S-LINER/H-LINER) classified according to the four emission-line diagrams discussed so far, both before and after the removal of SF line contributions. For the [NII]/H$\alpha$ BPT, objects in our sample are classified into SF (below \citet{kauffmann2003} line), composite (between \citet{kauffmann2003} and \citet{kewley2001b} lines), and Sy2s and LINERs (above the \citet{kewley2001b} line and split using the demarcation from \citet{schawinski2007}). For the [SII]/H$\alpha$ BPT, [OI]/H$\alpha$ BPT, and [OIII]/[OII] vs. [OI]/H$\alpha$ diagram, we present classifications into SF, LINER, and Sy2, as defined by \citet{kewley2006}. The tables provide an easy way for readers to estimate how their samples may compare to the different groups.

        \begin{figure}[t!]
        \begin{center}
            \epsscale{1.1}
            \plotone{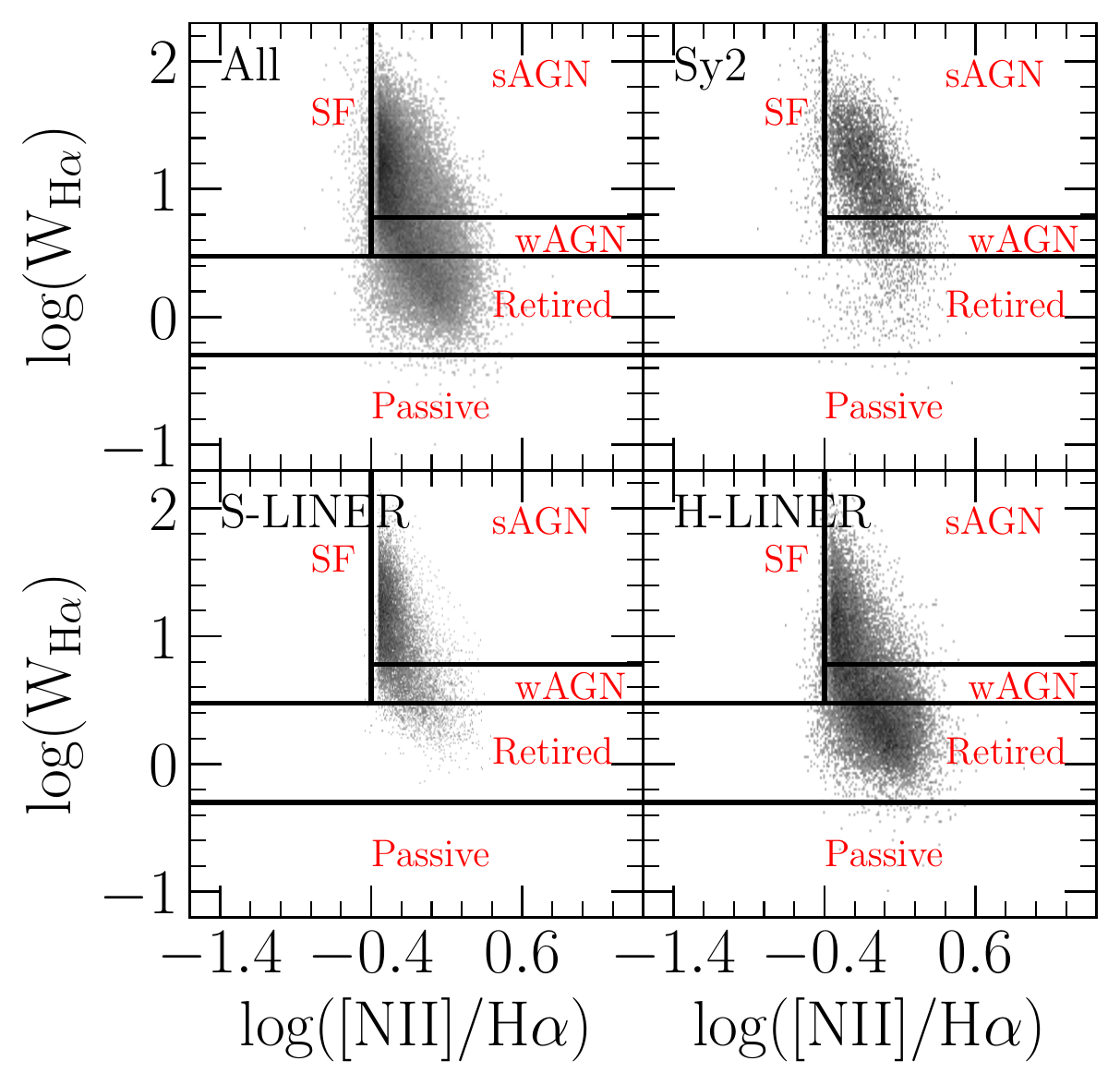}  
            \caption{Seyferts and LINERs on the H$\alpha$ equivalent width versus [NII]/H$\alpha$ emission line diagnostic from \citet{cidfernandes2011}, also known as the WHAN diagram. Lines are not corrected for SF contributions. We show the demarcations from \citet{cidfernandes2011} into regions labeled ``SF'', ``sAGN''(Strong AGN), ``wAGN'' (Weak AGN), ``Retired'', and ``Passive''. Sy2 and S-LINERs are primarily among the strong AGN group. H-LINERs spread across strong AGN, Weak AGN and the Retired groups.
            \label{fig:whan}}
        \end{center}
        \end{figure}

        \begin{figure}[t!]
        \begin{center}
            \epsscale{1.1}
            \plotone{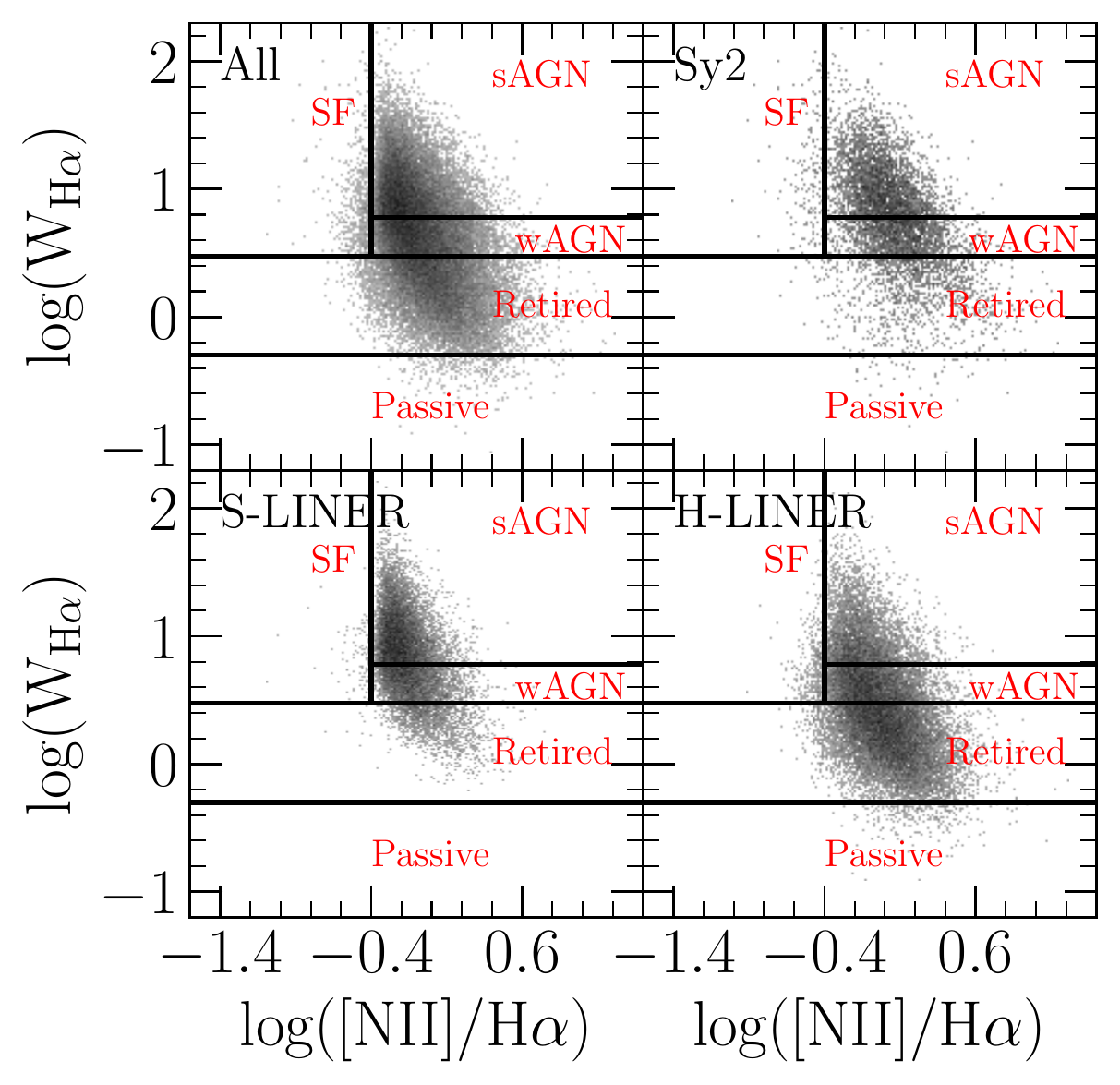}  
            \caption{Pure Seyferts and LINERs (without the SF contribution to emission lines) on the WHAN diagram. The equivalent width generally decreases due to the removal of the SF component, but Sy2 and S-LINERs are still primarily among the strong AGN group, though that portion has decreased compared to the pre-subtraction version. H-LINERs still spread across strong AGN, weak AGN and the Retired groups, but have more among the retired region and less in the strong AGN region compared to the pre-subtraction version.
            \label{fig:whan_sub}}
        \end{center}
        \end{figure}    
Finally, we explore the WHAN diagram \citep{cidfernandes2011} which does utilize [NII]/H$\alpha$, but in combination with the equivalent width of H$\alpha$. This diagnostic diagram was introduced in order to include a greater proportion of all galaxies compared to the BPT diagram, as it only requires information from two emission lines instead of four. WHAN distinguishes between star-forming galaxies, AGNs (strong and weak), retired galaxies (``fake'' AGNs) and passive (lineless) galaxies. \citet{cidfernandes2011} found that the equivalent width of H$\alpha$ could be used as a proxy for the ratio of the extinction-corrected H$\alpha$ luminosity with respect to the luminosity expected from the photoionization by stellar populations older than 10$^8$ yr, and thus galaxies with lower H$\alpha$ equivalent widths are suggested to have their ionization dominated by older stellar populations (retired galaxies), unrelated to an AGN. WHAN splits AGNs into strong and weak also based on H$\alpha$ equivalent width, where the division is supposed to mimic the Seyfert/LINER split. 

We show the WHAN diagram before and after the removal of the SF components in Figures \ref{fig:whan} and \ref{fig:whan_sub}. We compute the decontaminated equivalent widths as 
\begin{equation}
\mathrm{W}_{\mathrm{H}\alpha,\ \mathrm{corrected}} = \frac{f_{\mathrm{H}\alpha,\ \mathrm{corrected}}}{f_{\mathrm{H}\alpha,\ \mathrm{original}}} \cdot \mathrm{W}_{\mathrm{H}\alpha,\ \mathrm{original}}    
\end{equation}
where $f$ is the flux. Both before and after the subtraction of the SF component, the Sy2 and S-LINERs are principally within the strong AGN region, with each leaking into the weak AGN and retired portions as well. The portion in the latter two of these categories increases after the decontamination, and this is a more dramatic effect for the Sy2s. S-LINERs, H-LINERs, and Sy2s are not generally found to lie within the passive region, likely because of our requirement to have S/N$>$2 in all 7 lines. A larger fraction of H-LINERs are present in the retired region compared to the other groups, both before and after the removal of the SF component, but they  are also found within the strong and weak AGN categories. In Table \ref{tab:whan_pure}, we provide the percentages of Sy2/S-LINER/H-LINER in different WHAN categories so that the relative portions of each can be more easily determined. As described in Section \ref{sec:ion}, we have found the three groups generally differ in terms of their ionization parameter, a quantity closely linked to the `strength' of the S/L, yet these groups overlap significantly in the WHAN diagram. It appears that the WHAN diagram does not differentiate Seyferts and non-Seyferts very well. This is most notable in comparing the Sy2s and S-LINERs, which overlap the most in terms of their H$\alpha$ equivalent widths. While the groups do tend to decrease in the equivalent width of H$\alpha$ after the decontamination, there are no clear separations based on the decontaminated quantities, in contrast to what we find for the other diagnostic diagrams wherein the Sy2s and H-LINERs tended to occupy regions similar to those suggested by the Seyfert and LINER demarcations of \citet{kewley2006} based on the observed line ratios.

To summarize the findings from this section, we confirm that the [NII]/H$\alpha$ excess that gives rise to the S/L branch in the regular BPT diagram is the most sensitive indicator of the presence of non-SF ionizing sources, and the fact that some of the [NII] BPT S/L branch galaxies, even after the SF removal may not be distinguishable from purely star-forming ones in other, non-nitrogen diagrams does not warrant their exclusion from AGN/LINER studies. We find the WHAN diagram less useful for distinguishing between Seyferts and LINERs, because many LINERs have high H$\alpha$ equivalent widths, even after the SF removal.

    \begin{deluxetable*}{c|cccc|ccc|ccc|ccc}
     \tablecaption{Percentages of our Sy2/S-LINER/H-LINER in various classification diagrams according to \citet{schawinski2007} and \citet{kewley2006} demarcations, before the removal of SF contributions to lines.  \label{tab:class_pre}
     }
\tablehead{\colhead{S/L Group} &\multicolumn{4}{c}{[NII]/H$\alpha$ BPT}& \multicolumn{3}{c}{[SII]/H$\alpha$ BPT} & \multicolumn{3}{c}{[OI]/H$\alpha$ BPT} &\multicolumn{3}{c}{[OIII]/[OII] vs. [OI]/H$\alpha$}\\
 \omit & \colhead{SF}& \colhead{Composite} &\colhead{Sy2} &\colhead{LINER} & \colhead{SF} &\colhead{Sy2} &\colhead{LINER} &\colhead{SF} &\colhead{Sy2} &\colhead{LINER} &\colhead{SF} &\colhead{Sy2} &\colhead{LINER}}  
    \startdata
  Sy2     & $<1$ & 22 & 69 & 9 & 25 & 68 & 7 & 18 & 77 & 5 & 26 & 62 & 12\\
  S-LINER &26 & 65 & $<1$ & 8 & 88 & $<1$ & 12 & 76 & 2 & 22 & 88 & 1 & 11\\
  H-LINER &3 & 52 & 5 & 39  & 47 & 6 & 47 & 30 & 15 & 55 & 44 & 5 & 51\\
  \hline
  All     & 10 & 52 & 14 & 24  & 57 & 14 & 29 & 43 & 21 & 36 & 56 & 12 & 32 \\
    \enddata
    \end{deluxetable*}

    \begin{deluxetable*}{c|cccc|ccc|ccc|ccc}
     \tablecaption{Percentages of our Sy2/S-LINER/H-LINER in various classification diagrams according to \citet{schawinski2007} and \citet{kewley2006} demarcations, after the removal of SF contributions to lines.    \label{tab:class_pure}
     }
\tablehead{\colhead{S/L Group} &\multicolumn{4}{c}{[NII]/H$\alpha$ BPT}& \multicolumn{3}{c}{[SII]/H$\alpha$ BPT} & \multicolumn{3}{c}{[OI]/H$\alpha$ BPT} &\multicolumn{3}{c}{[OIII]/[OII] vs. [OI]/H$\alpha$}\\
 \omit & \colhead{SF}& \colhead{Composite} &\colhead{Sy2} &\colhead{LINER}  & \colhead{SF} &\colhead{Sy2} &\colhead{LINER} &\colhead{SF} &\colhead{Sy3} &\colhead{LINER} &\colhead{SF} &\colhead{Sy2} &\colhead{LINER}}  
     \startdata
  Sy2     & $<1$ & 4 & 80  & 16 &7 & 76 &17 & 4 & 85 & 11 & 7 & 71 & 22\\
  S-LINER & 10 & 76 & $<1$ & 14 & 81 & $<1$ & 19 & 61 & 2 & 37 & 80 & 1 & 19\\
  H-LINER &2 &38 & 6 & 54  & 30 & 6 &64 & 12 & 15 & 0.73 & 22 & 8 & 70\\
  \hline
  All     & 4 & 46 & 15 & 35  & 43 &15 &42 & 27 & 22 & 51 & 39 & 16 & 45 \\
    \enddata
    \end{deluxetable*}
    
 \begin{deluxetable*}{c|ccccc}
      \tablecaption{Percentages of Sy2/S-LINER/H-LINER classified according to the WHAN diagram, based on their pre-subtraction [NII]/H$\alpha$ and H$\alpha$ equivalent width. \label{tab:whan_pre}
     }
     \tablehead{\colhead{AGN Group}  &\multicolumn{5}{c}{WHAN} \\ \omit &\colhead{Pure SF} &\colhead{Strong AGN} &\colhead{Weak AGN} &\colhead{Retired Galaxies} &\colhead{Passive Galaxies}}
    
         \startdata
 Sy2 & 4 & 75 & 14 & 7 & $<1$\\
 S-LINER &  $<1$ & 73 & 20 & 7 & 0.0\\
 H-LINER & 1 & 33  & 24 & 42 & $<1$\\ 
 \hline
 All & 1 & 53 & 21 & 25 & $<1$\\
     \enddata
 \end{deluxetable*}

\begin{deluxetable*}{c|ccccc}
    \tablecaption{Percentages of Sy2/S-LINER/H-LINER classified according to the WHAN diagram, based on their pure (SF-subtracted) [NII]/H$\alpha$  and H$\alpha$ equivalent width. \label{tab:whan_pure}
    }
   \tablehead{\colhead{S/L Group}  &\multicolumn{5}{c}{WHAN} \\ \omit &\colhead{Pure SF} &\colhead{Strong AGN} &\colhead{Weak AGN} &\colhead{Retired Galaxies} &\colhead{Passive Galaxies}}
    
       \startdata
 Sy2 & 2 & 50 & 28  & 19 &  $<1$ \\
 S-LINER & 1 & 57  & 30  & 12 & 0   \\
 H-LINER & 2 & 19 & 25  & 54 &  1   \\
 \hline
 All & 2 & 36 & 27  & 34 &  $<1$ \\
   \enddata
\end{deluxetable*}


\vspace{-35mm}
    \section{Discussion} \label{sec:discussion}
    With this study, we are able to address a variety of topics relating to the morphology of the S/L branch of the BPT diagram, as well as provide some insights regarding the physical properties of galaxies on it. We compare our method for the removal of SF contribution to previous studies which had similar aims and discuss the implications for the mixing and non-mixing scenarios in Section \ref{sec:dis_mix}. We then discuss the nature of LINERs in Section \ref{sec:dis_liners} and the evidence for a new class of LINERs in Section \ref{sec:dis_two_liners}. Lastly, we consider the host properties of Seyferts and LINERs in Section \ref{sec:dis_hosts}.

    \subsection{Removal of star formation contribution to emission lines and the mixing model}\label{sec:dis_mix}
    
As we pointed out in Section \ref{sec:intro}, the mixing scenario has been prevalent in the literature for at least the last two decades. An early example of the mixing scenario picture is \citet{kewley2001a}, who studied emission line diagrams of far-infrared-selected galaxies using the MAPPINGs \citep{mappings} photoionization code to compute AGN models and starburst models. Based on these models, they constructed a mixing track (sequence) which connects the points they consider to be 100\% starburst (the SF model highest in [NII]/H$\alpha$ and lowest in [OIII]/H$\beta$, i.e., the maximum starburst) and 100\% AGN (the AGN model with highest [NII]/H$\alpha$). While they acknowledged that the objects lying along what now would be recognized as an S/L branch could also be excited by a softer ionizing radiation field from an AGN (essentially, the non-mixing scenario), they considered it likely that the diversity of line ratios seen on the BPT diagram was the result of mixing of different amounts of SF versus AGN contributions, pointing out that the infrared emissions of galaxies had been shown to be of a composite nature. 

Obviously, the verification of the mixing scenario requires the disentanglement of the contributions due to SF on the S/L branch without simply assuming the mixing. Several studies have previously attempted this, using either empirical or modeling approaches. \citet{jones2016} performed this assessment by assuming that the Eddington ratios of AGNs were distributed as a Schechter function and then adding a simulated [OIII] contribution to non-AGNs that match the Seyfert/LINERs by redshift, D4000 index and velocity dispersion. They then set the line ratios of the simulated objects in a way that replicates the observed number distribution of objects on the S/L branch. From the line ratios and the non-AGN emission line fluxes, they determine an AGN contribution fraction. For their sample of Seyfert/LINERs with younger hosts, \citet{jones2016} find that the objects sharing the same position on the S/L branch, especially those close to the \citet{kauffmann2003} demarcation line, have a wide range of SF contributions (from zero to $\sim 100$\%), in excellent agreement with our results. They find more uniformly low SF fractions ($<20$\%) near the top of the Seyfert/LINER sub-branches, whereas we find a large range ($<70$ \%) even there. Note, that the method we employ does not make assumptions about the underlying accretion rate distribution and allows for the scenario where SF contributes different fractions to different BPT lines, which may explain some of the differences between our results.
    
    \citet{tanaka2012a} developed a scheme to account for the SF contributions by fitting stellar populations to the stellar continuum to derive SFRs and masses, from which they predict the [OIII] anf [OII] line fluxes produced by SF. The prediction is based on a calibration for these lines with respect to H$\alpha$ and the galaxy's stellar mass as a proxy for the metallicity. They compare the expected SF fluxes and observed fluxes for these lines and identified AGNs (or, more broadly, the objects with non-SF ionizing source) as objects which contained emission in excess of the SF contributions, which are referred to as oxygen-excess AGNs. The oxygen-excess AGNs are then used to test the merits of various AGN selection techniques, including the X-ray and radio selections. It is reasonable to assume that a substantial portion of genuine AGNs, especially low metallicity or low ionization AGNs, will not be identified using the oxygen-excess method. Similarly, the relatively low 1.5-sigma threshold used in identifying the excess will likely misidentify some SF galaxies as AGNs, contaminating the AGN sample. As the method requires an empirical calibration for estimating the SF emissions in [OIII] and [OII], it may not be effective in accounting for the SF emission in other lines.

   \citet{thomas2018, thomas2019}  tackle the problem of SF contamination, in particular in Seyferts, by fitting the observed emission lines with a combination of SF (HII) models and AGN (NLR) models produced using the Bayesian parameter estimation software NebulaBayes \citep{nebulabayes}. They conclude that the mixing dominates, at least for Seyferts. This can be seen in Figure 2 of \citet{thomas2019}, where the mixing fraction, which they define as the fraction of H$\beta$ produced by the narrow-line region ($f_{\textrm{NLR}}$) is shown to be a tight function of the distance along  the Seyfert 2 branch on the BPT diagram ($d=0$ being the intersection with the \citealt{kewley2001b} maximum starburst line). They have a low NLR fraction of  $\sim0.15$ (SF dominates) at $d=0$ which increases sharply as $d$ increases, with a small scatter throughout. Their nearly pure Seyferts ($f_{\mathrm{NLR}}>0.8)$ are restricted to the very top of the Sy2 branch ($d=0.75$). In contrast, we find, also for Seyferts, that whereas $f_{\textrm{NLR}}$ does on average rise with $d$, there is an enormous scatter in  $f_{\textrm{NLR}}$ at any $d$. For example, at $d=0$ we already see a high average NLR fraction ($\sim$0.5), with a large scatter from $\sim0.2$ to $\sim0.7$. We find that pure Sy2 ( $f_{\mathrm{NLR}} \sim 1$) can be found already at $d=0$. Our results imply that the intrinsic diversity is more important than mixing. Consequently, the split of the S/L branch of the BPT diagram using the maximum starburst line has little practical significance because of the wide range of star formation contributions anywhere on the branch.

   Why do the two approaches produce such different results? While NebulaBayes allows a great diversity of physical properties, and \citet{thomas2018, thomas2019} take ionization parameter, ISM pressure, and metallicity to vary, some other parameters in their analysis are kept fixed. Most critically, they assume a single peak energy (and thereby a hardness) for all Seyferts. As shown in Figure 2 of \citet{thomas2018}, relaxing this condition and/or using peak energies lower than their adopted value of 45 eV leads to a much wider range of NLR fractions at a given location on the branch. Furthermore, their adopted peak energy may be too high for the black hole masses typical in AGNs, though this also depends on the Eddington ratio. Figure 3 of \citet{thomas2016} suggests $E_{\textrm{peak}} \sim 5$ eV for $\log (L/L_{\mathrm{edd}})=-2$ (the median value for the Sy2s found using the bolometric correction on [OIII] luminosity from \citet{kh09} and with BH masses from M-$\sigma$) and $\log M_{\mathrm{BH}}=7.5$ (slightly lower than the median for Sy2s at 7.71 but the result is similar with the grid line for $\log M_{\mathrm{BH}}=8$). Similarly, \citet{trump2011} propose $E_{\textrm{peak}} < 4$ eV. Fundamentally, the challenge with the modeling efforts to remove SF contributions is that the hardness of the ionization field and the mixing fraction are degenerate.

    \citet{ji2020b} approach the problem of SF contamination by allowing for the mixing of SF and AGN flux in Bayesian likelihood estimations of model metallicities and ionization parameters of central MaNGA spaxels. They also propose a 3D extension of the BPT diagram which includes [SII]/H$\alpha$, and derive a re-projection with axes termed $P_{1}$, $P_{2}$, $P_{3}$. In particular, they point out that $P_{1}$ maximizes the gap between AGN and non-AGN models. According to these models, the galaxies lying in the gap must then be the result of mixtures. We inspected our decontaminated line ratios on $P_{2}$ versus $P_{1}$ diagram and find that even though the removal of SF somewhat reduces the spread of $P_1$ values for Sy2s and H-LINERs, and centers the Sy2s on the $f_{\textrm{AGN}}$ =0.9 line (where  $f_{\textrm{AGN}}$ is the emission due to an AGN in H$\alpha$), there is still a $\sim 1$ dex range of $P_1$ values in each category, again suggesting that the intrinsic diversity seems to be present. Indeed, their Figure 7 suggests that having a wider range of power-law slopes of ionizing SEDs than their adopted model with only one slope would have resulted in a wider intrinsic $P_1$ spread without the need for mixing.

    \subsection{The nature of LINERs}\label{sec:dis_liners}
    
    There is a long-standing debate regarding whether LINERs are predominantly powered by an AGN or some other source. In our analysis so far we did not have to assume any particular mechanism. Whatever is powering the LINERs, the majority of them are unarguably contaminated by SF that needs to be removed. Furthermore, if the diffuse ionized gas contributes to the LINER emission that mostly comes from some other principal source, then its contribution will also be removed by our matching procedure, since we expect it to be correlated with our matching parameters. Here, we do not intend to provide a comprehensive review of this extensive topic, but only present some potential insights based on our analysis. 
    
    Early studies of LINERs have preferred the AGN interpretation (e.g., \citet{heckman1980, ho1993c}) based on the presence of strong X-ray emission and/or broad H$\alpha$ lines, among other clues. Others have suggested that the LINER emission is powered by post-AGB stars \citep{binette1994, stasinska2008, sarzi2010, cidfernandes2011, yan2012, singh2013, belfiore2016, gomes2016, jones2017dig, zhang2017,  byler2019} or shocks \citep{dopita1995, dopita1996, kewley2001b, rich2010, rich2011, rich2014, davies2017, molina2018, dagostino2019a, dagostino2019b}, though it should be pointed out that some of these studies focus on passive (red) LINERs, which are by no means the majority of all LINERs. See \citet{ho2008,heckman2014} for a more thorough overview of LINERs.  IFU studies have confirmed the large spatial extent of the LINER region, one of the original reasons to doubt their association with AGN \citep{guo2019}. On the other hand, even Sy2 have emission regions with AGN-like lines extending out to $\sim$10 kpc \citep{chen2019nlr}, so it is unclear if the extent of the emitting region is a sufficient argument in itself for or against the AGN mechanism. Recent studies using X-ray selected AGNs find similar fractions of X-ray AGNs among Sy2s and LINERs \citep{agostino2019} and that X-ray variability is common among LINERs \citep{marquez2017}, though this does not necessarily mean that the AGN is responsible for the observed optical emission line ratios through photoionization. Based on the energetics of low accretion rate AGN \citep{eracleous2010}, and the observed radial profile of ionization parameter \citep{yan2012, belfiore2016}, the AGN may not emit enough ionizing photons to produce the line ratios observed on the scales probed by the SDSS spectroscopic fibers. Instead, it could be the case that AGN-driven jets or outflows are providing the energy to ionize the gas through shocks.  Recent spatially resolved observations of nearby AGNs with HST have found regions of Sy2-like line ratios surrounded by LINER-like line ratios with associated radio jets or outflow signatures, suggesting that they can be a result of AGN-related activity \citep{ma2021}. However, shock models are still unable to explain the hot temperatures observed in LINERs \citep{yan2018}.
    
One argument that we have not seen put forward in the literature is that if the LINERs were the result of the stellar populations alone, one would expect them to be ubiquitous among the galaxies of a certain mass, age (or sSFR), gas content and central mass concentration, and consequently all such galaxies would have to be LINERs. Our analysis shows that this is not the case. Non-LINER galaxies with host properties similar to LINERs exist in all cases. Specifically, the median match distances in multi-dimensional space of host properties (Eq. \ref{eq:dist}) for Sy2s, S-LINERs, and H-LINERs are all very similar. We discuss this in more detail in Appendix \ref{app:dists}. The existence of similar galaxies with and without LINER lines may suggest a transient rather than a secular phenomenon. Furthermore, LINERs are found in hosts with a wide range of star formation (Section \ref{sec:dis_hosts}), not just low sSFR, so an explanation tied specifically to an older stellar population seems incomplete. An alternative explanation could be that the stellar population giving rise to LINER lines exhibits some sort of drastic variability. \citet{lauer2005} suggested that a stellar-based LINER phenomenon could have a cyclical nature through a process where dust clouds, which hint at the presence of gas, form and then fall in to the central region of the galaxy where they are destroyed over a timescale of $\sim10^{8}$ yr. The associated infalling gas would be ionized by hot evolved stars. 

    If LINERs are AGNs, then their emission line properties may be due to a distinct accretion state. It has been suggested that the hardness of the radiation field increases as the Eddington ratio decreases for SMBHs with log($M_{\mathrm{BH}}$)$>7$ \citep{cann2019}. We find that the H-LINERs also have the largest SMBH masses ( $<\log(M_{\mathrm{BH}})>$=7.87 versus 7.71 and 7.69 for Sy2 and S-LINERs), as suggested by the stellar velocity dispersion, which, combined with their low ionization parameter, suggests that the H-LINERs would have lower Eddington ratios than Sy2s, and thereby a harder ionizing radiation field, which we see with the [OI]/[SII] ratios. If the Eddington ratios for H-LINERs are indeed quite low, the accretion may be in the form of a radiatively inefficient accretion flow \citep[RIAF, ][]{yuan2004} rather than a traditional optically thick, geometrically thin \citet{shakurasunyaev1973} disk, and which may explain why they have distinct emission line properties from that of the S-LINERs or the Sy2s. RIAFs would not provide a strong ionizing continuum to the nearby interstellar medium, but can still produce significant amounts of X-ray emission which can result in large partially ionized zones and thereby LINER-like emission line ratios. A problem with this picture is that the S-LINERs appear to have similar Eddington ratios to those of the H-LINERs, and so an additional component (whether the accreting gas is cold versus hot) may be necessary to explain their difference, which we discuss more in \ref{sec:dis_hosts}. At low accretion rates, AGN-driven jets could provide additional energy to the nearby ISM through shocks, though uncertainties remain. Indeed, recent modelling efforts have disfavored shocks as a dominant ionization source for LINERs (e.g. \citealt{yan2018,ji2020a}), and we find that neither the H-LINERs nor S-LINERs have extremely high electron densities, which one might expect for a shock-driven gas.
    
To conclude, our work does not require us to assume the nature of LINERs. We are drawing attention to what may be another element of the puzzle, that having a LINER contribution is by no means inevitable in a population of galaxies that have similar stellar populations, gas content and structure.  

    \subsection{Two Types of LINERs?}\label{sec:dis_two_liners}
    
    In this work, we find that the LINER population selected using the [NII] BPT may be composed of two sub-groups, which mostly differ by the hardness of their ionizing radiation. In previous works, it has been noted that there are LINER-like objects with distinct line ratios from the traditional LINERs, but not much work has been done to characterize them. \citet{filippenko1992} and \citet{shields1992} suggested weak [OI]/H$\alpha$ LINERs could be powered entirely by early O stars, though such a scenario seems unlikely to explain the line ratios of S-LINERs as most are not found in extreme starbursts (Section \ref{sec:dis_hosts}). \citet{ho1993b, ho1993c} and \citet{ho1997} suggested that the weaker [OI]/H$\alpha$ in what they called ``transition objects'' is due to SF contamination. Our S-LINERs extend to lower [OI]/H$\alpha$ than their transition objects, even when we remove the SF contributions, so while the two groups may overlap, our S-LINERs are a more comprehensive category than transition objects of Ho and collaborators. \citet{kewley2006} suggested that the transition objects are part of the regular LINERs, which again would be plausible if they only appear different because of the SF contamination. On the other hand, we find that LINERs form a more diverse population even when SF is removed. 

Recently we learned that \citet{yesufho2020} performed a clustering of SDSS galaxies based on the values of the principal components obtained from a combination of stellar population, structural and emission line measurements, and including only one line ratio ([OIII]/[OII]). They identify five clusters, of which two are exclusively star-forming (C0 and C1), whereas three have Sy/LINER contributions. Remarkably, the latter three clusters (C2, C3 and C4) qualitatively agree with our S-LINERs, Seyferts and H-LINERs, respectively. Like our S-LINERs, their C2 galaxies mostly lie in the so called composite region of the BPT diagram, while partially overlapping with the star-formers in the [OIII]/H$\beta$ vs.\ [OI]/H$\alpha$ diagram. Where we differ is in the interpretation. We see all S-LINERs as having an extra ionization with respect to the star formers, whereas \citet{yesufho2020}, because of their adoption of the AGN classification based on the [OIII]/H$\beta$ vs.\ [OI]/H$\alpha$ diagram, see them as predominantly star forming. Nevertheless, it is very encouraging that what we refer to as S-LINERs were recognized as a category distinct from the traditional LINERs using an independent methodology.

Based on our results and the extensive reading of the literature, our principal message is not to discard the objects that lie near the base of the S/L branch on account that they coincide with the SF galaxies in non-[NII] diagnostic plots. Rather, they should be treated as a softer ionization continuation of the usual LINER population. Indeed, \citet{ho1993a} demonstrated almost two decades ago that softer power-law sources (steeper than $-2$) do result in line ratios which lie within the SF locus on the [OI] and [SII] diagnostic diagrams. It is unclear why many subsequent studies assume much smaller ranges of models, or even just one, relatively hard, set of models.

            \begin{figure}[t!]
        \begin{center}
            \epsscale{1}
            \plotone{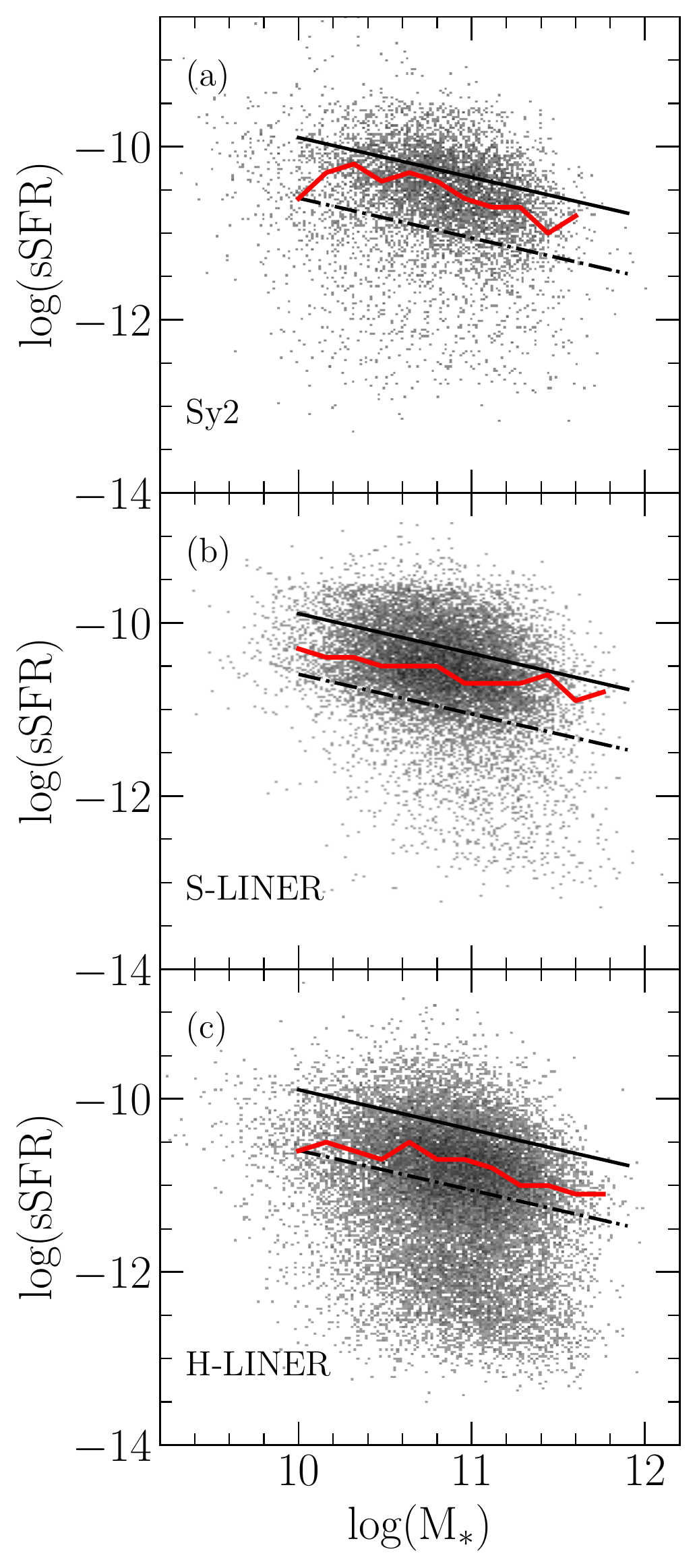}
            \caption{sSFR-M$_{*}$ diagram for Sy2s in top panel (a) , S-LINERs in middle panel (b), and H-LINERs in bottom panel (c). Solid black line and dash-dotted black line show the local main sequence and its lower bound, respectively. Solid red line shows the mode of each distribution which is determined by computing a histogram of sSFRs (with bin sizes of 0.1 dex) for each mass bin of 0.2 dex and taking the sSFR bin with the most counts. Sy2s and S-LINERs are primarily in hosts with ongoing star formation, whereas H-LINERs are in hosts which have a wide range of star formation, including essentially quiescent galaxies (log sSFR$<-12$) but the most common among them lies somewhere between the main sequence and the cutoff of the green valley. \label{fig:ssfrm_agns}}
        \end{center}
        \end{figure}    
         
        \begin{figure}[t!]
        \begin{center}
            \epsscale{1.2}
            \plotone{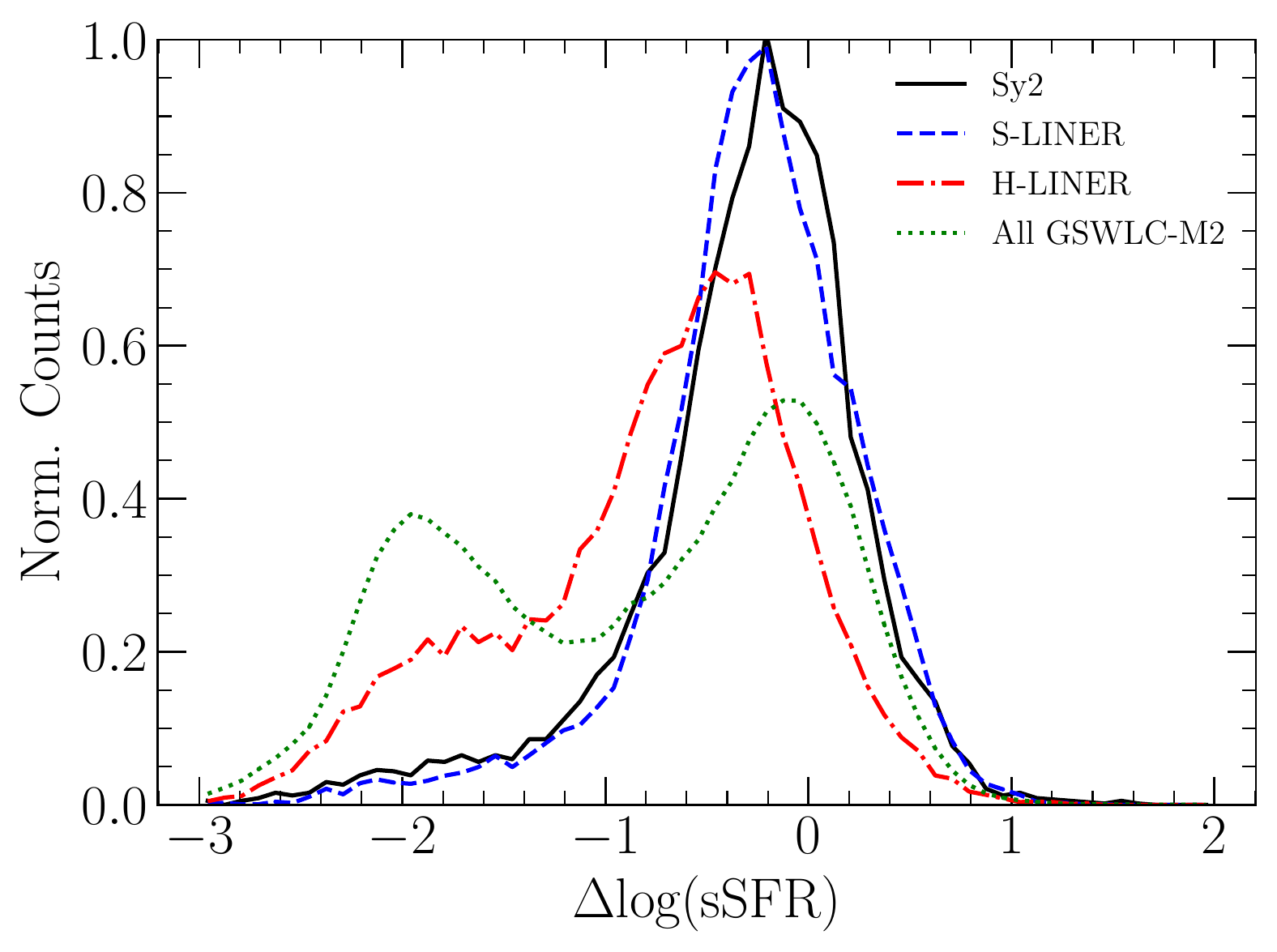}
            \caption{Normalized histograms of distance from the star-forming main sequence ($\Delta$log(sSFR)) for each of the three S/L types and for all galaxies from GSWLC-M2 shown as a reference. As shown in the sSFR-M$_{*}$ diagram, Sy2s and S-LINERs predominantly lie in hosts that are on or near the star-forming main sequence ($\Delta$log(sSFR)=0) and are remarkably similar to each other in their hosts. The distribution of H-LINERs also peaks between the main sequence and the green valley ($\Delta$log(sSFR)$<-0.7$) but has a wider tail towards quenched galaxies.  \label{fig:delta_ssfr_agns_hists}}
        \end{center}
        \end{figure}

    \subsection{Host properties of Seyfert/LINERs} \label{sec:dis_hosts}
    
    Another aspect to consider for understanding the distinct emission line properties of Seyfert/LINERs is their host's star formation, since it relates to the amount of fuel available for AGN accretion and/or the gas available to be ionized. We show the sSFR-M* diagram for the three types of objects on the S/L branch in Figure \ref{fig:ssfrm_agns} and histograms of $\Delta$log(sSFR) (the distance from the star forming main sequence) in Figure \ref{fig:delta_ssfr_agns_hists}. In the latter, we also show the distribution for all galaxies in GSWLC-M2 (including the S/Ls) as a reference. The first thing to note is that all three types have a similar stellar mass range, which allows them to be potentially evolutionary connected. 

The Sy2s and S-LINERs are typically found to lie on the star-forming main sequence, and their distribution in $\Delta$log(sSFR) is remarkably similar, 

The Sy2s are typically found to lie on the star-forming main sequence (slightly offset downwards from its ridge), which is expected since they are accreting at the highest rates, requiring the presence of significant supplies of cold gas. The S-LINERs are also primarily on the main sequence, so if they were AGN they would also be accreting cold gas, but at lower rates than the Sy2s. This is indeed seen in the analysis of \citet{yesufho2020} whose cluster C2 galaxies are qualitatively similar to our S-LINERs.  The S-LINERs could still have their NLR powered by accretion disk photons but with a lower peak energy than that of the Sy2s, which would simultaneously result in the S-LINERs having a lower ionization parameter. S-LINERs and Sy2 could therefore be more closely connected than S-LINERs and H-LINERs ion the sense that they represent different levels of activity of otherwise similar population of galaxies. Indeed we see in Figure \ref{fig:delta_ssfr_agns_hists} that their distribution in $\Delta$log(sSFR) is remarkably similar. This resonates with the findings of \citet{yesufho2020} that their C2 and C3 (Seyfert) clusters have comparable gas contents and stellar populations. They also propose that the two groups may be evolutionary linked. The difference in our pictures is that (if they are indeed AGN) we see S-LINERs as \emph{all} being at a lower level of activity, rather than having a smaller \emph{fraction} of active objects. As discussed in Section \ref{sec:dis_two_liners}, this difference arises from \citet{yesufho2020} choice to use [OI]/H$\alpha$ as the basis for calling something an AGN or not.

A problem with the picture that connects S-LINERs and Sy2 is that we might expect a more gradual gradient in the ionization parameter among S-LINERs where they connect to the lower end of the Sy2s, but there is instead a significant jump in the ionization parameter between them (Figure \ref{fig:bpt_U_ccode}). Alternatively, the S-LINER vs.\ Sy differences may hint at a transition in the accretion mode, perhaps to one with an inner RIAF and an outer accretion disk (i.e., as described in Figure 9 of \citealt{trump2011}).

In contrast to S-LINERs and Sy2, H-LINERs span a wide range of sSFRs, from those on the main sequence to the green valley and the red cloud. Still, the most common position for the H-LINERs is in the lower portion of the main sequence, just 0.1 dex lower in sSFR than the former. This may seem contrary to the usual picture of the Seyfert vs.\ LINER dichotomy in terms of SF. We suspect that the notion that LINERs are predominantly found in quiescent hosts is a consequence of basing the classification on [OI] or [SII] diagrams, or even on the [NII] BPT, but requiring high [NII]/H$\alpha$ thresholds, which are more commonly found in galaxies with low sSFR \citep{cidfernandes2010}. Furthermore, we need not lose from sight the fact that the massive galaxies on the main sequence are optically red (like M31), even though they are far from being quiescent \citep{salim2014}.

If H-LINERs are AGN powered, it might be the case that their AGN has regulated its own accretion through heating of the nearby ISM (perhaps during a previous Sy2-like phase) but that the physical scale of such an effect is not large enough to significantly quench star formation on the main sequence.  On the other hand, it may be that the AGN has more recently changed in character and a larger portion of the galaxy will eventually become quenched as a result of the AGN activity. A third possibility, is that the host galaxy's consumption of the gas itself is the primary driver of quenching, as has been suggested by \citet{yesufho2020}. It may be the case that the primary distinguishing factor in the different emission line properties of the S-LINERs (plus Sy2s) and H-LINERs is in the accretion of cold versus hot gas, leading to distinct accretion flows (i.e., a RIAF) and ionizing properties \citep{hardcastle2007}. 
  
    We have also found that the H-LINERs have the lowest oxygen abundances among our three groups (Figure \ref{fig:bpt_logoh_ccode}), which is puzzling but may be because many of them are in hosts which are already quenched, and so they did not have enough time to build up significant amount of metals. A problem with this picture is that even at similar masses and sSFRs, H-LINERs appear to have lower oxygen abundances than both Sy2s and S-LINERs. It may be the case that outflows (potentially AGN-driven) carried significant portions of their metals out of the host, and the lower abundances could help explain the hot temperatures observed in LINERs as they would be less efficient at cooling. A more nuanced investigation into the properties of low-ionization gas and a description of the difficulties in reproducing their observed line ratios with models are given in \citet{yan2018}.  
    
                \begin{figure}[t!]
        \begin{center}
            \epsscale{1.2}
            \plotone{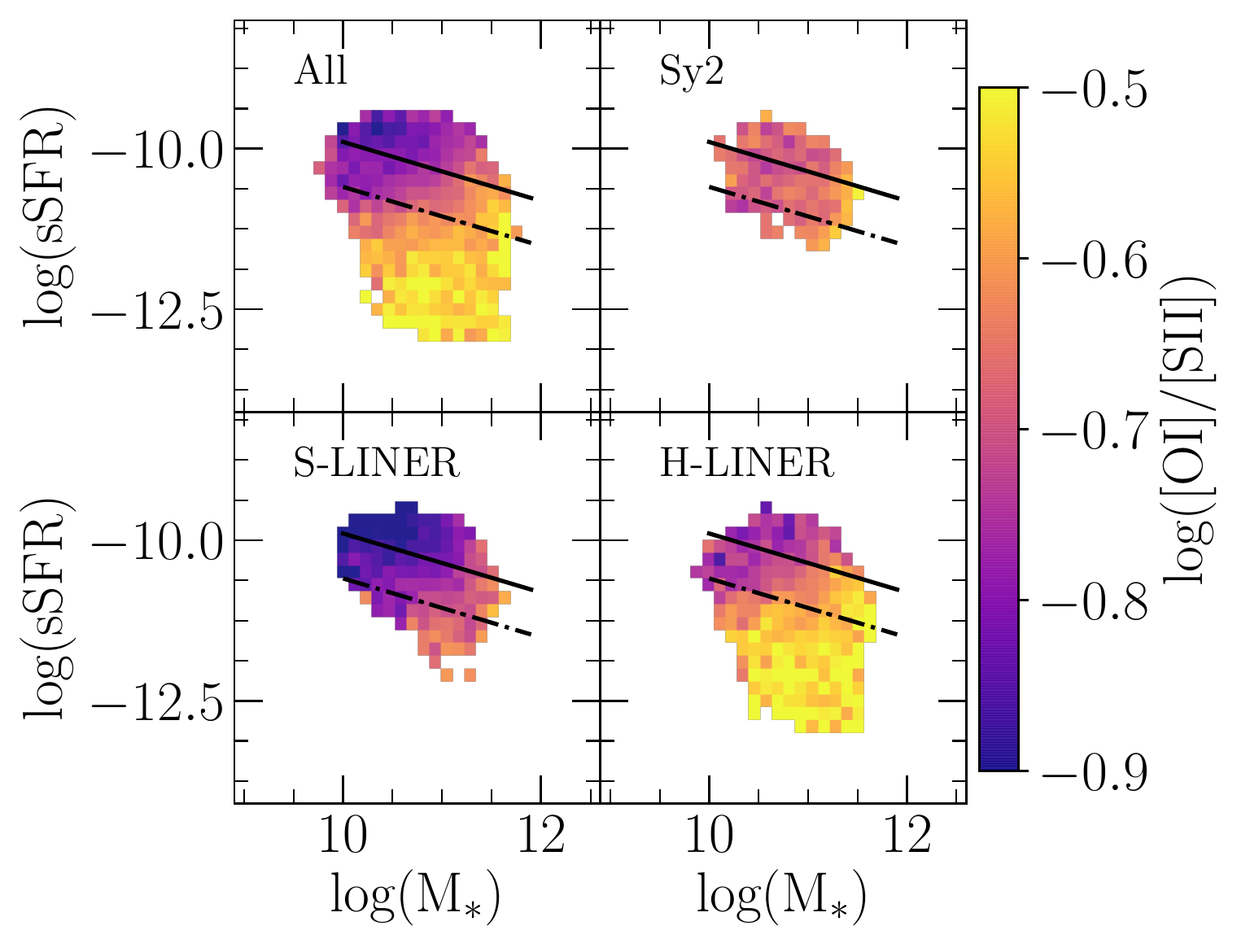}
            \caption{sSFR-M$_{*}$ diagram of Seyfert/LINERs, color-coded by the hardness ([OI]/[SII]). Solid black line and dash-dotted black line show the local main sequence and its lower bound, respectively. The hardness of S-LINERs and H-LINERs shows a correlation with their host properties, whereas the same is not found for the Sy2s.  \label{fig:ssfrm_agns_hardness}}
        \end{center}
        \end{figure}    
\citet{yesufho2020} proposed a scenario where galaxies with and without strong AGN activity evolve into the LINER region. As pointed out, based on their locations on the [OI]/H$\alpha$ diagram as well as their [OIII] luminosities, their clusters C2, C3, and C4 are largely similar to our S-LINERs, Sy2s, and H-LINERs, and so a meaningful comparison is possible, with a caveat that they base their classification on observed lines. \citet{yesufho2020} suggest that quenching is not related to the activity of the AGN, and that their clusters evolve from C2 or C3 to C4 when the host sufficiently consumes its available gas and thus quenches, a claim based on their finding that the gas reservoirs and the ability to form stars in the hosts of C3 (Seyferts) are not instantaneously affected by the AGN activity. The results of our work agree with their data, as we similarly find Sy2s (and S-LINERs) are largely in galaxies with ongoing star formation. However, it is possible that the Sy2s would self regulate and that they would already convert to H-LINERs before their hosts would have enough time to quench and so a causal link may be missed because of the different timescales for the nebular emissions adjusting to a change in the power source versus the quenching. In this case, the host star formation of the C3 (Sy2s) would not be sufficient to rule out the role of AGNs in quenching
    
Recent work by \citet{chen2020} suggests the cumulative amount of radiated energy needed to begin measurable quenching is approximately four times the halo-gas binding energy, which might be accomplished in a relatively short amount of time if an AGN’s accretion rate is high enough. Additionally, S-LINERs could slowly convert into H-LINERs if they have a steady, low accretion-rate AGN which over a long time imparts enough energy into the host galaxy to begin quenching. Contrary to this, it could be the case that the AGN's role is not so significant and that the quenching is mostly driven by the host's consumption of gas \citep{yesufho2020}. To explore the extent to which AGN-driven (active) or host-driven (passive) quenching may be applicable, we look at the sSFR-M* diagram color-coded by the hardness of the ionizing radiation in Figure \ref{fig:ssfrm_agns_hardness}, as the hardness might vary with the amount of cold gas present, which is dependent on the host properties. We find trends of increasing hardness along the star-forming main sequence for S-LINERs and H-LINERs, with the latter increasing towards the red cloud. This would be sensible if the hardness was linked to whether or not the host is accreting cold versus hot gas. The dependence of hardness on host properties for S-LINERs and H-LINERs hints at a passive quenching through the host's consumption of gas in star formation. Most interestingly, however, is the fact that the Sy2s do not show a clear trend in hardness as one moves across the sSFR-M* diagram, and so perhaps they are in such a high accretion mode due to a recent inflow and the new reservoir of cold gas may wipe out any former dependence of the hardness on the host properties. Then, after the Sy2 phase they may be more like H-LINERs. Such a scenario may help explain why some H-LINERs are still on the star-forming main sequence. 

Evolutionary scenarios are outside of the scope of this work, but such studies would also benefit from being able to isolate the intrinsic Seyfert/LINER emission.


 \section{Conclusions}
In this study, we statistically remove the host SF (and diffuse ionized gas contributions, if present) from the emission lines of SDSS galaxies lying on the Seyfert/LINER branch of the [NII] BPT diagram, with the aim to establish the extent and the physical drivers of intrinsic diversity in their line ratios.

Our conclusions are the following: 
\begin{enumerate}

\item The extent of the Seyfert/LINER branch in the observed BPT diagram of galaxies in the local universe is primarily the result of the diverse physical conditions of the NLRs or nuclear ISM, and is not primarily due to the various levels of SF contamination (the mixing scenario). This holds for both the Seyfert and the LINER sub-branches. As a result, when the SF contributions are taken out, the morphology of the Seyfert/LINER branch does not change appreciably.

\item While the average SF contribution drops along the observed Seyfert/LINER branch, there is a very wide range of SF contributions at any position in the branch, both within and above what is referred to as the ``composite'' region. Galaxies from the ``composite'' region should therefore not be excluded from AGN/LINER studies, as many of them are intrinsically different from objects nearer the top of the branch.

\item While not a strong confounding factor, the SF contamination does produce an average shift of 0.2 dex on the BPT diagram and should ideally be removed in order to improve the accuracy of studies of Seyferts/LINERs using SDSS emission lines. 

\item Once the SF contribution is taken out, the position of Seyferts and LINERs on the BPT diagram is to first order governed by ionization parameter ($U$, along the Seyfert/LINER branch) and the metallicity ($Z$, horizontally across the Seyfert/LINER branch), in accordance with the model grids. However, Seyfert and LINERs have a relatively wide range of values of ionization field hardness, which is not captured by simple 2-parameter models, making the correspondence between the location on the Seyfert/LINER branch of the BPT diagram and $U$ and $Z$ somewhat fuzzy.

\item Consequently, additional line ratios, such as [OI]/H$\alpha$ or [SII]/H$\alpha$, are needed to better differentiate Seyferts and LINERs and reveal their range of hardnesses, even when the SF contributions are removed.

\item A new classification based on the clustering of seven emission lines with removed SF reveals the existence of a third class of objects on the Seyfert/LINER branch, lying near its base and extending the traditional LINER category towards lower [OI]/H$\alpha$ ratios. The traditional and newly identified population of LINER-like galaxies on average differ in the hardness of the ionization field, the new population being softer.

\item Hosts of Seyferts and soft LINERs are predominantly star-forming galaxies, leaning somewhat into the green valley, whereas the hosts of hard LINERs have a wide range of sSFRs, from star-forming to quiescent. When the full census of LINERs is considered (i.e., including those from the  ``composite'' region), LINERs are overwhelmingly star formers. 

\item Hosts of both classes of LINERs have their doppelgangers (in terms of the stellar populations and the dust content) among the non-LINERs,  The existence of similar galaxies with and without LINER lines may suggest a transient, rather than a secular phenomenon.

\end{enumerate}

\section{Acknowledgements} \label{sec:acknowl}
This research made use of Astropy, a community-developed core Python package for Astronomy \citep{astropy2013}.
CJA performed part of this work with support as a Master's Fellowship recipient from the Indiana Space grant Consortium and acknowledges John J. Salzer and David Carr for thoughtful discussions. The construction of GSWLC used in this work was funded through NASA awards NNX12AE06G and 80NSSc20K0440. We thank the anonymous referee for their comments as they helped to better contextualize the results described and to improve the clarity of the work.

Funding for SDSS-III has been provided by the Alfred P. Sloan Foundation, the Participating Institutions, the National Science Foundation, and the U.S. Department of Energy Office of Science. The SDSS-III web site is http://www.sdss3.org/.

SDSS-III is managed by the Astrophysical Research Consortium for the Participating Institutions of the SDSS-III Collaboration including the University of Arizona, the Brazilian Participation Group, Brookhaven National Laboratory, Carnegie Mellon University, University of Florida, the French Participation Group, the German Participation Group, Harvard University, the Instituto de Astrofisica de Canarias, the Michigan State/Notre Dame/JINA Participation Group, Johns Hopkins University, Lawrence Berkeley National Laboratory, Max Planck Institute for Astrophysics, Max Planck Institute for Extraterrestrial Physics, New Mexico State University, New York University, Ohio State University, Pennsylvania State University, University of Portsmouth, Princeton University, the Spanish Participation Group, University of Tokyo, University of Utah, Vanderbilt University, University of Virginia, University of Washington, and Yale University.

\bibliographystyle{aasjournal}
\bibliography{refs}

\appendix
\section{Alternative matching without a S/N requirement on subtracted fluxes}\label{app:no_sn}
\begin{figure}[t!]
        \begin{center}
            \epsscale{1.2}
            \plotone{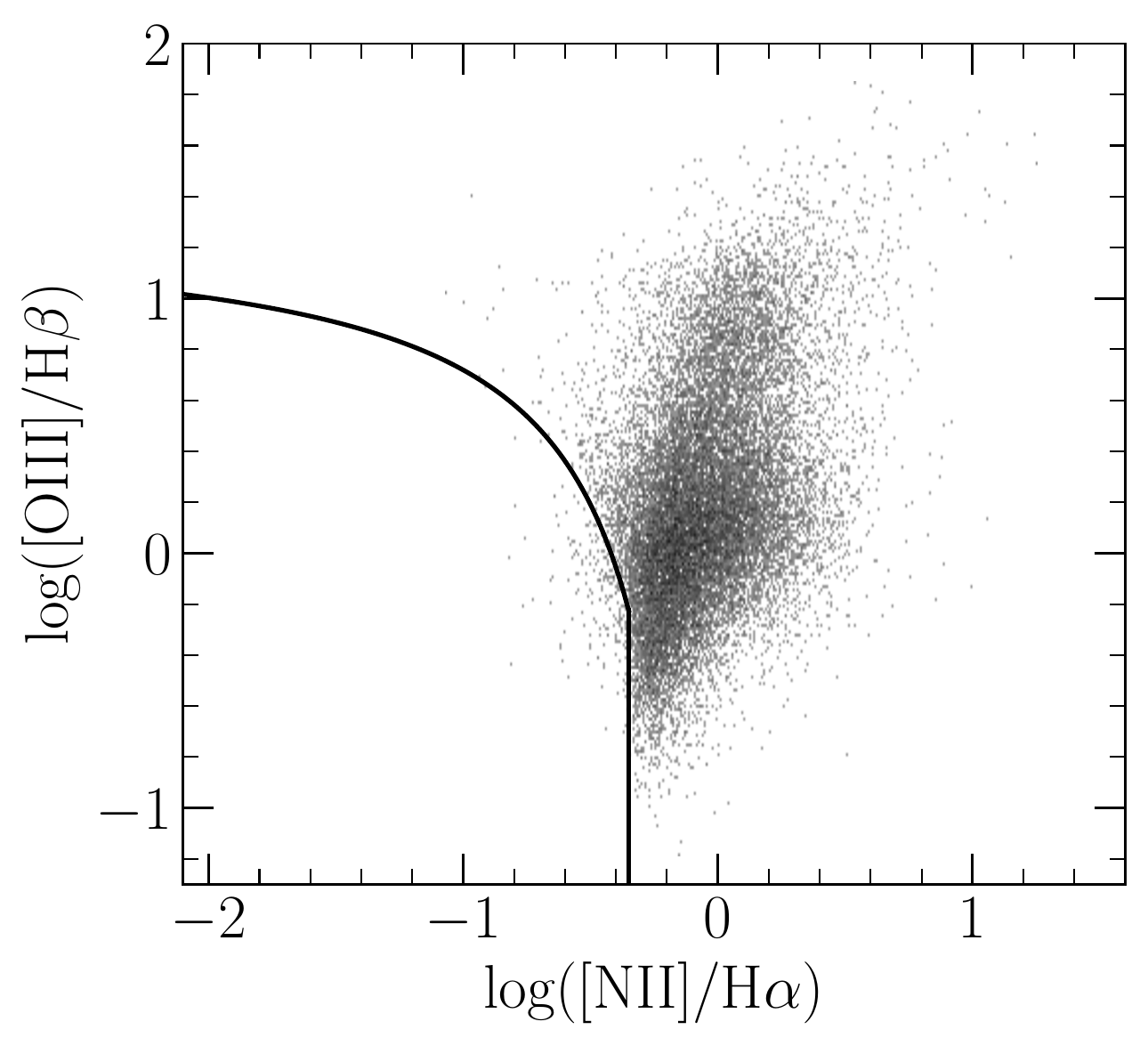}     
            \caption{BPT diagram for pure Seyferts/LINERs  (emission lines without SF or stellar contribution) when the subtracted fluxes are not required to have S/N$>2$ during the matching procedure.   \label{fig:bpt_no_sn_req}}
        \end{center}
        \end{figure}

In this work, we matched Seyfert/LINERs to non-S/Ls based on global properties and required that the subtracted fluxes for the pure S/Ls have S/N$>2$, i.e., that the AGN component is significant, based on the idea that this is required for an object to belong to such a distinct feature such as the S/L branch in the first place Here we check if this requirement of positive, S/N$>2$ fluxes would bias the ultimate results in some capacity. We have performed the matching procedure without enforcing this requirement and present the pure Seyfert/LINERs on the BPT diagram in Figure \ref{fig:bpt_no_sn_req}, and, as is necessary, keep only the objects which have S/N$>2$ in all four BPT lines (22\%). It might be expected that the strongest Seyferts/LINERs (their line ratios being high in [OIII]/H$\beta$ and/or [NII]/H$\alpha$) would be the ones which preferentially remain, but this does not happen. Similar extent of line ratios and similar S/L morphologies are observed in Figure \ref{fig:bpt_no_sn_req} as in Figure \ref{fig:bpt_gen}(b).

We also find that the grid of displacements is similar to that shown in Figure \ref{fig:bpt_gen}(c), so there is no clear dependence on the initial BPT position in determining whether or not a Seyfert/LINER will still have a fainter match, suggesting the main results in this work are not comparatively biased. Finally, the observed trends in ionization parameter, oxygen abundance, electron density, and hardness are effectively the same as what we find in the main results of this work, though we do not show 

We conclude that by requiring the fluxes to be S/N$>2$ as part of the matching procedure, we effectively improve the match quality (even if it has a higher $d$), as the match ought to have less flux in each line, and ultimately, we increase the number of objects which can be studied, allowing us to more thoroughly characterize the trends and results presented in this work.

\section{Alternative matching using fiber-based quantities} \label{app:altmatch}
\begin{figure}[t!]
        \begin{center}
            \epsscale{1.2}
            \plotone{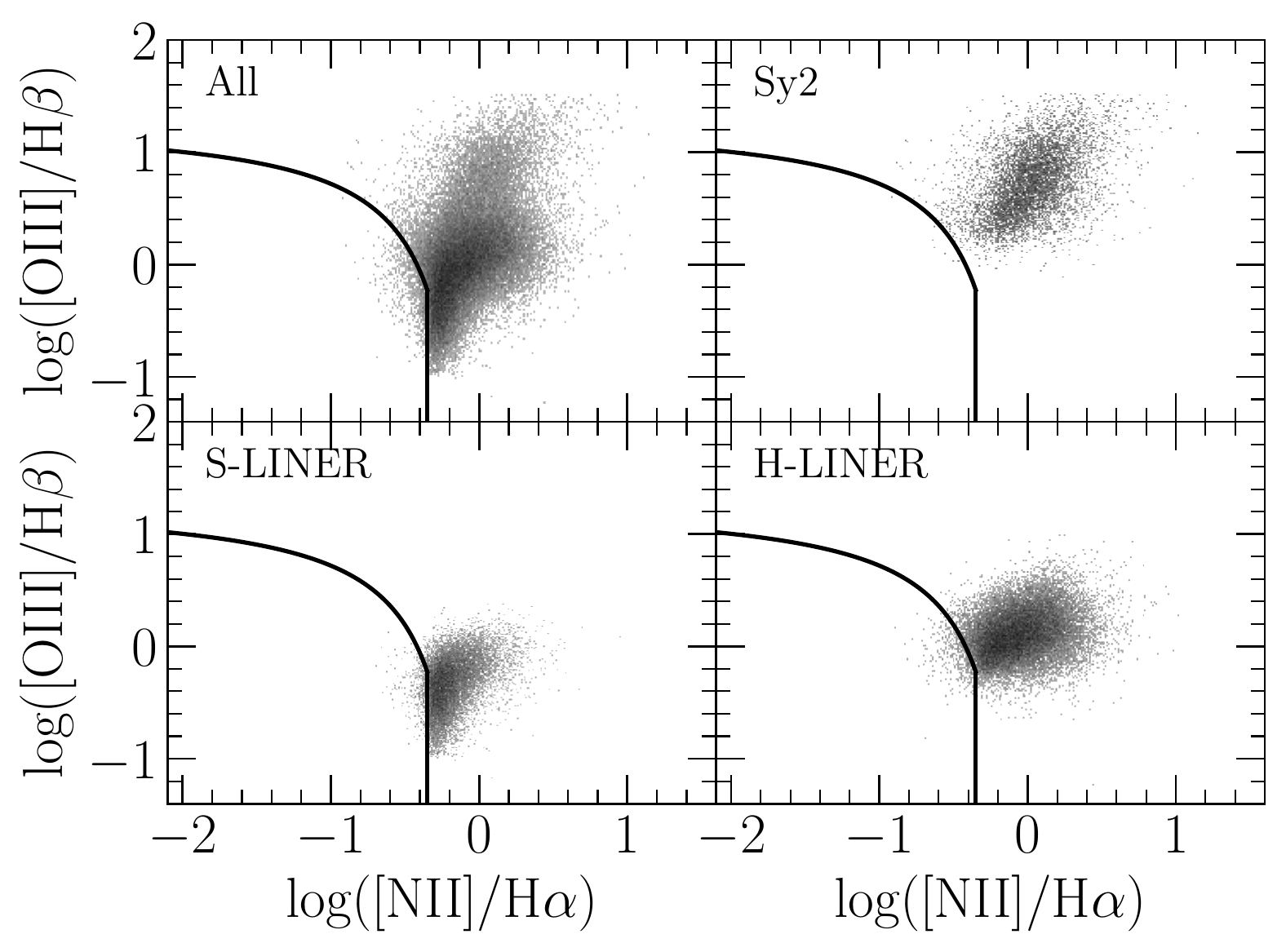}     
            \caption{BPT diagrams where the star formation component in Seyferts/LINERs is removed with a matching procedure based on fiber-based quantities related to host star formation properties: D4000 and H$\delta$ Lick index. The morphology of the S/L branch is similar to that seen in Figure \ref{fig:bpt_clusters}.   
            \label{fig:bpt_d4hd}}
        \end{center}
        \end{figure}

\begin{figure}[t!]
        \begin{center}
            \epsscale{1.2}
            \plotone{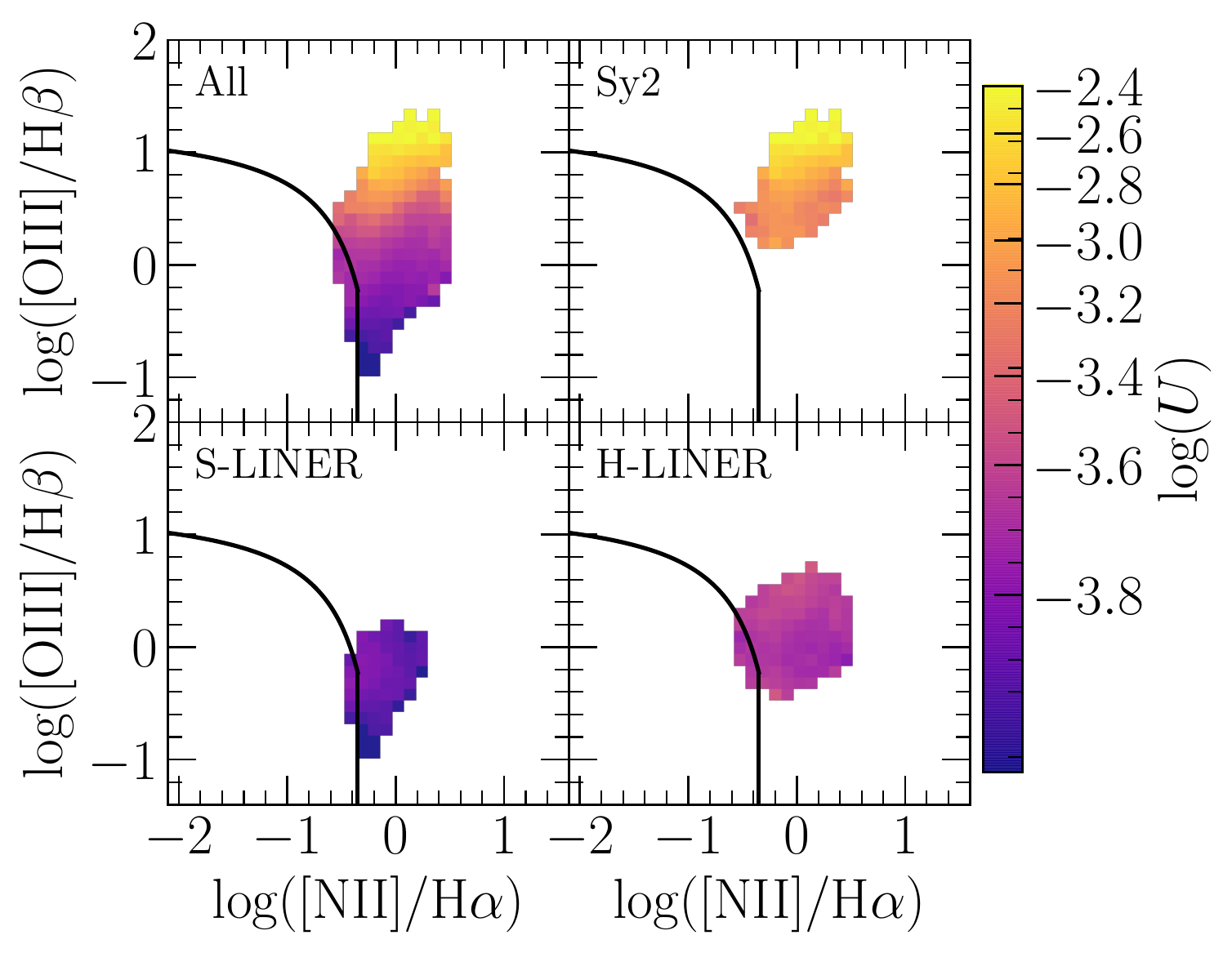}     
            \caption{BPT diagrams for pure Seyferts/LINERs (emission lines without SF or stellar contribution) as determined using fiber-based quantities related to star formation and color-coded by $\log(U)$. \label{fig:bpt_U_ccode_sub_d4hd}}
        \end{center}
        \end{figure}

\begin{figure}[t!]
        \begin{center}
            \epsscale{1.2}
            \plotone{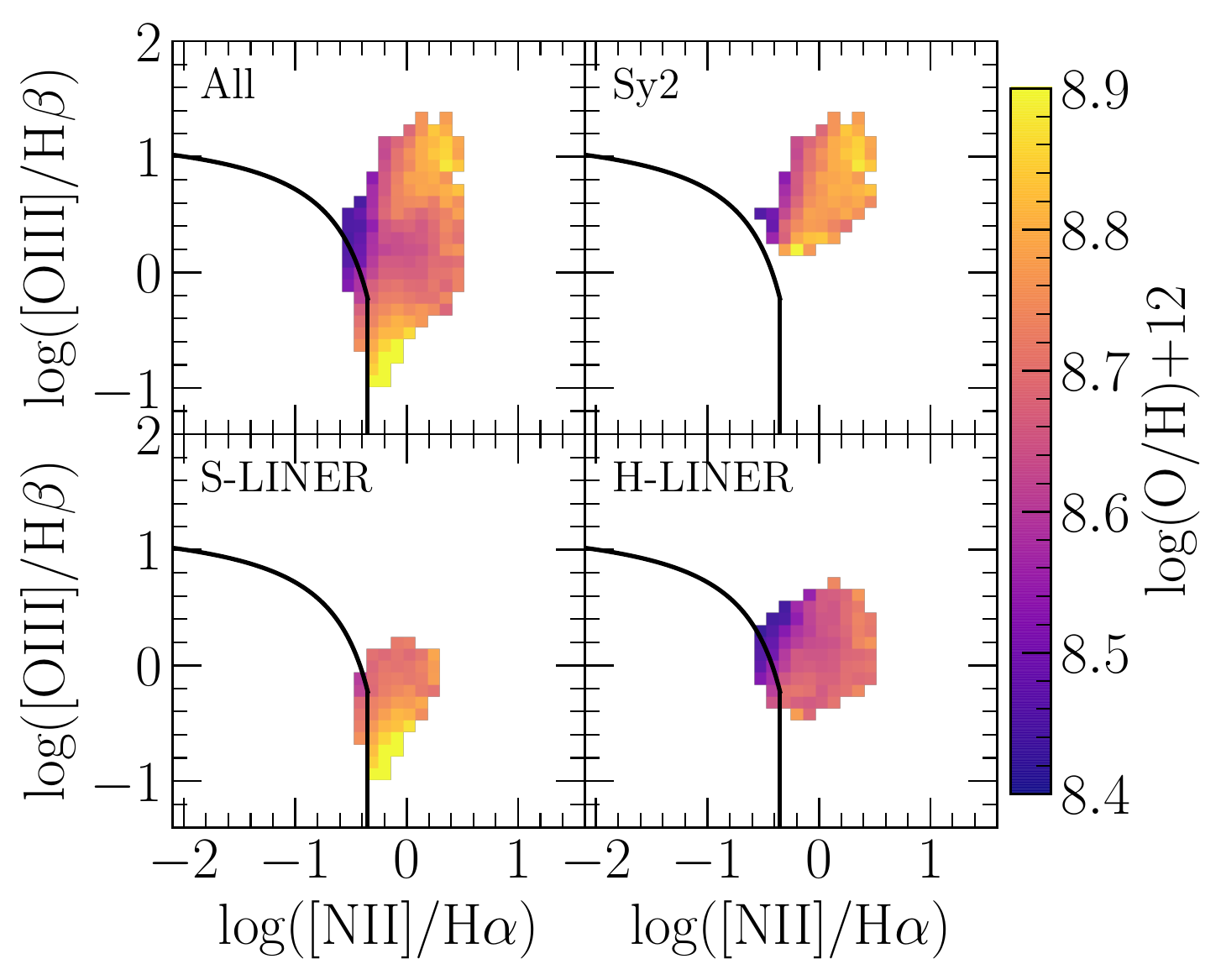}     
            \caption{BPT diagrams for pure Seyferts/LINERs  (emission lines without SF or stellar contribution) as determined using fiber-based quantities related to star formation and color-coded by oxygen abundance. \label{fig:bpt_logoh_ccode_sub_d4hd}}
        \end{center}
        \end{figure}

In this work, we have matched Seyfert/LINERs to non-S/L objects based on redshift, mass contained in the spectroscopic fiber, as well as global quantities which are derived via the SED fitting: star formation rate, the stellar mass, and the stellar continuum dust attenuation. As it is possible that in some cases the majority of star formation in a disk galaxy is taking place in the disks rather than in the central regions primarily probed by the 3$''$ SDSS spectroscopic fiber, we investigated the alternative matching procedure with D$_{4000}$ and H$\delta$ Lick index instead of the global SFR. We find relatively little difference between the results. We show the BPT diagram produced via this method in Figure \ref{fig:bpt_d4hd}. We split Seyfert/LINERs into the same three different subgroups using the method described in Section \ref{sec:agntypes}. The different types occupy similar regions to those found with the SFR-based matching procedure. The emission line properties of the three groups are qualitatively similar to that found with the SFR-based matching, as can be seen by BPT color-coded by the ionization parameter in Figure \ref{fig:bpt_U_ccode_sub_d4hd} and by oxygen abundance in Figure \ref{fig:bpt_logoh_ccode_sub_d4hd} in comparison to Figures \ref{fig:bpt_U_ccode} and \ref{fig:bpt_logoh_ccode}. In this alternative matching scheme, the use of the fiber stellar continuum dust attenuation ($\tau_{V}$ from MPA/JHU) instead of the global value $A_{V}$ does not change the main results.

\section{Quality of Multi-Dimensional Matching} \label{app:dists}
\begin{figure}[t!]
        \begin{center}
           \epsscale{1.2}
            \plotone{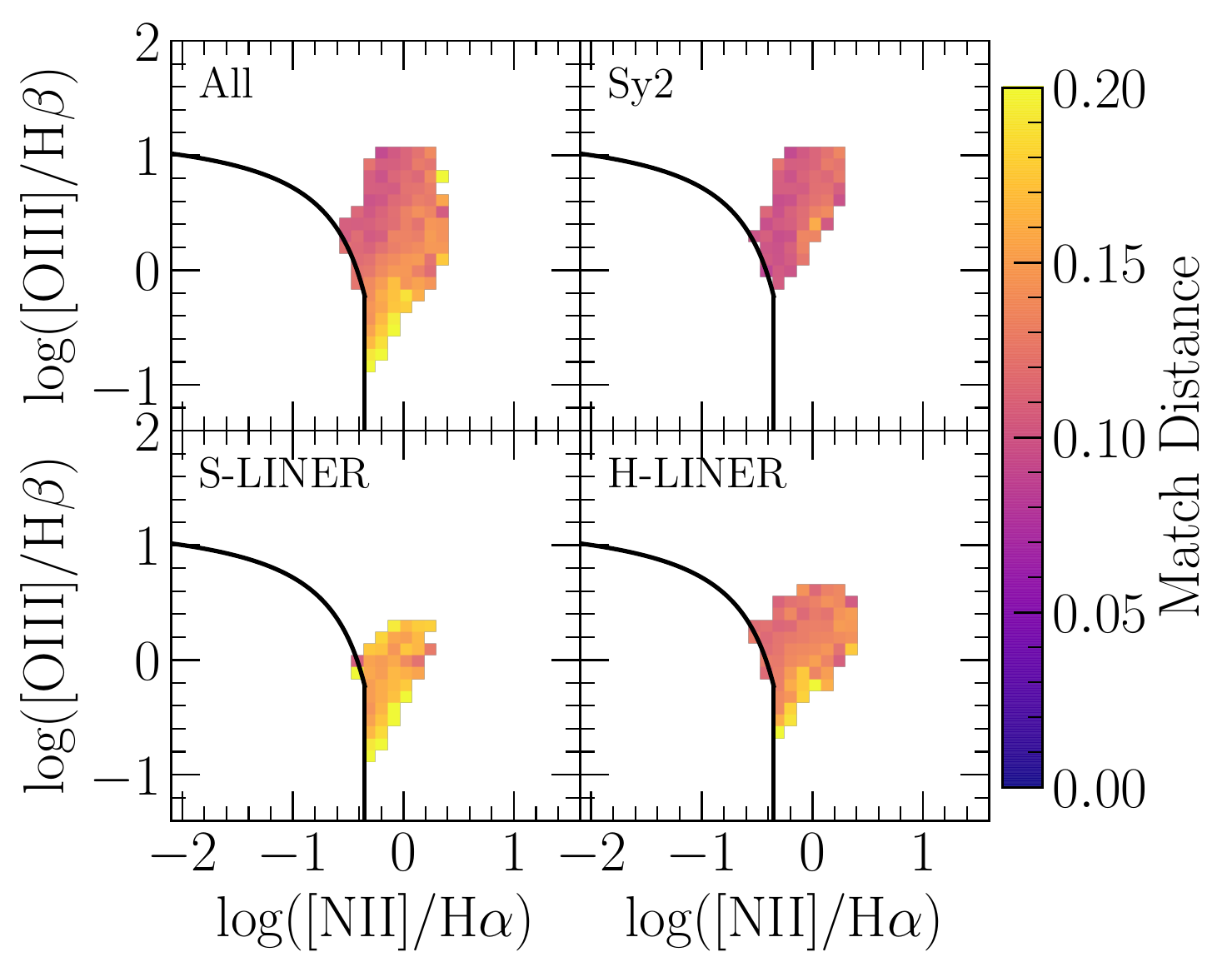}            
            \caption{BPT diagrams prior to removal of SF component, color-coded by the average distance to the non-S/L match.\label{fig:match_dists}}
        \end{center}
        \end{figure}

We present the BPT diagrams color-coded by the average match distances in Figure \ref{fig:match_dists}. The match distances for H-LINERs are not significantly higher than Sy2s, suggesting the hosts for each are similarly well matched. The median match distances for Sy2s, S-LINERs, and H-LINERs are 0.11, 0.16, and 0.13. To put these match distances in perspective, we also perform self-matching for the target sample, finding which other S/L is the most similar using the distance metric described in Section \ref{sec:matching}. The median distance for the S/L self-matching is 0.05. While the median match distance we find for the SF matches is $\sim2-3$ times that, it should be noted that the pure S/L fluxes must have S/N$>2$, increasing the distance. Indeed, if  the S/N$>2$ requirement was not implemented, the typical distance was closer to that for the S/L self-matching. It may be the case that lower-quality matches (higher $d$) potentially bias the results described here. To test this, we have implemented thresholds in $d$, restricting the analysis to those with $d$ less than 0.1, 0.15, and 0.2. We find no substantial difference in the morphology in the S/L branch, suggesting the results are robust.

We also perform our nominal matching to the second best candidate, from which we determine that the random errors in the decontaminated line ratios log([NII]/H$\alpha$) and log([OIII]/H$\beta$) must be $\lesssim$0.12 and $\lesssim$0.20 dex, respectively.

\section{Ionization parameter and metallicity trends using observed lines}\label{app:presub}
\begin{figure}[t!]
        \begin{center}
            \epsscale{1.2}
            \plotone{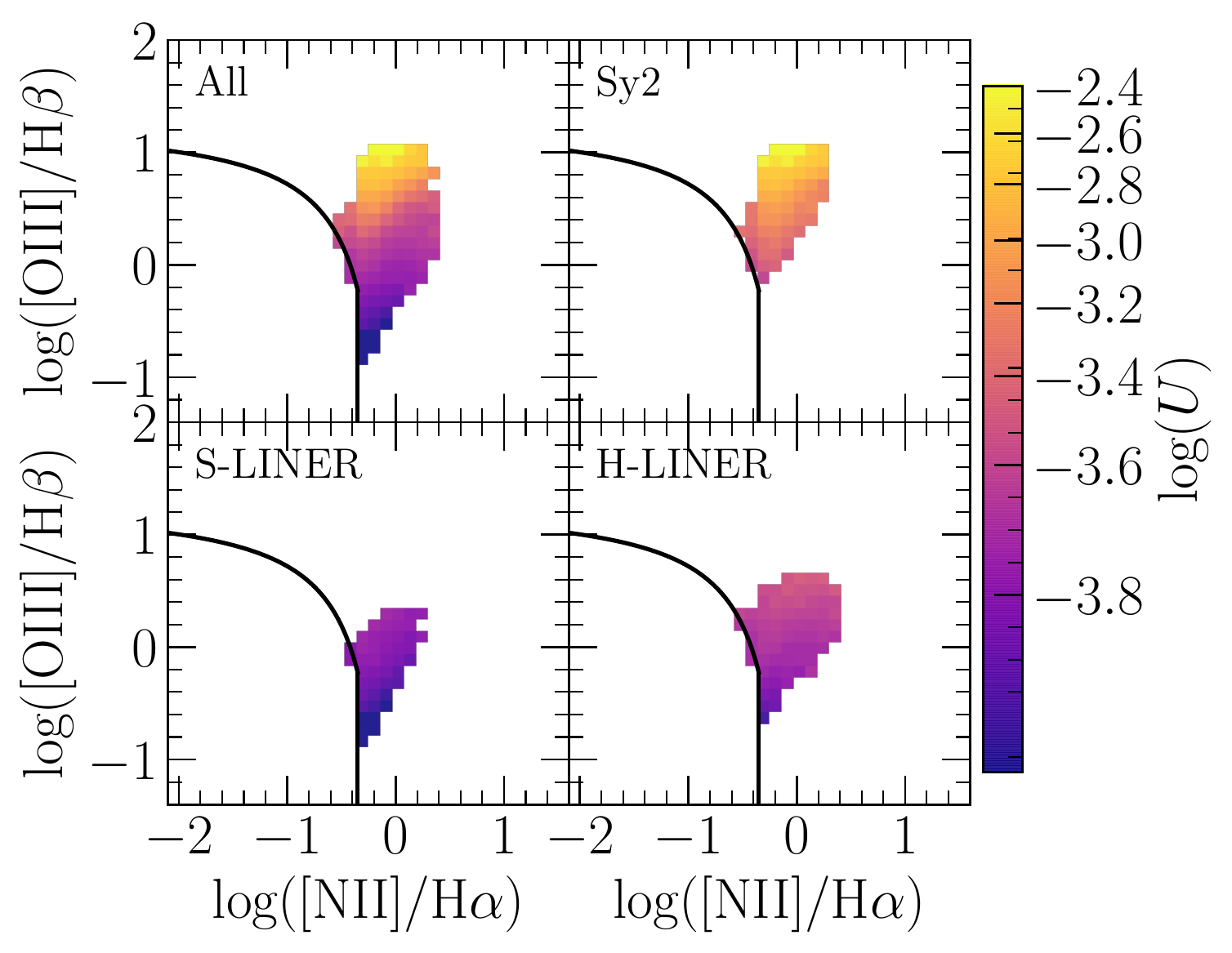}     
            \caption{ BPT diagrams color-coded by log($U$), a proxy for ionization parameter, using the observed lines contaminated by SF. The trends for log($U$) appear to be tilted such that it increases transverse to the direction of the S/L branch, as opposed to the version with SF removed where it relatively cleanly increases based just on [OIII]/H$\beta$. \label{fig:bpt_U_ccode_presub}}
        \end{center}
        \end{figure}

\begin{figure}[ht!]
        \begin{center}
            \epsscale{1.2}
            \plotone{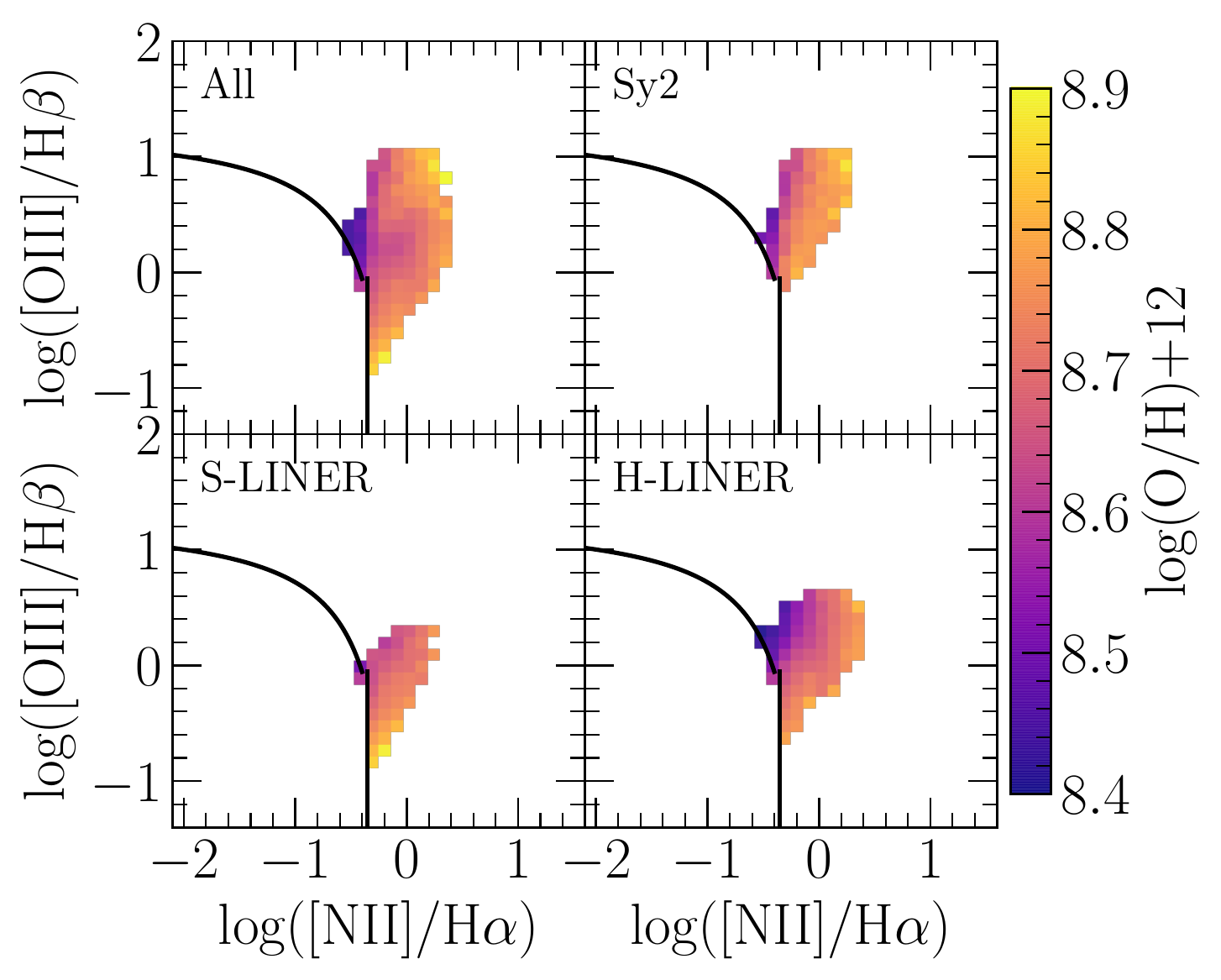}     
            \caption{Pre-subtraction of the observed lines contaminated by SF. We do not see as clean of trends for the Sy2s as in the version with SF removed and no discernible trend is present for the S-LINERs. The trend appears relatively similar for the H-LINERs. \label{fig:bpt_logoh_ccode_presub}}
        \end{center}
        \end{figure}

Here we look at the ionization parameter and metallicity trends using observed lines, i.e, without SF removal (Figures \ref{fig:bpt_U_ccode_presub} and \ref{fig:bpt_logoh_ccode_presub}). For log($U$), the Sy2 again have log($U$) values higher than S-LINERs and H-LINERs, but the increase upwards with [OIII]/H$\beta$ is not as clean and appears to be tilted more than in SF-removed version. There are not clear differences for the S-LINERs and H-LINERs before and after SF removal for log($U$) and they appear to occupy the same ranges of log($U$) as their pure counterparts. When we look at the trends with log(O/H) for the pre-subtraction branch, we find much starker differences with respect to the pure version. We do not see as strong of a dependence for the Sy2s and the log(O/H) values for all three groups are higher than the pure counterparts, likely because of the additional contributions from star-formation to the [NII] and [OII] lines used in the calculation of log(O/H)+12. It is possible that the proposed lack of metal-poor AGNs (e.g., \citet{groves2006}, \citet{kawasaki2017}) is in part a result of an inability to remove SF contributions.

\end{document}